\documentclass[aps,pra,twocolumn,nofootinbib,superscriptaddress,preprintnumbers,floatfix,10pt]{revtex4-2} 

\usepackage[english]{babel} 
\usepackage[utf8]{inputenc}
\usepackage[T1]{fontenc}
\usepackage{lmodern}

\usepackage{amsmath,amsfonts,amssymb}
\usepackage{braket}
\usepackage{siunitx}

\usepackage{graphicx}
\usepackage{booktabs}
\usepackage{tabularx}
\usepackage{longtable}
\usepackage{array}

\usepackage[colorlinks=true, linkcolor =blue, citecolor=green!70!black]{hyperref}

\newcommand{\I}{\mathrm{i}}   
\newcommand{\E}{\mathrm{e}}   
\newcommand{\xit}{\tilde{\xi}} 
\newcommand{\taut}{\tilde{\tau}}
\newcommand{\varphit}{\tilde\varphi}
\newcommand{\rmT}{\mathrm{T}}
\newcommand{\rmR}{\mathrm{R}}
\DeclareMathOperator{\He}{H}
\DeclareMathOperator{\artanh}{artanh}
\DeclareMathOperator{\Fhyp}{{}_{2}F_{1}}

\usepackage{xcolor}
\usepackage{soul} 
\soulregister\ref7
\soulregister\eqref7
\soulregister\cite7

\begin{document}

\title{Alternating Squeezing and Photon Subtraction for Optical Cat-State Generation}

\author{Johanna Conrad}
\email{johanna.conrad@uni-jena.de}
\affiliation{Institute of Applied Physics, Abbe Center of Photonics, Friedrich-Schiller-University Jena, Albert-Einstein-Str. 6, 07743 Jena, Germany}
\affiliation{Fraunhofer Institute for Applied Optics and Precision Engineering IOF, Albert-Einstein-Str. 7, 07745 Jena, Germany}

\author{Fabian Steinlechner}
\affiliation{Institute of Applied Physics, Abbe Center of Photonics, Friedrich-Schiller-University Jena, Albert-Einstein-Str. 6, 07743 Jena, Germany}
\affiliation{Fraunhofer Institute for Applied Optics and Precision Engineering IOF, Albert-Einstein-Str. 7, 07745 Jena, Germany}

\author{Ren\'{e} Sondenheimer}%
\email{rene.sondenheimer@uni-jena.de}
\affiliation{Institute of Condensed Matter Theory and Optics, Friedrich-Schiller-University Jena, Max-Wien-Platz 1, Jena, 07743, Germany}
\affiliation{Fraunhofer Institute for Applied Optics and Precision Engineering IOF, Albert-Einstein-Str. 7, 07745 Jena, Germany}

\begin{abstract}
Schrödinger cat states are important non-Gaussian resources for quantum information processing and quantum-enhanced metrology. Although photon subtraction from squeezed vacuum states has proven useful for generating such states in the optical domain, scaling to large amplitudes while maintaining high fidelity and heralding probability remains challenging. To address these obstacles, we analyze alternating squeezing and photon-number-subtraction operations. Rather than optimizing for a single heralding outcome, we formulate cat-state generation as a fidelity-constrained multibranch problem, jointly optimizing the total probability of all photon-number outcomes satisfying a prescribed fidelity threshold at one common experimental setting. Already for plain photon subtraction, this provides a more operational measure of the usable generation probability and can favor parameter configurations that differ from single-branch optima. Considering experimental constraints on available squeezing and photon-number-resolving capability, we demonstrate improvements in fidelity and success probability compared with plain photon subtraction, without necessarily increasing the required squeezing resources. Without restrictions on photon-number resolution, we find that the optimal architecture of the alternating scheme reduces to one of two single-subtraction endpoint configurations, depending on the available squeezing resources. When photon-number resolution is additionally limited, genuinely alternating branches can become advantageous. We derive analytical expressions for the conditional states, cat-state fidelities, and heralding probabilities, enabling efficient multibranch optimization and a systematic analysis of resource trade-offs, sensitivity to experimental parameter deviations, and heralding loss. Together, these results identify the architecture best suited to a given resource budget.
\end{abstract}

\maketitle

\section{Introduction} 
Quantum superposition allows systems to occupy linear combinations of distinguishable states. Schrödinger cat states provide a paradigmatic example consisting of superpositions of two coherent states $\ket{\alpha}$ and $\ket{-\alpha}$ \cite{dodonov1974}. While the superposition is nonclassical, the individual coherent states are often regarded as the most classical-like pure states of the quantum harmonic oscillator~\mbox{\cite{Glauber1963,Glauber1963-2,Sudarshan1963}}. Furthermore, cat states are non-Gaussian resource states that enable capabilities inaccessible with Gaussian states and Gaussian operations alone \cite{Eisert2002,Niset2009}. These include schemes for error correction in optical quantum computing \cite{Bergmann2016}, access to entanglement and correlations beyond a Gaussian description \cite{Gomes2009}, entanglement distillation and swapping \cite{Eisert2002,DellAnno2019,Takahashi2010}, and improved continuous-variable teleportation and cloning protocols \cite{DellAnno2007,Nha2010,He2024}. Additionally, in quantum metrology, cat states can enhance interferometric phase sensitivity \cite{Yurke1986,Hudelist2014,Knott2016}. These applications have motivated extensive efforts to generate optical cat states with high fidelity and high heralding probability.

The first proposal of generating optical cat states via photon subtraction was made in 1997 \cite{daknaGeneratingSchrodingercatlikeStates1997}, and experimentally demonstrated by Ourjoumtsev et al. in 2006 \cite{ourjoumtsevGeneratingOpticalSchrodinger2006}. This scheme relies on subtracting a well‑defined number of photons from a squeezed‑vacuum state. The conditionally generated state is a superposition of squeezed Fock states that can closely resemble an optical cat state. Especially for $|\alpha|<1$, these conditional states can approximate so-called Schrödinger kitten states with fidelities exceeding $0.99$ \cite{ourjoumtsevGeneratingOpticalSchrodinger2006}.  
However, even in the absence of optical loss, experimentally realistic resources remain finite and depend on the platform and measurement approach, with representative demonstrated squeezing values of $|\xi|=1.72$ ($\SI{15}{\decibel}$) in the cw regime \cite{vahlbruchDetection15DB2016} and $|\xi|=0.92$ ($\SI{8}{\decibel}$) in the pulsed regime \cite{Kalash2025}, as well as a finite number of subtracted photons (e.g., four \cite{Endo2025}), causing the fidelity to decrease rapidly with the size of the cat state. 
Under these constraints, the fidelity falls below $0.95$ once $|\alpha|>2$, illustrating how finite-order photon subtraction and limited squeezing restrict the generation of high‑quality, large-amplitude cat states. Moreover, the experimental constraints also restrict the heralding probabilities of generating large-$\alpha$ cat states. Even with large multi-photon subtraction enabled by novel photon number-resolving (PNR) detection technologies \cite{Sauer2023,Schapeler2024,Stasi2023,Hao2024,Schmidt2018,Li2024,DeLucia2024}, the heralding probability decreases sharply with the number of subtracted photons. 
Consequently, various extensions of the plain photon subtraction scheme and alternative setups have been proposed and demonstrated.

We can group purely optical efforts into approaches based on conditional measurements and post-selecting the successful events \cite{daknaGeneratingSchrodingercatlikeStates1997,ourjoumtsevGeneratingOpticalSchrodinger2006, Neergaard-Nielsen2006, Gerrits2010, Endo2025, Takahashi2008, Kuts2022, Podoshvedov2023, Lee2012, Ourjoumtsev2007, ulanovLosstolerantStateEngineering2016, Huang2015, Takase2021, Mikheev2019, Chen2024}, and schemes that amplify small, pre-prepared cat states \cite{Lund2004, Adam2010, etesseExperimentalGenerationSqueezed2015, Sychev2017, Sychev2018, Eaton2019}. Deterministic methods based on light-matter interaction have also been explored typically at the cost of increased experimental complexity and stringent requirements on coupling and loss \cite{Wang2005, Hacker2019, Li2024}. Therefore, they are only briefly mentioned here.

In plain photon subtraction \cite{daknaGeneratingSchrodingercatlikeStates1997,ourjoumtsevGeneratingOpticalSchrodinger2006, Neergaard-Nielsen2006, Gerrits2010, Endo2025} and its extensions \cite{Takahashi2008,Kuts2022, Podoshvedov2023}, the non-Gaussianity in the output state is created by the non-Gaussian operation of photon subtraction, e.g., implemented by splitting off a small fraction of the field on a beam splitter and heralding on a Fock state projection in the auxiliary arm. 
Subsequent studies have focused on achieving higher state fidelities through ancilla-assisted photon subtraction \cite{Takahashi2008} and cascaded photon subtraction stages \cite{Kuts2022,Podoshvedov2023}. 
Lee et al. proposed a single-photon interferometric scheme, where heralded single photons, generated via SPDC, interfere at the same crystal and are subsequently subjected to a single-photon subtraction \cite{Lee2012}. This scheme implements a superposition operation and can be considered a specialized instance of the scheme presented in this manuscript.

Further approaches exchange the roles of the (Gaussian) input and (non-Gaussian) operation and utilize Fock states as a non-Gaussian input that are split on a beam splitter and conditioning one of the output modes by a quadrature measurement as a Gaussian operation in the auxiliary arm \cite{Ourjoumtsev2007, ulanovLosstolerantStateEngineering2016}. 
Other proposals involve mixing two quantum states on a beam splitter, e.g., squeezed states \cite{Huang2015, Su:2019vdv, Takase2021}, or squeezed states and Fock states \cite{Mikheev2019, Kuts2022, Chen2024, Nauth:2026jyy, luo2025}. Alternative schemes combine photon addition and subtraction operations on coherent or displaced photon-number states to generate large cat state superpositions \cite{Podoshvedov2011}. In practical applications of cat states, the heralding probability of the conditioning measurement is a critical limiting factor. Generalized photon subtraction \cite{Takase2021, takaseGenerationSchrodingerCat2022} offers a potential solution, demonstrating a generation-rate improvement of more than three orders of magnitude for large cat states ($|\alpha| \geq  2$) compared to plain photon subtraction. More recently, Winnel et al. proposed optical protocols for the deterministic preparation of large-amplitude squeezed cat states using Gaussian operations and photon-number-resolving measurements~\cite{Winnel:2023tzb}. Their first scheme employs Fock states as non-Gaussian input resources, allowing squeezed-cat states to be prepared without postselection on specific measurement outcomes. Their second scheme also permits arbitrary states of definite photon-number parity, including squeezed vacuum, although obtaining sufficiently large cat amplitudes can require substantial input resources or postselection of favorable outcomes.

Another group of schemes that has emerged focuses on enlarging Schrödinger kitten states with small $|\alpha|$ \cite{Adam2010}. A prominent example is the cat breeding protocol \cite{Lund2004} which has been experimentally demonstrated in Refs.~\cite{etesseExperimentalGenerationSqueezed2015,Sychev2017, Sychev2018}. In 2019, a photon catalysis scheme was presented, which starts from a coherent state and uses sequential mixing with single-photon Fock states and $n$\nobreakdash-photon subtraction \cite{Eaton2019}.

The previously discussed approaches have led to important advances in optical cat state generation. However, when restricting the input states to deterministically available Gaussian states, in particular squeezed vacuum, preparing large-amplitude cat states ($|\alpha|>2$) with simultaneously high fidelity and sufficient heralding probabilities remains challenging and often requires significant experimental resources.

Previous studies of cat-state generation by plain photon subtraction have typically characterized or optimized a specified photon-number heralding outcome. Multiple PNR outcomes have been exploited in related contexts, including sequential photon subtraction that distributes a fixed total subtraction order among several detectors \cite{Podoshvedov2023}, adaptive non-Gaussian state engineering \cite{Crescimanna:2025lwf}, and multioutcome optimization in general Gaussian-PNR circuits \cite{Ismailzadeh:2026yye}. Here, we formulate cat-state generation as a fidelity-constrained multibranch problem. For one common experimental setting, all photon-number outcomes whose conditional states satisfy a prescribed fidelity threshold are accepted, and their probabilities are jointly optimized to maximize the total success probability. Already for plain photon subtraction, this provides a more operational measure of the usable generation probability and can favor parameter configurations that differ from single-branch optima. 
We are not aware of any previous work employing such a fidelity-constrained multibranch optimization.

Starting from plain photon subtraction of squeezed vacuum, we interleave squeezing operations with beam-splitter-based PNR conditioning, leading first to a nonlinear-interferometer-type configuration and, more generally, to alternating sequences of squeezing and photon subtraction. We derive analytical expressions for the resulting conditional states, their fidelities with target cat states, and the corresponding heralding probabilities. These expressions enable an efficient multibranch optimization of the fidelity--success-probability trade-off under experimentally relevant squeezing constraints and allow us to assess the associated photon-number resolution requirements. We find that genuinely alternating subtraction stages are not universally advantageous. Without an explicit restriction on photon-number resolution, the optimal architecture in the investigated lossless regimes reduces to one of two single-subtraction endpoint configurations, depending on the available squeezing resources. When photon-number resolution is additionally limited, as is the case in present-day experiments, genuinely alternating branches can become advantageous. Moreover, we analyze the sensitivity to experimental parameters and heralding loss, and identify the configurations and heralding patterns favored in different resource regimes.

The paper is structured as follows: in Sec.~\ref{sec:PhotonSubtraction}, we briefly review the plain photon subtraction scheme applied to squeezed vacuum states, which serves as a baseline for the more complex schemes. Moreover, we introduce the fidelity-constrained multibranch optimization. We extend this scheme by adding an additional squeezing stage, forming a nonlinear-interferometer-type setup in Sec.~\ref{sec:SqSuSq}. In Sec.~\ref{sec:AlternatingScheme}, we further generalize the protocol by introducing a second photon-subtraction stage after the second squeezer which naturally extends to longer sequences of alternating squeezing and subtraction operations. The performance of the different schemes and the robustness of their optimized parameter settings are compared in detail in Sec.~\ref{sec:Robustness}. We conclude in Sec.~\ref{sec:Conclusion}.

\section{Plain Photon Subtraction}
\label{sec:PhotonSubtraction}
In the first part of this section, we review the main steps of calculating the output state after photon subtraction from single-mode squeezed vacuum following Dakna et al.~\cite{daknaGeneratingSchrodingercatlikeStates1997}, and highlight equivalent formulations of this state and their utility for further analysis. Then, we examine the fidelity between the output state and cat states as well as the probability to herald this state. Finally, we analyze the fidelity--success-probability trade-off.

We consider an experimentally concrete realization of photon subtraction in the optical domain, implemented by mixing the signal mode with vacuum on a beam splitter and conditioning on an $m$-photon detection event in the reflected arm using photon-number-resolving (PNR) detection. For beam-splitter transmittance $T<1$, the resulting conditional map deviates from an ideal application of $\hat{a}^m$ (often interpreted as subtracting $m$ photons) by including an additional attenuation associated with the transmittance. While the common $\hat{a}^m \hat{\rho} \hat{a}^{\dagger m}$ description is approached in the limit $T\to 1$, the heralding probability simultaneously vanishes as no photons are tapped off into the reflected port. Hence, photon subtraction is realized operationally as a beam-splitter-based conditional preparation with PNR heralding rather than as an exact $\hat{a}^m$ map in an all-optical setting. Throughout this manuscript, we therefore use the term ``photon subtraction'' in this broader sense.

\begin{figure}[t]
    \centering
    \includegraphics[width=0.5\linewidth]{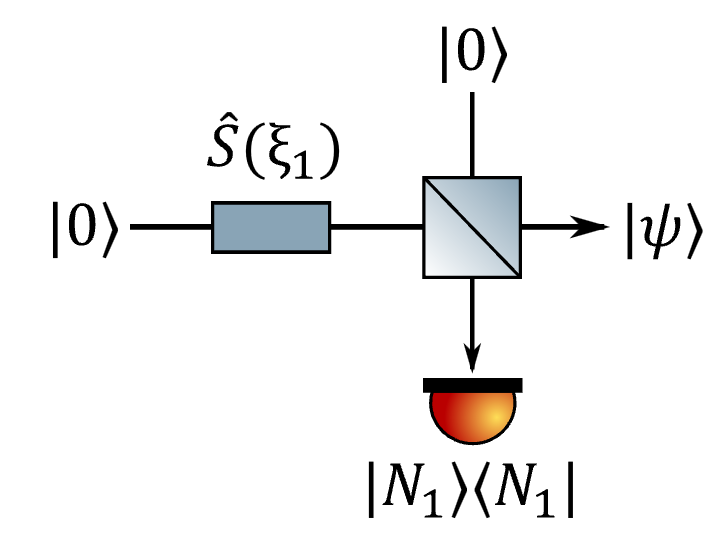}
    \caption{Setup of plain photon subtraction. A squeezed-vacuum input with squeezing parameter $\xi_1$ is mixed with vacuum on a beam splitter of transmittance $T_1$ and $N_1$ photons are detected in the reflected arm.}
    \label{fig:setup_single}
\end{figure}

The basic beam-splitter-based photon-subtraction scheme for a single-mode squeezed-vacuum input is depicted in Fig.~\ref{fig:setup_single}. Conditioning on the detection of $N_1$ photons in the reflected output mode prepares a non-Gaussian state in the transmitted mode, which we characterize in the following. First, we specify the squeezed-vacuum input state in terms of the single-mode squeezing operator defined as
\begin{align}
    \hat{S}(\xi) = \exp\left[\tfrac{1}{2} \left( \xi^* \hat{a}^2 - \xi \hat{a}^{\dagger 2} \right) \right],
\end{align}
with complex squeezing parameter $\xi = |\xi| \E^{\I\varphi}$ and $\hat{a}$ and $\hat{a}^\dagger$ denote the bosonic annihilation and creation operators, respectively. Using its normal-ordered version, ${\hat{S} = \E^{-\frac{1}{2}\tau \hat{a}^{\dagger 2}}\E^{\ln{(1-|\tau|^2)}\frac{1}{2}(\hat{a}^\dagger \hat{a}+\frac{1}{2})}\E^{\frac{1}{2}\tau^*\hat{a}^2}}$ with ${\tau = \E^{\I\varphi}\tanh{|\xi|}}$, it immediately follows a convenient description of a single-mode squeezed vacuum state via 
\begin{align}
 \ket{\xi} = \hat{S}(\xi) \ket{0} = \sqrt[4]{1-|\tau|^2}\; \exp\left(-\tfrac{1}{2}\tau \hat{a}^{\dagger 2}\right) \ket{0}.
\end{align}
Second, a general beam splitter is characterized on the mode level by the unitary scattering matrix \cite{Campos1989} 
\begin{align}
U_{\mathrm{BS}} =
\begin{pmatrix}
\sqrt{T}\,\E^{\I\phi_{\mathrm T}} & \sqrt{1-T}\,\E^{\I\phi_{\mathrm R}} \\
-\sqrt{1-T}\,\E^{-\I\phi_{\mathrm R}} & \sqrt{T}\,\E^{-\I\phi_{\mathrm T}}
\end{pmatrix},
\end{align}
with transmittance $T$, reflectance $1-T$, and phases $\phi_\mathrm{T}$ and $\phi_\mathrm{R}$. 
For consistency with the multistage schemes considered in the following sections, we label the squeezing and beam-splitter parameters of the present setup by the subscript $1$.

Mixing single-mode squeezed vacuum with vacuum at a beam splitter leads to a correlated two-mode Gaussian state in the output modes, $\ket{\xi_1}_{a_0}\ket{0}_{b_0} \to \ket{\Psi}_{a_1,c_1}$. Here, we denote the input modes by $a_0$ and $b_0$ while the output modes are $a_1$ for the transmitted signal and $c_1$ for the reflected signal. Rearranging the resulting exponent and using the generating function of Hermite polynomials, we obtain the output state
\begin{align}\label{eq:stateH}
   \ket{\Psi}_{a_1,c_1} \!&= \sqrt[4]{1-|\tau_1|^2}
    \exp\left[-\tfrac{1}{2} \tau_1 T_1\E^{2\I\phi_{\rmT_{1}}} \hat{a}_1^{\dagger 2}\right] \notag \\
    &\quad \times \sum_{n=0}^\infty \Bigg[ \frac{1}{n!} 
    \He_n \Big(\sqrt{\tfrac{\tau_1 T_1}{2}}\E^{\I\phi_{\rmT_{1}}} \hat{a}_{1}^{\dagger}\Big)  \notag \\
    &\qquad\qquad \times \left(\sqrt{\tfrac{\tau_1 (1-T_1)}{2}} \E^{\I\phi_{\rmR_{1}}} \hat{c}_{1}^{\dagger}\right)^n \Bigg] \ket{0}_{a_1,c_1} .
\end{align}
Conditioning on an $N_1$-photon detection event in mode $c_1$ yields 
${}_{c_1}\!\!\braket{N_1|\Psi}_{a_1,c_1} \sim \He_{N_1}\big( [\tfrac{1}{2}\taut_1]^{1/2} \hat{a}_1^{\dagger}\big) \hat{S}(\xit_1) \ket{0}_{a_1}$ with $\taut_1 = \E^{\I \phi_{\rmT_{1}}}T_1\tau_1$, and hence an effective squeezing parameter $\xit_1 = \E^{\I (\varphi_1 + \phi_{\rmT_{1}})} \artanh(T_1 \tanh\xi_1) \equiv \E^{\I \varphit_1} |\xit_1|$. Since $0<T_1<1$, it follows that $|\taut_1|<|\tau_1|$, i.e., the effective squeezing is reduced compared to the initial squeezing. Moreover, the effective squeezing parameter is bounded by the beam-splitter transmittance, $|\taut_1|<T_1$, as ${|\tau_1|<1}$. That the conditioned state depends only on the phase combination $\varphi_1+\phi_{\rmT_{1}}$ simply reflects the fact that the beam-splitter phase can be absorbed into the squeezing phase, or vice versa. Thus, we choose $\phi_{\rmT_{1}}=0$ without loss of generality in the following.

Owing to the matched parameter structure by the same effective squeezing parameter $\xit_1$ that appears in the Hermite-polynomial argument (associated with $\taut_1$) and in the squeezing operator, the Hermite polynomial $\He_{N_1}\big( [\tfrac{1}{2}\taut_1]^{\frac{1}{2}}\, \hat{a}_1^{\dagger}\big)$ and $\hat{S}(\xit_1)$ can be reordered up to an overall prefactor. Thus, the conditional state can also be expressed as a squeezed Hermite-polynomial state, ${}_{c_1}\!\!\braket{N_1|\Psi}_{a_1,c_1} \sim \hat{S}(\xit_1) \He_{N_1}\big( [\tfrac{1}{2}\taut_1]^{\frac{1}{2}}\, \hat{a}_1^{\dagger}\big) \ket{0}_{a_1}$.
In this ordering, the norm of the state can be evaluated directly, which yields the heralding probability of the conditional measurement,
\begin{align}
 P_{N_1} &= || {}_{c_1}\!\!\braket{N_1|\Psi}_{a_1,c_1} ||^2 \notag \\
 &= \sqrt{\frac{1-\frac{|\taut_1|^2}{T^2}}{1-|\taut_1|^2}} \bigg[\frac{1-T_1}{T_1} \frac{|\taut_1|^{2}}{1-|\taut_1|^2} \bigg]^{N_1} \notag \\ 
 &\quad \times \!\Fhyp\!\! \bigg(\frac{1-N_1}{2},-\frac{N_1}{2},1,\frac{1}{|\taut_1|^{2}}\bigg),
\label{eq:ProbPS}
\end{align}
with hypergeometric function $\Fhyp$. 
Finally, we can express the normalized output state in mode $a_{1}$ as
\begin{align}
 \ket{\psi_{N_1}} &= \sqrt{\frac{(1-|\taut_1|^2)^{N_1}}{N_1! (2|\taut_1|)^{N_1} \Fhyp\!\Big(\frac{1-N_1}{2},-\frac{N_1}{2},1,\frac{1}{|\taut_1|^{2}}\Big)}} \notag\\
 &\quad \times \He_{N_1}\!\Big(\sqrt{\tfrac{1}{2}\taut_1} a_1^{\dagger}\Big) \hat{S}(\xit_1)  \ket{0}. 
\label{eq:psiNCPS}
\end{align}

At this point, we would like to emphasize an important distinction. The conditional output state itself, and thus state-dependent quantities such as photon-number statistics and fidelities with other states, depend on the initial squeezing parameter $\xi_1$ and the beam-splitter transmittance $T_1$ only through the single effective squeezing parameter $\tilde{\xi}_1$. By contrast, the heralding probability retains an explicit dependence on both $\xi_1$ and $T_1$. Hence, there is a continuous family of parameter pairs $(\xi_1,T_1)$ that generate the same conditional output state for fixed $\tilde{\xi}_1$, but with different heralding probabilities.

\subsection{Fidelity with cat states}

The resulting conditional states are non-Gaussian by construction (for $N_1 > 0$) and exhibit a definite parity structure. For even (odd) $N_1$, the operator $\He_{N_1}\!\big( [\tfrac{1}{2}\tilde{\tau}_1 ]^{\frac{1}{2}} \, \hat{a}_1^\dagger\big)$ contains only even (odd) powers of $\hat{a}_1^\dagger$, and the corresponding state therefore has support only on even (odd) Fock states. It is therefore natural to compare these states to even (odd) Schr\"odinger cat states, defined as
\begin{align}
 \ket{\mathrm{cat}_{\pm}}
  = \sqrt{ \frac{\E^{|\alpha|}}{2(\E^{|\alpha|} \pm \E^{-|\alpha|})} } \, \bigl(\ket{\alpha}\pm\ket{-\alpha}\bigr).
\end{align}

The similarity between photon-subtracted squeezed vacuum states and Schr\"odinger-cat-like states was first pointed out in the context of their phase-space representations \cite{daknaGeneratingSchrodingercatlikeStates1997}. 
This similarity can also be inferred at the operator level in an asymptotic fashion. To this end, we first express the even and odd cat states as
\begin{align}
  \ket{\mathrm{cat}_{+}}
  = \frac{\cosh(\alpha \hat{a}^\dagger)}{\sqrt{\cosh |\alpha|^2}} \ket{0}, \;\;
  \ket{\mathrm{cat}_{-}}
  = \frac{\sinh(\alpha \hat{a}^\dagger)}{\sqrt{\sinh |\alpha|^2}} \ket{0}.
\end{align}
Using a heuristic large $N_1$ expansion for the operator generating the conditional state in Eq.~\eqref{eq:psiNCPS}, we obtain
\begin{align}
  \He_{N_1} (x) \, \E^{-x^2}
  &\sim \frac{N_1!}{\big(\frac{N_1}{2}\big)!} \cos\big(\sqrt{2N_1}x \big) \E^{-\tfrac{x^2}{2}}
\end{align}
for even $N_1$ with $x^2 = \frac{\taut_1}{2} \hat{a}^{\dagger 2}$, and
\begin{align}
  \He_{N_1} (x) \, \E^{-x^2}
  &\sim \frac{2^{N_1} \big(\frac{N_1 - 1}{2}\big)!}{\sqrt{\pi}} \sin\big(\sqrt{2N_1}x \big) \E^{-\tfrac{x^2}{2}}
\end{align}
for odd $N_1$. Comparing these asymptotic expressions with the hyperbolic-function representation of even and odd cat states suggests that, in the large-$N_1$ limit, the conditional states approximate specific squeezed cat states with scaling relation 
\begin{align}
|\alpha| \sim \sqrt{N_1|\tilde{\tau}_1|}
\label{eq:LargeNScaling}
\end{align}
and relative phase $\frac{\tilde{\varphi}_1}{2} - \varphi_{\alpha} = \frac{\pi}{2}$. We refer to Ref.~\cite{Zimmermann-InPreparation} for further details on this connection. Accordingly, the output states may be interpreted either in terms of cat states or, more specifically, as squeezed cat states. While the former interpretation is already motivated by the definite parity structure of the conditional states, the latter is further justified by the asymptotic operator correspondence. Nevertheless, in the following we only focus on fidelities with standard (unsqueezed) cat states, both to maintain direct comparability with the plain photon-subtraction literature and to avoid introducing an additional optimization over the target-state squeezing. Furthermore, in the strict large-$N_1$ limit, the scaling relation implies that, for a fixed target cat size $|\alpha|$, the effective squeezing parameter must scale as $|\taut_1| \sim |\alpha|^2/N_1$ and therefore tends to zero as $N_1$ increases. At the same time, this corresponds to the regime in which the associated squeezed cat approaches an ordinary (nonsqueezed) cat state.

To quantitatively compare the conditional output state of the plain photon-subtraction protocol with an even or odd cat state, we evaluate the fidelity ${F = |\braket{\mathrm{cat}_{\pm}|\psi_{N_{1}}}|^{2}}$ which reads, 
\begin{align}
 F &= \frac{1 \pm (-1)^{N_1}}{\E^{|\alpha|^2} \pm \E^{-|\alpha|^2}} \frac{(1-|\taut_1|^2)^{N_1+\frac{1}{2}}}{N_1! (2|\taut_1|)^{N_1}\!\Fhyp\!\! \Big(\frac{1-N_1}{2},-\frac{N_1}{2},1,\frac{1}{|\taut_1|^{2}}\Big)} \notag \\
 &\quad \times \E^{-\Re(\taut_1 \alpha^{* 2})} 
 \Bigg| \He_{N_1}\Bigg(\sqrt{\frac{\taut_1}{2}} \alpha^* \Bigg) \Bigg|^{2},
\label{eq:fidelityPS}
\end{align}
where $\Re(z)$ denotes the real part of $z$. 
This resulting fidelity captures two characteristic properties of the conditional output state. First, the prefactor $1\pm(-1)^{N_1}$ encodes the parity structure. For even $N_1$, the fidelity with the odd cat state vanishes and vice versa. 
Second, the fidelity depends only on the relative phase between $\taut_1$ and $\alpha^2$, $\varphit_1 - 2\varphi_\alpha$. This is not surprising as it reflects the gauge freedom to absorb one of the two phases into a redefinition of the mode, i.e., an appropriate choice of the phase-space coordinate system.

To determine the optimal relative phase, we note that the phase-dependent part of the fidelity is the product of two nonnegative functions, $\E^{-|\taut_1||\alpha|^2\cos(\varphit_1 - 2\varphi_\alpha)}$ and $\big| \He_{N_1} \big(\big[\frac{|\taut_1 \alpha^2| \E^{\I (\varphit_1 - 2\varphi_\alpha)}}{2}\big]^{\frac{1}{2}} \big) \big|$. Thus, it is sufficient to show that both are maximized for the same value of the relative phase. 
For the exponential term it immediately follows that the maximum is attained for $\varphit_1 - 2\varphi_\alpha = \pi$. 
To demonstrate that this relative phase also maximizes the Hermite-polynomial term, we use the sum representation
\begin{align}
 &\bigg| \He_{N_1} \! \bigg(\sqrt{\frac{\taut_1}{2}} \alpha^{*} \bigg) \bigg|  \notag \\
 &\quad = \bigg| \sum_{k=0}^{\lfloor \frac{N_1}{2}\rfloor} 
 \frac{(-1)^k N_1!  \big( 2|\taut_1 \alpha^2| \E^{\I (\varphit_1 - 2\varphi_\alpha)} \big)^{\frac{N_1-2k}{2}}}{k!(N_1-2k)!} \bigg|  \notag \\
 &\quad = \bigg| \sum_{k=0}^{\lfloor \frac{N_1}{2}\rfloor} 
 \frac{N_1!  \big( \sqrt{2|\taut_1|}\, |\alpha| \big)^{N_1-2k} }{k!(N_1-2k)!}\E^{\I k [\pi - (\varphit_1 - 2\varphi_\alpha)]} \bigg|.
\end{align}
As the coefficients of the oscillatory phase terms are strictly nonnegative in the last line, this function is maximized when all summands have the same phase which is realized for $\varphit_1 - 2\varphi_\alpha = \pi$. 
Since both nonnegative phase-dependent contributions to the fidelity are maximized for $\varphit_1 - 2\varphi_\alpha = \pi$, the fidelity itself is maximized for this relative phase. 
This particular phase choice can also be motivated in a natural way from a phase-space perspective as it aligns the orientation of the conditional state with that of the target cat state. We therefore restrict to this phase choice in the remainder of this section.

\begin{figure}[t]
    \centering
    \includegraphics[width=0.9\linewidth]{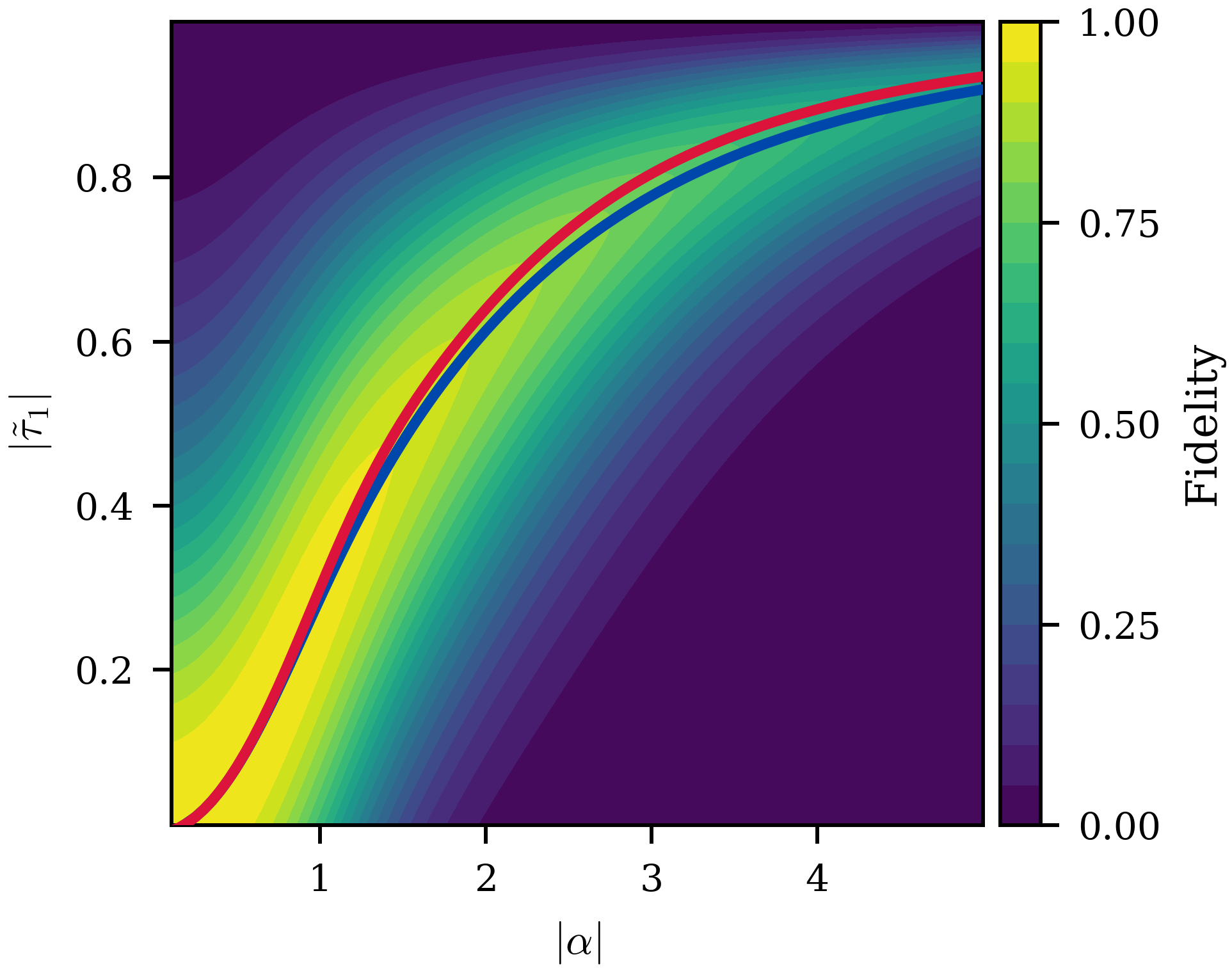}
    \caption{Fidelity of an even cat state over cat state size $|\alpha|$ and input squeezing $|\tilde{\tau}_1|$ for subtracting $N_1 = 2$ photons. The red line shows the resource-driven and the blue line the target-driven optimum.}
    \label{fig:FidLandscapePS2}
\end{figure}

\begin{figure}[t]
    \centering
    \includegraphics[width=0.9\linewidth]{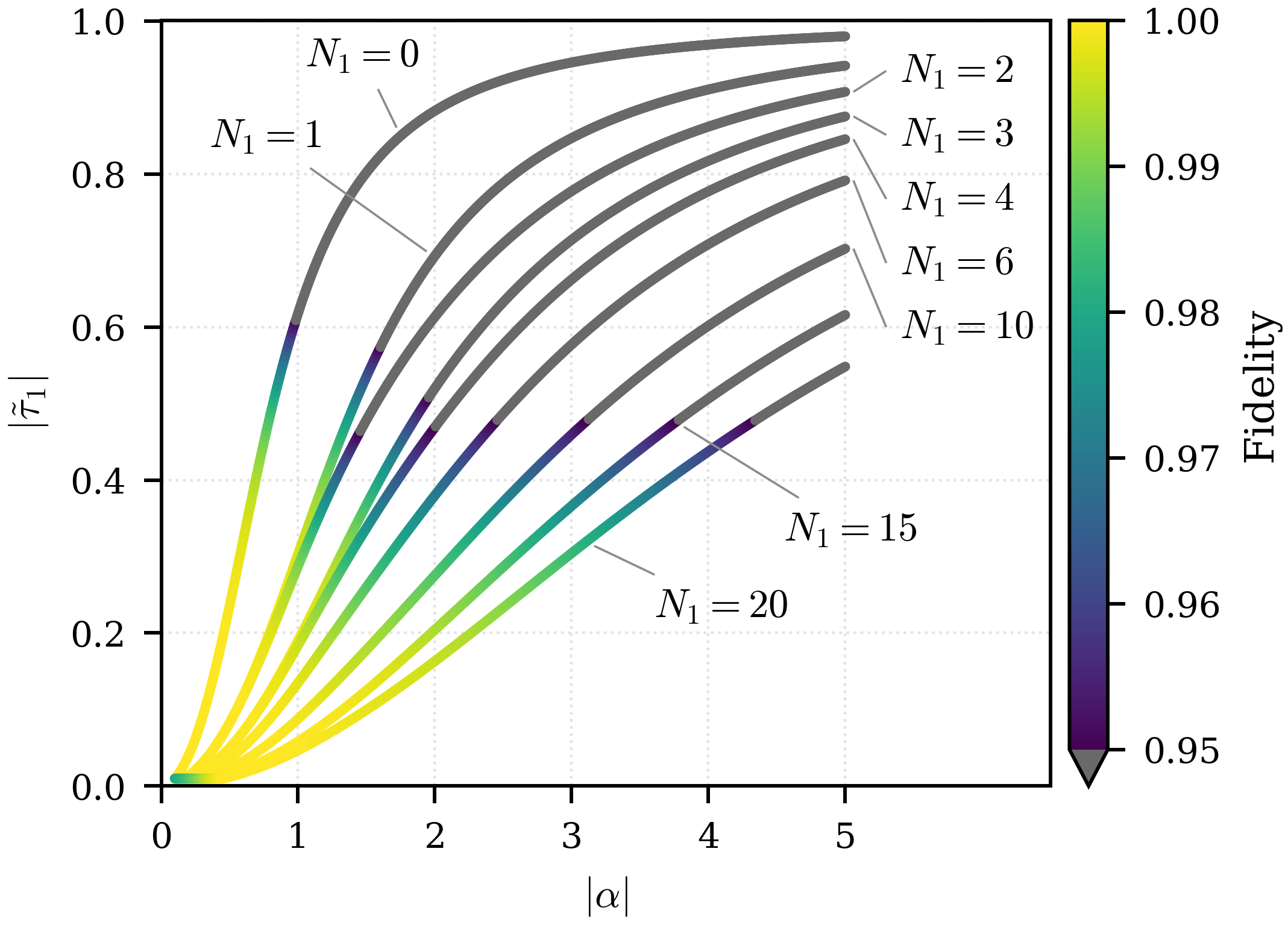}
    \caption{Fidelity optima between cat states and photon-subtracted squeezed vacuum states for the target-driven view for a variation of subtracted photon numbers $N_1$. Grey lines describe fidelities below $0.95$.}
    \label{fig:ridgecurvesPS}
\end{figure}

For a fixed cat amplitude $|\alpha|$, the fidelity now depends only on the effective squeezing parameter $|\taut_1| = \tanh|\xit_1|$ and the number of subtracted photons $N_1$. For fixed $N_1$, the fidelity landscape in the $(|\alpha|,|\tilde{\tau}_1|)$-plane is characterized by a single elongated high-fidelity ridge. We exemplarily show the case for $N_1 = 2$ in Fig.~\ref{fig:FidLandscapePS2}. The ridge bends towards larger effective squeezing as the target cat amplitude $|\alpha|$ increases, reflecting that the optimal approximation of larger cat states occurs at larger values of $|\tilde{\tau}_1|$ for fixed $N_1$. As $N_1$ increases, the ridge shifts systematically towards larger $|\alpha|$, see Fig.~\ref{fig:ridgecurvesPS}.

For low subtraction numbers $N_1$, fidelity maximization can be performed analytically in this simple scheme. We first consider the even-parity sector and start with its lowest-order contribution, $N_1=0$, for illustrative purposes. The optimal effective squeezing maximizing the fidelity for a target cat state amplitude then satisfies 
\begin{align}
 |\taut_1| = \frac{\sqrt{1+4|\alpha|^4}-1}{2|\alpha|^2}.
 \label{eq:OptSqPS0}
\end{align}
 Thus, the maximally achievable fidelity for each cat size is 
\begin{align}
 F = \frac{\sqrt{\sqrt{1+4|\alpha|^4}-1}}{\sqrt{2} |\alpha|^2 \cosh|\alpha|^2}\, \E^{\frac{\sqrt{1+4|\alpha|^4}-1}{2}}.
\end{align}
Alternatively, we may switch from a target-driven view to a resource-driven view by addressing the question which cat size is naturally supported by a given (squeezing) resource. For this latter view, we keep the effective squeezing parameter fixed and study variations with respect to $|\alpha|$, obtaining $|\alpha|^2 = \artanh|\taut_1| = |\xit_1|$ and 
\begin{align}
 F = (1-|\taut_1|^2)\, \E^{|\taut_1| \artanh|\taut_1|}
\end{align}
for the maximal achievable fidelity for fixed resource $|\taut_1|$. 
For $N_1 = 0$, however, conditioning corresponds to a vacuum projection in the reflected mode. Consequently, all involved operations are Gaussian such that the conditional output state remains Gaussian as can also be inferred from Eq.~\eqref{eq:psiNCPS}, which reduces to the squeezed-vacuum state $\hat{S}(\xit_1) \ket{0}$. Although this state can exhibit a high overlap with an even cat state in the low-$|\alpha|$ regime, this result has to be interpreted with care as the state lacks genuinely non-Gaussian cat-state features such as Wigner negativity. The large overlap in this regime is mainly due to the fact that both states are then dominated by the vacuum component, with a small admixture of a two photon Fock state. Higher even Fock components are even further suppressed. Thus, $N_1=0$ provides a useful analytical baseline, but not yet a faithful approximation of an optical cat state in a non-Gaussian sense.

Thus, the first genuinely non-Gaussian even-parity case is $N_1=2$. Optimizing the corresponding fidelity from a target-driven perspective, i.e., varying the effective squeezing strength $|\taut_1|$ for a fixed cat-state amplitude, leads to a fourth-order polynomial in $|\taut_1|$. Although this polynomial can in principle be solved analytically, the explicit expression offers little physical insight. A more compact relation is obtained in implicit form by solving for the cat-state amplitude, which yields
\begin{align}
 |\alpha|^2 = \frac{\sqrt{9+24|\taut_1|^4(1+2|\taut_1|^2)^2}-3(1-2|\taut_1|^2-4|\taut_1|^4)}{2|\taut_1|(1+|\taut_1|^2-2|\taut_1|^4)}.
\end{align}
For a given effective squeezing value $|\taut_1|$, this gives the cat size for which that squeezing is optimal in the target-driven sense. 
Conversely, we may take the resource-driven perspective and optimize with respect to $|\alpha|$ for fixed $|\taut_1|$. This leads to a transcendental equation in $|\alpha|^2$. However, since this equation remains quadratic in the parameter $|\tilde{\tau}_1|$, it can also be recast into an implicit relation 
\begin{align}
|\taut_1| &= \frac{\sqrt{8+(1-|\alpha|^2 \tanh|\alpha|^2)^2} - 3}{2|\alpha|^2}  + \frac{\tanh|\alpha|^2}{2},
\end{align}
giving the squeezing for which that cat amplitude is optimal in the resource-driven sense.

Similar inverse constructions can still be carried out for $N_1=4$ and $N_1=6$ for the target-driven and the resource-driven case, but the resulting expressions rapidly become too cumbersome to be of practical use, and beyond these cases no closed-form solutions are available. Nevertheless, the corresponding optimization conditions remain numerically tractable, since in the inverse formulation they reduce to polynomial equations of degree $\frac{N_1}{2}+1$ either in $|\alpha|^2$ or $|\tilde{\tau}_1|$. For $N_1=2$, we plot the target- and the resource-driven optimum as blue and red curve in Fig.~\ref{fig:FidLandscapePS2}, respectively. We generally observe that the two curves become increasingly similar as $N_1$ increases. 
The curves in Fig.~\ref{fig:ridgecurvesPS} correspond to the target-driven fidelity-maximizing ridges for different $N_1$. These curves provide a compact representation of the optimal trade-off between achievable cat size and required effective squeezing. Their overall behavior is qualitatively consistent with the heuristic scaling relation given in Eq.~\eqref{eq:LargeNScaling}. At fixed $|\alpha|$, the optimal $|\taut_1|$ decreases with increasing $N_1$, whereas at fixed effective squeezing $|\taut_1|$ the optimal cat amplitude increases with increasing $N_1$.

In the odd sector, the optimization of the lowest subtraction orders can also be performed analytically. For $N_1 = 1$, maximizing from a target-driven perspective leads to
\begin{align}
 |\taut_1| = \frac{\sqrt{9+4|\alpha|^4}-3}{2|\alpha|^2}.
\end{align}
For the resource-driven perspective, we obtain a transcendental equation for $|\alpha|^2$, which nevertheless admits a simple inverse solution,
\begin{align}
 |\taut_1| = \coth|\alpha|^{2} - \frac{1}{|\alpha|^2}.
\end{align}
Similar inverse constructions can still be carried out for $N_1=3$ for both, the target- and resource-driven perspective by solving quadratic equations. In general and similar to the even case, the two different optimizations reduce to polynomial equations of degree $\frac{N_1+1}{2}$ in both $|\alpha|^2$ and $|\tilde{\tau}_1|$ in the inverse formulations.

\subsection{Heralding probability and fidelity-probability trade-off}
Besides the fidelity, the probability of heralding the conditional output state is a key figure of merit. In the present parametrization, it is convenient to express this quantity in terms of the effective output squeezing parameter $|\tilde{\tau}_1|$ and the beam-splitter transmittance $T_1$ with physically allowed parameter regime ${0<|\tilde{\tau}_1|<T_1<1}$, see Eq.~\eqref{eq:ProbPS}. This representation is particularly useful because it separates the description of the conditional output state from the choice of its physical implementation. For a fixed cat-state amplitude, the fidelity depends only on $|\tilde{\tau}_1|$, whereas the heralding probability retains an additional dependence on $T_1$. For fixed $|\tilde{\tau}_1|$, there is thus a continuous family of parameter pairs $(|\tau_1|,T_1)$ satisfying $|\tilde{\tau}_1|=T_1|\tau_1|$ that generate the same conditional output state, while generally differing in their heralding probability. Equivalently, one may revert to the experimentally more direct variables $(|\tau_1|,T_1)$ when discussing the required input squeezing resource.

For fixed $|\tilde{\tau}_1|$, the dependence of $P_{N_1}$ on $T_1$ given by Eq.~\eqref{eq:ProbPS} can be understood from its behavior at the boundaries of the physically allowed interval. 
For vacuum heralding, i.e., $N_1=0$, the probability is monotonically increasing in $T_1$ and therefore maximal in the limit $T_1 \to 1$. 
For all nontrivial subtraction orders $N_1\ge 1$, by contrast, the heralding probability vanishes both for $T_1\to |\tilde{\tau}_1|$ and for $T_1 \to 1$. In the first limit, the required input parameter $|\tau_1|=|\tilde{\tau}_1|/T_1$ approaches unity, corresponding to the limit of infinite input squeezing. In this regime, the squeezed-vacuum photon-number distribution becomes arbitrarily broad, with a diverging mean photon number, such that the probability associated with any fixed finite Fock component vanishes. This is reflected by the square-root prefactor in Eq.~\eqref{eq:ProbPS} that suppresses the heralding probability in this case. In the second limit too little light is reflected to herald a nonzero subtraction event. Thus, optimal state generation requires balancing sufficient tapping to herald the subtraction event against the increased input squeezing required to maintain the same effective output parameter. Hence, for every fixed $|\tilde{\tau}_1|$ and $N_1\ge 1$, an interior optimum in $T_1$ exists. Maximizing with respect to $T_1$ yields
\begin{align}
 T_{1,\mathrm{opt}} = \frac{\sqrt{|\taut_1|^4+4N_1(N_1+1)|\taut_1|^2} -|\taut_1|^2 }{2N_1}.
\label{eq:ToptSucProbPS}
\end{align}
This allows us to determine the maximized heralding probability along the fidelity-optimal ridges. We first select the optimal effective squeezing $|\taut_1|$ for a target cat-state amplitude by maximizing the fidelity, while the heralding probability retains the additional dependence on $T_1$ being optimized via Eq.~\eqref{eq:ToptSucProbPS}.

\begin{table}[t]
\centering
\setlength{\tabcolsep}{8pt}
\begin{tabular}{@{} c c c c c c @{}}
        $N_1$ & $F_{\mathrm{max}}$ & $P$ & $|\tilde{\tau}_1|$ & $T$ & $|\xi_1|$ \\
        \hline\hline
        \multicolumn{6}{c}{Even cat state} \\
        \hline
        2  & 0.7690 & 0.1050 & 0.7777 & 0.8133 & 1.9013 \\
        4  & 0.8698 & 0.0558 & 0.6622 & 0.6876 & 1.9870 \\
        6  & 0.9151 & 0.0380 & 0.5772 & 0.5963 & 2.0590 \\
        8  & 0.9399 & 0.0289 & 0.5116 & 0.5266 & 2.1212 \\
        10 & 0.9552 & 0.0233 & 0.4595 & 0.4715 & 2.1763 \\
        \hline\hline
        \multicolumn{6}{c}{Odd cat state} \\
        \hline
        1 & 0.6823 & 0.1813 & 0.8471 & 0.8918 & 1.8310 \\
        3 & 0.8338 & 0.0725 & 0.7141 & 0.7439 & 1.9443 \\
        5 & 0.8968 & 0.0452 & 0.6165 & 0.6384 & 2.0241 \\
        7 & 0.9293 & 0.0328 & 0.5423 & 0.5592 & 2.0910 \\
        9 & 0.9485 & 0.0258 & 0.4841 & 0.4975 & 2.1495 
\end{tabular}
\caption{Fidelity--heralding-probability trade-off for approximating an $|\alpha|=3$ even or odd cat state using photon-subtracted squeezed vacuum. For each subtraction number $N_1$, we list the maximal achievable fidelity $F_{\mathrm{max}}$, the optimized probability for the given fidelity, and the required input parameters in terms of the effective squeezing parameter $|\tilde{\tau}_1|$, the beam-splitter transmittance $T_1$, and the modulus of the input squeezing $|\xi_1|$.}
\label{tab:PS_tradeoff}
\end{table}

In general, the heralding probability may also be studied as a function of $|\taut_1|$ for fixed transmittance $T_1$. In this case, $P_{0}$ decreases monotonically with $|\taut_1|$, whereas for all $N_1\ge 1$ the probability vanishes both for $|\taut_1|\to 0$ and for $|\taut_1|\to T_1$. Physically, small $|\taut_1| \ll T_1$ corresponds to insufficient squeezing to populate the relevant higher Fock components, while $|\taut_1|\to T_1$ again implies a divergent input squeezing. Numerically, the heralding probability exhibits a unique maximum at an intermediate value of $|\taut_1|$ for all subtraction orders tested up to $N_1=50$. As $N_1$ increases, the maximizing value of $|\taut_1|$ shifts towards the boundary $|\taut_1|=T_1$ (and thus, larger input squeezing), while the corresponding maximal heralding probability decreases. Generally, the heralding probability decreases markedly with increasing $N_1$. Hence, higher-order subtraction favors larger effective squeezing, but at the cost of rapidly suppressed heralding rates, limiting the generation of large-amplitude cat states. At the same time, the probability is strongly suppressed for small initial squeezing.

\begin{figure}[t!]
 \centering
 \includegraphics[width=0.99\linewidth]{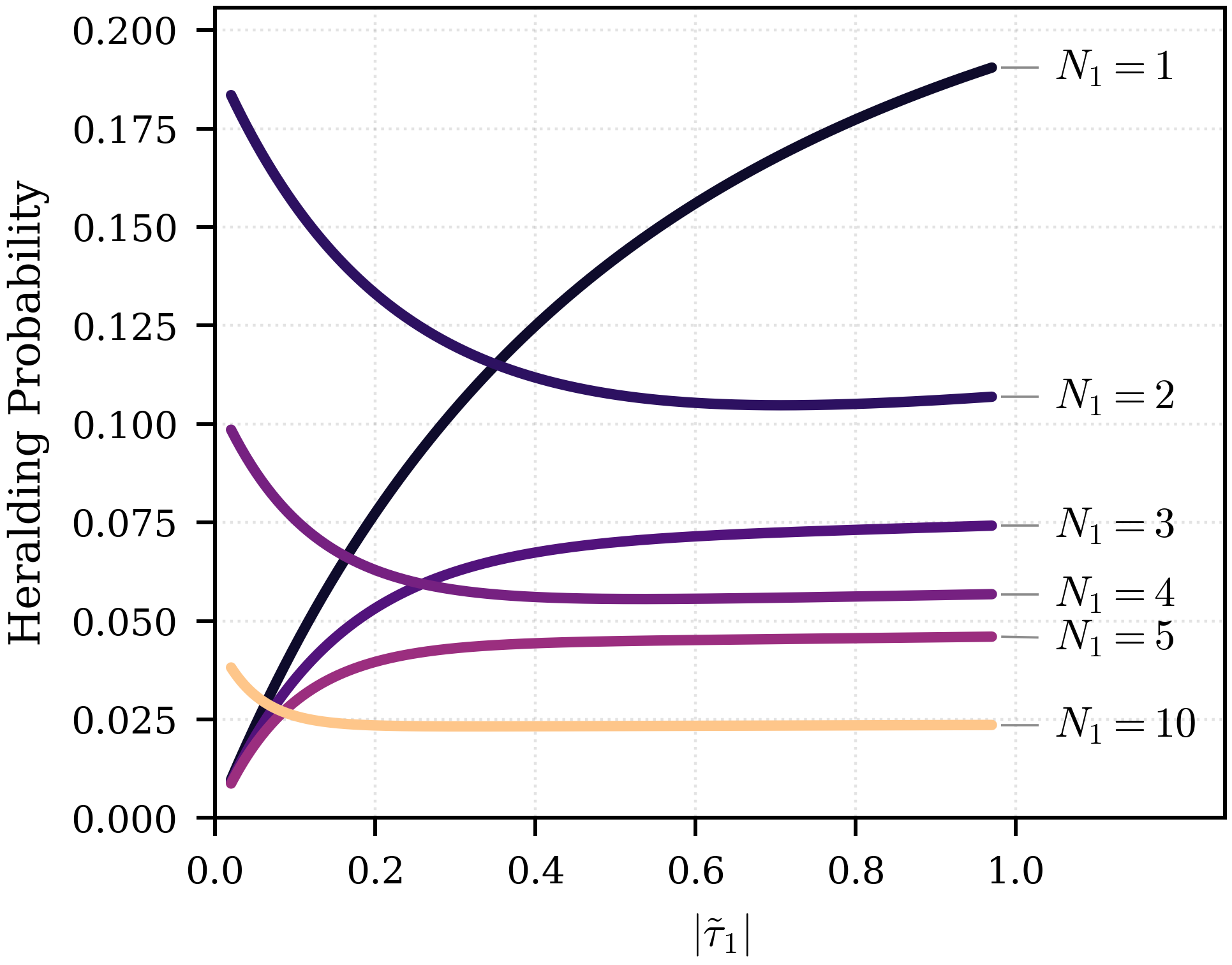}
\caption{Heralding probability as a function of the effective squeezing parameter $|\tilde{\tau}_1|$. Colors encode different photon subtraction numbers $N_1$ between $1$ and $10$.}
\label{fig:prob_vs_tau}
\end{figure}

For any fixed target cat size, this leads to a trade-off between fidelity and heralding probability. Increasing $N_1$ improves the achievable fidelity in general, but at the cost of a decreasing heralding probability. For specific heralding patterns, this trade-off is illustrated in Tab.~\ref{tab:PS_tradeoff}, which lists, for different $N_1$, the maximal achievable fidelity with an $|\alpha| = 3$ cat state, the corresponding maximal heralding probability to herald the corresponding state, and the associated optimal parameter settings. 
We have chosen $|\alpha|=3$ as a representative example. The same qualitative behavior is found for other target amplitudes. In particular, the input squeezing required to maximize the fidelity increases with $N_1$, and, for fixed $N_1$, it also increases with the target amplitude. Interestingly, the optimized heralding probability remains nearly unchanged when varying $|\alpha|$ at fixed $N_1 \geq2$. Although the heralding probability does not depend explicitly on $\alpha$, it acquires an implicit dependence through the fidelity-optimal value of $|\tilde{\tau}_1|$. However, substituting the optimal transmittance from Eq.~\eqref{eq:ToptSucProbPS} into Eq.~\eqref{eq:ProbPS} and examining the resulting probability as a function of $|\tilde{\tau}_1|$ for fixed $N_1$ shows that it varies only weakly over a broad parameter range for $N_1 \geq 2$, see Fig.~\ref{fig:prob_vs_tau}. In the large-$|\taut_1|$ plateau, the optimized heralding probability decreases asymptotically for large $N_1$ as $P_{N_1} \sim \frac{1}{\sqrt{2\pi\E}\,N_1}$.

The different initial trends of the heralding probabilities for even and odd $N_1$ in Fig.~\ref{fig:prob_vs_tau} can be understood from the even photon-number parity of squeezed vacuum. For fixed nonzero input squeezing $|\tau_1|$, the limit $|\taut_1|\to 0$ corresponds to vanishing transmission into the retained mode, such that the heralding detector effectively measures the original squeezed vacuum and only even photon-number outcomes are possible. Importantly, small $|\taut_1|$ does not necessarily imply weak input squeezing $|\tau_1|$, since the effective squeezing parameter also depends on the beam-splitter transmission. Increasing $|\taut_1|$ allows photons to be distributed between the retained and heralding modes. This enables odd heralding outcomes, whose probabilities initially increase from zero, while the probabilities of even outcomes initially decrease as photons are diverted into the retained mode.

In practice, the maximum available squeezing becomes a key limiting resource. Thus, we perform the same analysis implementing the constraint $|\xi_1| < |\xi_{1,\mathrm{max}}|$. 
As long as $|\tau_{1,\mathrm{max}}| = \tanh|\xi_{1,\mathrm{max}}|$ exceeds $|\tilde{\tau}_{1,\mathrm{opt}}|$ which optimizes the fidelity for fixed $|\alpha|$ and $N_1$, the maximal achievable fidelity is unaffected. The squeezing constraint nevertheless lower bounds the admissible transmittance for a given $|\taut_1|$ through $T_1 \geq \frac{|\taut_1|}{|\tau_{1,\mathrm{max}}|}$, and may therefore prevent choosing the optimal transmittance $T_{1,\mathrm{opt}}$ given by Eq.~\eqref{eq:ToptSucProbPS} that maximizes the heralding probability. 
This is the case whenever $|\taut_{1,\mathrm{opt}}| \leq |\tau_{1,\mathrm{max}}| \leq \frac{|\taut_{1,\mathrm{opt}}|}{T_{1,\mathrm{opt}}}$. 
In this regime, we can preserve the maximal achievable fidelity by saturating the maximal available input squeezing and adjust the transmittance such that $|\taut_{1,\mathrm{opt}}| = T_1 |\tau_{1,\mathrm{max}}|$, which yields the highest possible heralding probability under this fidelity constraint. 
Nevertheless, the resulting transmittance satisfies $T_1 > T_{1,\mathrm{opt}}$, thereby reducing the heralding probability. In the limiting case ${|\tau_{1,\mathrm{max}}| \to |\taut_{1,\mathrm{opt}}|}$, one has $T_1 \to 1$ and hence $P_{N_1}\to0$. If a practically relevant heralding probability shall be maintained, one has to accept a smaller value of $|\taut_1| < |\taut_{1,\mathrm{opt}}|$, allowing for smaller $T_1$ closer to $T_{1,\mathrm{opt}}$, at the cost of reduced fidelity. 
If $|\tau_{1,\mathrm{max}}| < |\taut_{1,\mathrm{opt}}|$, the fidelity-optimal conditional state is no longer accessible, and the maximal achievable fidelity necessarily decreases.

\begin{figure}[t!]
 \centering
 \includegraphics[width=0.99\linewidth]{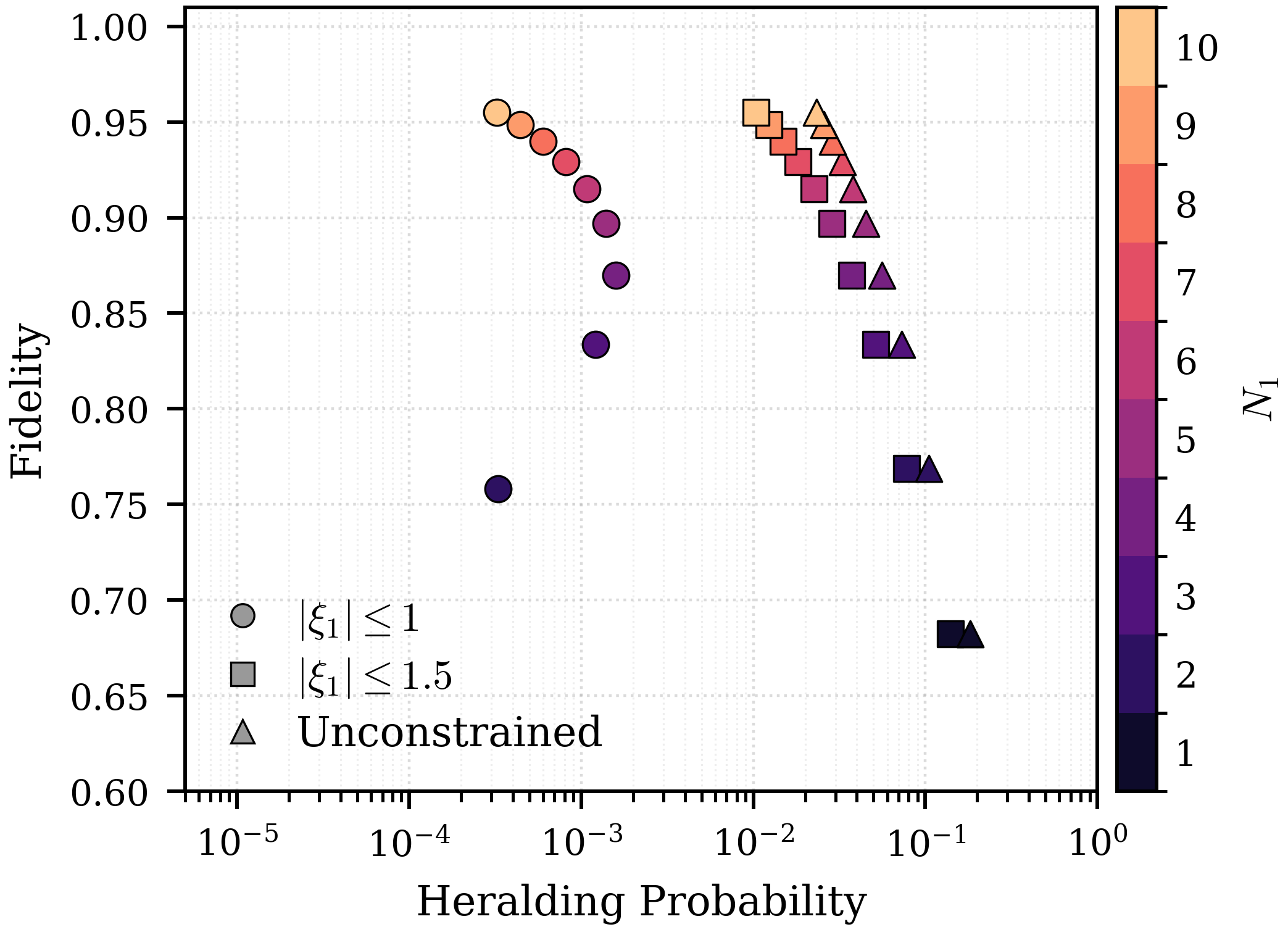}
\caption{Fidelity over heralding probability for plain photon subtraction of squeezed vacuum with a maximum beam splitter transmission of $0.99$ for $|\alpha| = 3$ cat states. We compare different maximum squeezing levels $|\xi_{\mathrm{max}}|$ (markers) and different photon subtraction numbers $N_1$ (color encoded).}
\label{fig:fid_vs_prob-ps}
\end{figure}

As an example, we show the fidelity-probability trade-off for $|\alpha| = 3$ in Fig.~\ref{fig:fid_vs_prob-ps}. Increasing the squeezing bound from $|\xi_{1,\mathrm{max}}| = 1$ to $|\xi_{1,\mathrm{max}}| = 1.5$ substantially improves the achievable fidelity-probability trade-off. Further increases provide only comparatively modest gains because the unconstrained optimal squeezing values lie only slightly above this range for reachable heralding events with current PNR technology, see Tab.~\ref{tab:PS_tradeoff}. For the cases $|\tau_{1,\mathrm{max}}| < |\taut_{1,\mathrm{opt}}|$, we have restricted the allowed transmittance domain to $T_1 < 0.99$ to retain nonnegligible heralding probabilities.

\subsection{Fidelity-constrained multibranch success probability optimization}
For a fair assessment of the fidelity-probability trade-off, multiple heralding branches must actually be considered simultaneously. Due to the inherently postselected nature of the cat-state generation protocol, this is also the natural perspective from an experimental point of view. This turns the optimization from a single-branch problem into a discontinuous multibranch acceptance problem, in which the total probability has to be maximized subject to a given fidelity constraint. For each detected photon number $N_1$, the conditional output state is characterized by its branch fidelity $F(\taut_1,N_1)$ and heralding probability $P_{N_1}(\taut_1,T_1)$. Given a chosen lower fidelity threshold $F_{\mathrm{min}}$, only those heralding outcomes are retained for which the branchwise condition $F(\taut_1,N_1)\geq F_{\mathrm{min}}$ is satisfied, while all remaining events are rejected. Accordingly, we define the set $\mathcal{S}(\taut_1) = \{N_1 \,|\, F(\taut_1,N_1) \geq F_{\mathrm{min}}\}$. In the following, we reserve the term heralding probability for the probability associated with a particular photon-count outcome. By contrast, we define the success probability as the total probability of all accepted heralding outcomes, being $\sum_{N_1\in S(\taut_1)} P_{N_1}(\taut_1,T_1)$ for a fixed experimental setting.
The maximal achievable success probability is then obtained by optimizing the common experimental parameters $(\taut_1,T_1)$,
\begin{align}
P_{\mathrm{succ}} = \max_{0<\taut_1< T_1 <1} \sum_{N_1\in S(\taut_1)} \!\!\!\! P_{N_1}(\taut_1,T_1),
\end{align}
where every branch contained in $S(\taut_1)$ individually satisfies $F(\taut_1,N_1)\geq F_{\mathrm{min}}$. Thus, the mixture of accepted conditional states also satisfies $F\geq F_{\mathrm{min}}$, even if the individual heralding outcomes are not retained.

A related multi-outcome optimization strategy was recently considered in Ref.~\cite{Ismailzadeh:2026yye}, where a common Gaussian-PNR circuit is optimized while aggregating the probabilities of several useful measurement outcomes. In the present setting, the accepted set is instead determined by a prescribed branchwise fidelity threshold and can contain different photon-subtraction orders, allowing the total success probability to be optimized directly for the target cat state.

\begin{figure}[t]
    \centering
    \includegraphics[width=0.99\linewidth]{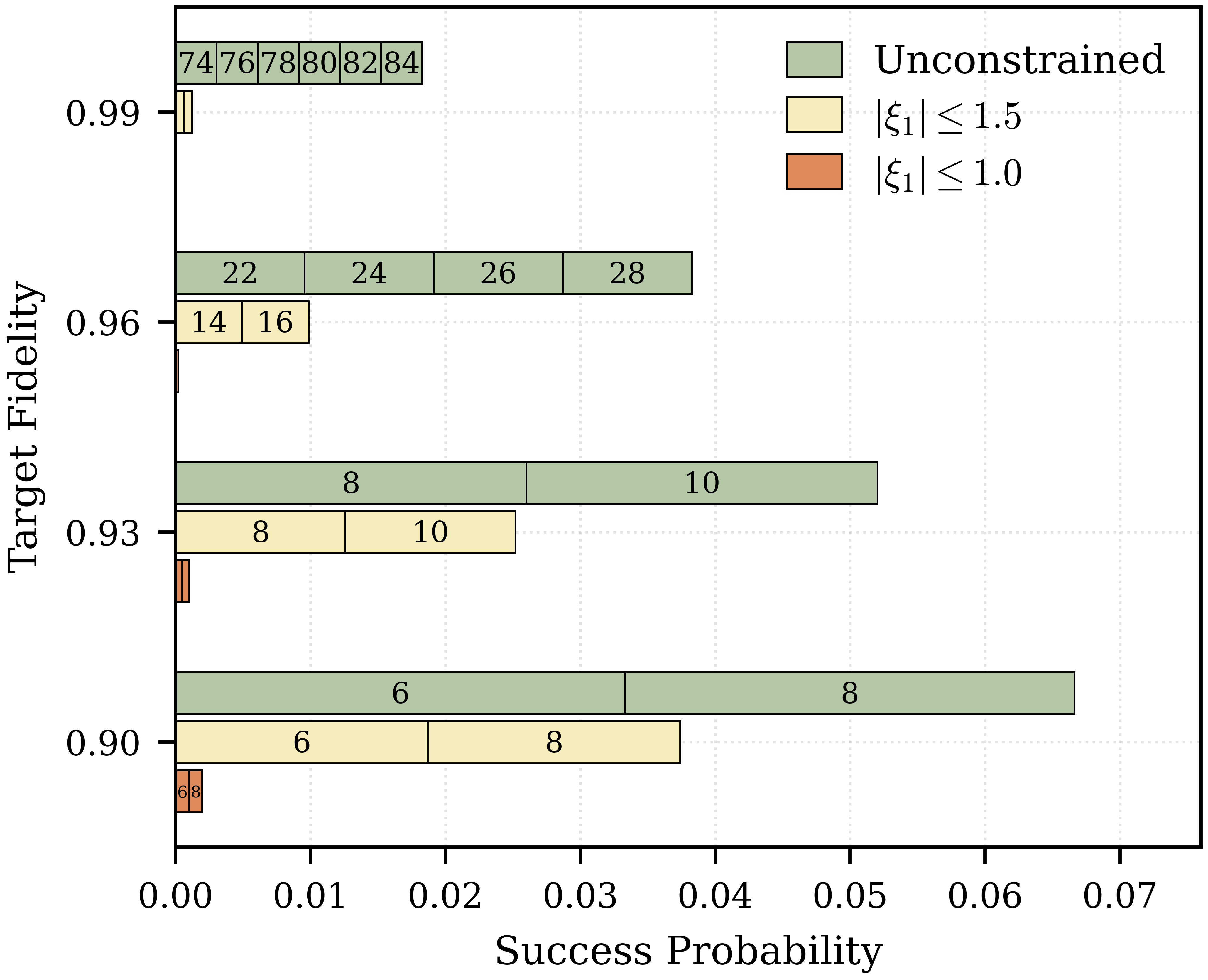}
    \caption{Fidelity--success-probability trade-off for multibranch even cat state generation with amplitude $|\alpha|=3$. The total optimized success probability is plotted as a function of the target fidelity constraint across different hardware limits on the squeezing parameter $|\xi_1|$. Discrete stacked bars decompose the cumulative success probability into the absolute contributions of individual heralding modes ($N_1 = 2,4,6,...$). Colors indicate distinct squeezing constraints ($|\xi_1| \le 1.0$, $|\xi_1| \le 1.5$, and the unconstrained parameter space).}
    \label{fig:ps_multibranchOptimization}
\end{figure}

We exemplarily show the fidelity--success-probability trade-off for even target cat states with $|\alpha| = 3$ in Fig.~\ref{fig:ps_multibranchOptimization}. The corresponding numerical results are summarized in Tab.~\ref{tab:PS-multibranch-even} in the \hyperref[app:numerical-results]{Appendix}, which includes additional fidelity thresholds beyond those displayed in the figure and the corresponding optimized parameter settings. We also provide corresponding results for odd cat states in Tab.~\ref{tab:PS-multibranch-odd}. The figure shows that the maximal total success probability decreases as the fidelity threshold is raised as expected. This decrease is not generally smooth, because the set of accepted photon-number outcomes changes discontinuously with $F_{\mathrm{min}}$. The individual segments indicate the contributions of the accepted heralding branches. As the fidelity requirement becomes more stringent, the optimization generally shifts toward larger photon-subtraction numbers, which provide higher-fidelity approximations but occur with lower probability.

Imposing a bound on the available squeezing strongly suppresses the maximal achievable success probability. Similar to the single-heralding-pattern optimization, the optimal solutions of the multibranch problem typically saturate the squeezing constraint for $|\xi_{1,\mathrm{max}}|=1.5$. At moderate fidelity thresholds around $0.9$, the success probability is reduced by roughly a factor of two relative to the unconstrained case, while at high thresholds, $F_{\mathrm{min}}\gtrsim 0.98$,  the suppression approaches an order of magnitude. By contrast, for $|\xi_{1,\mathrm{max}}|=1$, the success probability is suppressed by more than an order of magnitude across most of the considered fidelity range and drops to nearly negligible values for $F_{\mathrm{min}}\gtrsim 0.98$, see also Tab.~\ref{tab:PS-multibranch-even}.

Limiting the available squeezing also reduces the number of heralding branches that can be accepted simultaneously. In the unconstrained case, the optimal solution contains between two and seven branches whereas the constraint $|\xi_{1,\mathrm{max}}| = 1.5$ reduces this to two branches over the considered fidelity range. For $|\xi_{1,\mathrm{max}}| = 1$, only one or two branches remain, with a single accepted outcome at most of the higher fidelity thresholds. Thus, restricted squeezing below the unconstraint optimum (being in a range of $2.09$ to $2.94$ for the considered fidelities, $0.9\leq F_{\mathrm{min}} \leq 0.99$) not only lowers the total success probability but also suppresses the multibranch advantage obtained by combining several heralding outcomes.

As for the single-branch optimization, we use the ${|\alpha|=3}$ cat state only as a representative example. Repeating the multibranch optimization for $|\alpha|=2$ and $4$ reveals a clear resource scaling with the target cat-state amplitude. For all tested fidelity thresholds, increasing $|\alpha|$ shifts the optimal accepted heralding outcomes toward larger photon numbers and requires stronger input squeezing, while the corresponding optimized success probability decreases. Thus, generating larger cat states becomes progressively more demanding in terms of both squeezing and photon-number resolution.

Finally, the odd-parity case exhibits the same qualitative behavior as the even-parity case. In particular, the optimized success probabilities are nearly identical in the unconstrained regime. Finite squeezing constraints lead to somewhat larger differences between the even- and odd-parity cases for fixed $F_{\mathrm{min}}$, but no systematic advantage of either parity is observed over the full fidelity range.

\section{Combining Photon Subtraction and Nonlinear Interferometer}
\label{sec:SqSuSq}
Although the plain photon-subtraction scheme can, in principle, approximate cat states with arbitrarily high fidelity, its practical applicability is limited by small success probabilities, particularly under constraints on the available squeezing, even in the idealized setting neglecting losses, mode mismatch, and other imperfections. Thus, the trade-off between fidelity and success probability motivates suitable extensions of this plain photon-subtraction scheme that preserve its conceptual simplicity while offering additional freedom to shape the conditional output state.

\begin{figure}[t]
 \centering
 \includegraphics[width=0.6\linewidth]{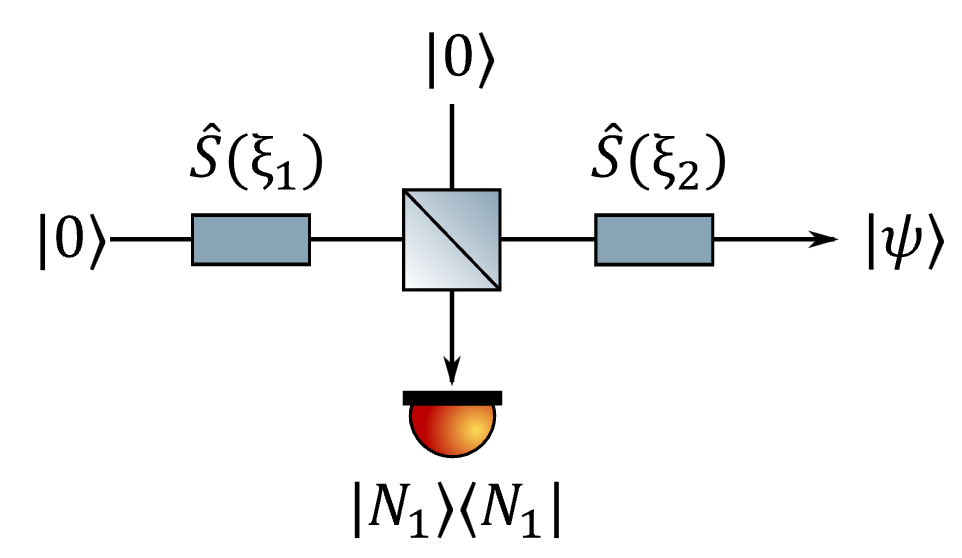}
 \caption{Setup of nonlinear-interferometer-type configuration. A squeezed-vacuum input with squeezing parameter $\xi_1$ is mixed with vacuum on a beam splitter of transmittance $T_1$ and $N_1$ photons are detected in the reflected arm. A second squeezing operation with squeezing parameter $\xi_2$ is subsequently employed in the transmitted arm.}
 \label{fig:setup_ni}
\end{figure}

A natural modification is to insert a second squeezing stage after the photon-subtraction event, see Fig.~\ref{fig:setup_ni}. 
The resulting setup may be viewed as a degenerate nonlinear-interferometer-type configuration with a photon-subtraction stage embedded between two squeezing operations. Originally, nonlinear interferometers were proposed by Yurke et al. designed to reach the Heisenberg limit for high-precision phase measurements \cite{Yurke1986}. It has been shown that such interferometers are more loss-tolerant regarding phase sensitivity compared to their passive counterparts \cite{Yurke1986,Jing2011,Hudelist2014}. The first experimental implementation by Zou et al. has shown the effect of induced coherence \cite{zou1991}, which has found applications in spectroscopy of linear and nonlinear materials \cite{Burlakov1998} as well as imaging \cite{lemos2014,Fuenzalida:2020bgu,GilaberteBasset-Video2021,Fuenzalida:2023anr,Basset:2023hlg}. Recently, photon-number-detection enhanced nonlinear interferomtery for phase sensitivity measurements has also been demonstrated \cite{Machado:2026aip}. A detailed analysis of different experimental configurations can be found in \cite{Chekhova2016}.

The second squeezer provides an additional degree of freedom that can further redistribute the weights of the different Fock-state contributions and thereby improve the overlap with a target cat state. At the same time, since the added squeezing stage corresponds to a unitary operation, it does not alter the probability of the heralding event itself.
Therefore, any improvement in heralding efficiency arises only in an implicit fashion. If the added squeezing stage allows a given target fidelity to be reached with a smaller subtraction number, then the heralding probability can remain substantially higher than in the plain photon subtraction scheme, where achieving a comparable fidelity would typically require larger $N_1$.

The conditional output state obtained in this extended scheme reads
\begin{align}
 \hat{S}(\xi_2) \ket{\psi_{N_1}} &= \sqrt{\frac{(1-|\taut_1|^2)^{N_1}}{N_1! (2|\taut_1|)^{N_1} \!\Fhyp\!\! \Big(\frac{1-N_1}{2},-\frac{N_1}{2},1,\frac{1}{|\taut_1|^{2}}\Big)}} \notag\\
 &\quad \times \frac{1}{\sqrt{(1+\taut_1 \tau_2^*)^{N_1}}}
 \He_{N_1} \Bigg( \sqrt{ \frac{\taut_1}{2} \frac{1-|\tau_2|^2}{1+\taut_1\tau_2^*} } \hat{a}^\dagger \Bigg)  \notag \\
 &\quad \times \hat{S}(\breve{\xi}_2) \ket{0},
\label{eq:psiNPSNI}
\end{align}
where $\breve{\xi}_2$ is defined via $\E^{\I \breve{\varphi}_2} \tanh|\breve{\xi}_2| = \breve{\tau}_{2} \equiv \frac{\taut_1 + \tau_2}{1+\taut_1\tau_2^*}$ which is the M\"obius transformation arising from the composition of two squeezing operations acting on the vacuum, $\hat{S}(\xi_2)\hat{S}({\xit}_1) \ket{0} = \hat{S}(\breve{\xi}_2) \ket{0}$.

\subsection{Fidelity with cat states}
The fidelity of this conditional output state with an even or odd cat state is given by
\begin{align}
 F &= \frac{1 \pm (-1)^{N_1}}{\E^{|\alpha|^2} \pm \E^{-|\alpha|^2}} 
\frac{(1-|\taut_1|^2)^{N_1+\frac{1}{2}} (1-|\tau_2|^2)^{\frac{1}{2}}}{N_1! (2|\taut_1|)^{N_1}\!\Fhyp\!\! \Big(\frac{1-N_1}{2},-\frac{N_1}{2},1,\frac{1}{|\taut_1|^{2}}\Big)}
 \notag \\
 &\quad \times\! \frac{\E^{-\Re\big(\frac{\taut_1 + \tau_2}{1+\taut_1 \tau_2^*} \alpha^{* 2}\big)} }{|1+\taut_1 \tau_2^*|^{N_1 + 1}}\,
 \Bigg| \He_{N_1}\!\Bigg(\sqrt{\frac{\taut_1}{2} \frac{1-|\tau_2|^2}{1+\taut_1 \tau_2^*}} \alpha^* \Bigg) \Bigg|^{2} \!\!,
\label{eq:fidelityS-PS-S}
\end{align}
which reduces to Eq.~\eqref{eq:fidelityPS} for $\tau_2 = 0$.

Before addressing the full parameter dependence of the fidelity, we first discuss particularities of the lowest-order heralding events in the even and odd parity sectors. For $N_1=0$ and $N_1=1$, the fidelity does not depend separately on the two squeezing parameters $\taut_1$ and $\tau_2$ but only on the combined squeezing parameter $\breve{\tau}_2$. Therefore, the additional squeezing operation does not improve the optimal achievable fidelity beyond what is already obtained in the plain photon subtraction scheme for these two heralding branches. This is expected for $N_1=0$, since the conditional projection is on the vacuum state and, thus, all operations are Gaussian. The output state is therefore still a squeezed vacuum state, with an effective squeezing parameter $\breve{\tau}_2$, see Eq.~\eqref{eq:psiNPSNI}. The fidelity with an even cat state then reduces to
\begin{align}
 F = \frac{1}{\cosh|\alpha|^2} \big(1-|\breve{\tau}_2|^2\big)^{\frac{1}{2}} \E^{-\Re(\breve{\tau}_2 \alpha^{*2})},
\end{align}
being consistent with Eq.~\eqref{eq:fidelityPS} for $N_{1}=0$ and replacing $\taut_1$ by $\breve{\tau}_2$.

For $N_1=1$, the heralded state is obtained by a single photon subtraction from a squeezed vacuum, which is equivalent to a single photon-added squeezed vacuum according to Eq.~\eqref{eq:psiNCPS}, see also \cite{Steuernagel:2024seq}. Using standard ladder-operator algebra, this state can also be written, up to normalization, as a squeezed single-photon state, $\hat{a}^{\dagger}\hat{S}(\xit_{1})\ket{0} \sim \hat{S}(\xit_{1})\hat{a}^{\dagger}\ket{0}$. The subsequent squeezing operation $\hat S(\xi_2)$ can be combined with $\hat S(\tilde\xi_1)$ into a single effective squeezing operation, up to a phase rotation. Since this phase shift acts on a single Fock component it contributes merely an unimportant global phase term. 
Furthermore, as the non-Gaussian part of the state is represented by a single creation operator only, there is no remaining structure that could retain separate information about $\taut_1$ and $\tau_2$. The fidelity with an odd cat state therefore depends only on the combined squeezing parameter $\breve{\tau}_2$ and takes the form
\begin{align}
 F = \frac{|\alpha|^2}{\sinh|\alpha|^2} \big(1-|\breve{\tau}_2|^2\big)^\frac{3}{2} \E^{-\Re(\breve{\tau}_2 \alpha^{*2})},
\end{align}
being again equivalent to the fidelity of the plain subtraction scheme.
This simple reduction no longer occurs for higher subtraction orders. For $N_1\geq 2$, the photon-subtracted squeezed state contains a nontrivial superposition of non-Gaussian components. The relative weights of these components depend on the squeezing associated with the subtraction process itself, while the subsequent squeezing transforms each component according to its photon-number structure. Hence, the two squeezing parameters cannot be combined into a single effective parameter. In this regime the fidelity depends separately on $\tilde{\tau}_1$ and $\tau_2$, giving rise to additional phase and modulus optimization.

To optimize the fidelity given in Eq.~\eqref{eq:fidelityS-PS-S} for the present scheme with $N_1\geq 2$, we first reconsider the role of the phase parameters. In contrast to the plain photon-subtraction scheme, where the fidelity depends only on the single relative phase $\tilde{\varphi}_1-2\varphi_\alpha$, the nonlinear-interferometer-type configuration involves a second independent relative phase associated with the squeezers, $\tilde{\varphi}_1-\varphi_2$. The full phase dependence is contained in the second line of Eq.~\eqref{eq:fidelityS-PS-S}.
Optimization with respect to the phase parameters is therefore more intricate than in the plain subtraction case. Instead of two phase-dependent factors depending on the same relative phase, the fidelity now contains three competing contributions that depend differently on the two relative phases. The denominator $|1+\tilde{\tau}_1\tau_2^*|^{N_1+1}$ is minimized for $\tilde{\varphi}_1-\varphi_2=\pi$. The Hermite-polynomial contribution is maximized when its argument is purely imaginary and the modulus of its argument is as large as possible. The latter is again realized when the denominator $|1+\tilde{\tau}_1\tau_2^*|$ is minimized, i.e., $\tilde{\varphi}_1-\varphi_2=\pi$. Then, $|\He_{N_1}|$ is maximized for $\tilde{\varphi}_1 - 2\varphi_\alpha=\pi$, following the same line of argument as in the plain photon-subtraction scheme. 
By contrast, the exponential factor is maximized whenever $\arg(\breve\tau_2) - 2\varphi_\alpha = \pi$ and therefore prefers the phase choice for which $|\breve\tau_2| = |\frac{\tilde{\tau}_1+\tau_2}{1+\tilde{\tau}_1\tau_2^*}|$ is maximal. This occurs for $\tilde{\varphi}_1-\varphi_2=0$, implying $\tilde{\varphi}_1-2\varphi_\alpha=\pi$. Hence, the optimization of the relative squeezing phase is governed by a competition between the factor $|\He_{N_1}|^2/|1+\tilde{\tau}_1\tau_2^*|^{N_1+1}$ favoring opposite squeezing directions, $\tilde{\varphi}_1-\varphi_2=\pi$, and the exponential contribution that favors aligned squeezing, $\tilde{\varphi}_1-\varphi_2=0$. In both cases the optimal relative phase between the first squeezer and the phase of $\alpha$ is $\tilde{\varphi}_1-2\varphi_\alpha=\pi$.

In order to better understand this potential competition, we consider the scale set by $|\alpha|$. For small $|\alpha|$, the impact of the exponential factor on the fidelity is comparatively weak, and the optimization of the relative squeezer phase is mainly governed by the algebraic denominator $|1 + \taut_1\tau_2^*|$. In this regime the fidelity is therefore maximized by opposite squeezing directions, as can also be verified directly from a small-$|\alpha|$ expansion. For large $|\alpha|$, by contrast, one might instead expect the exponential factor to dominate, implying aligned squeezers to be favored in this regime. 
However, this reasoning is incomplete for large but finite $|\alpha|$ as the fidelity optimization remains more subtle. The phase cannot be optimized from the exponential factor alone, because changing the relative phase also shifts the optimum with respect to the squeezing magnitudes. In fact, the optimum results from a balance between exponential enhancement and algebraic suppression near the boundary of the allowed parameter space. Consequently, it is possible for the opposite alignment $\varphit_1-\varphi_2=\pi$ to remain optimal even for large, but finite, cat amplitudes.

To make this transparent, we consider the leading large-$|\alpha|$ expansion of the fidelity,
\begin{align}
F \approx \frac{1 \pm (-1)^{N_1}}{\E^{|\alpha|^2} \pm \E^{-|\alpha|^2}} 
\frac{|\alpha|^{2N_1} (1-|\breve\tau_2|^2)^{N_1+\frac{1}{2}} \, \E^{- \Re(\breve\tau_2 \alpha^{*2})} }{N_1! \Fhyp\!\! \Big(\frac{1-N_1}{2},-\frac{N_1}{2},1,\frac{1}{|\taut_1|^{2}}\Big)},
\end{align}
where we reparametrized the fidelity in terms of $(\taut_1,\breve\tau_2)$ instead of $(\taut_1,\tau_2)$ for convenience. At fixed $\breve\tau_2$, the remaining dependence on $|\taut_1|$ is now entirely contained in the hypergeometric normalization. Since this hypergeometric function is a positive truncated polynomial in $|\taut_1|^{-2}$, the fidelity is maximized by taking the limit $|\taut_1|\to 1$. 
The optimal phase value of $\breve\tau_2$ is given by $\arg\breve\tau_2 = 2\varphi_\alpha + \pi$. Furthermore, the optimal modulus $|\breve\tau_2|$ can be analytically determined as well within the large-$|\alpha|$ expansion,
\begin{align}
|\breve\tau_2| &= \sqrt{1+ \frac{(2N_1+1)^2}{4|\alpha|^4}} - \frac{(2N_1+1)}{2|\alpha|^2} \notag \\
&= 1-\frac{2N_1 + 1}{2|\alpha|^2} +O(|\alpha|^{-4}) .
\end{align}
For fixed $\breve\tau_2$ and $\taut_1$, the second squeezing parameter is determined by
\begin{align}
\tau_2 = \frac{ \breve\tau_2(1-|\taut_1|^2) - \taut_1(1-|\breve\tau_2|^2)}{ 1-|\taut_1\breve\tau_2|^2}.
\end{align}
Therefore, $\tau_2$ has to approach its optimal value in a correlated way to $\taut_1$ in the boundary limit $|\taut_1| \to 1$ with $|\breve\tau_2|<1$ fixed, giving $\tau_2 \to -\taut_1$.
Thus, we obtain $|\tau_2|\to1$ and $\varphit_1 - \varphi_2 \to \pi$, implying that even for large cat amplitudes $|\alpha|$ the optimum is approached by opposite squeezing directions, even though the purely exponential contribution alone would suggest aligned squeezers. In summary, it is not appropriate to infer the optimal phase solely from the exponential factor, since the optimum simultaneously approaches the boundary $|\breve\tau_2|\to1$. In this regime the exponential enhancement is balanced by the vanishing prefactor $(1-|\breve\tau_2|^2)^{N_1+\frac{1}{2}}$. The resulting optimum therefore lies in a narrow region near the boundary of the physically allowed parameter space and favors opposite rather than aligned squeezing directions.

Numerical optimization confirms that the phase relations $\varphit_1-\varphi_2 = \pi$ and $\varphit_1-2\varphi_\alpha = \pi$ are optimal for all tested parameter sets. This agreement with both the small- and large-$|\alpha|$ analyses indicates that the asymptotic arguments capture the relevant optimal phase structure over a broad parameter range.

\begin{figure}[t]
    \centering
    \includegraphics[width=0.99\linewidth]{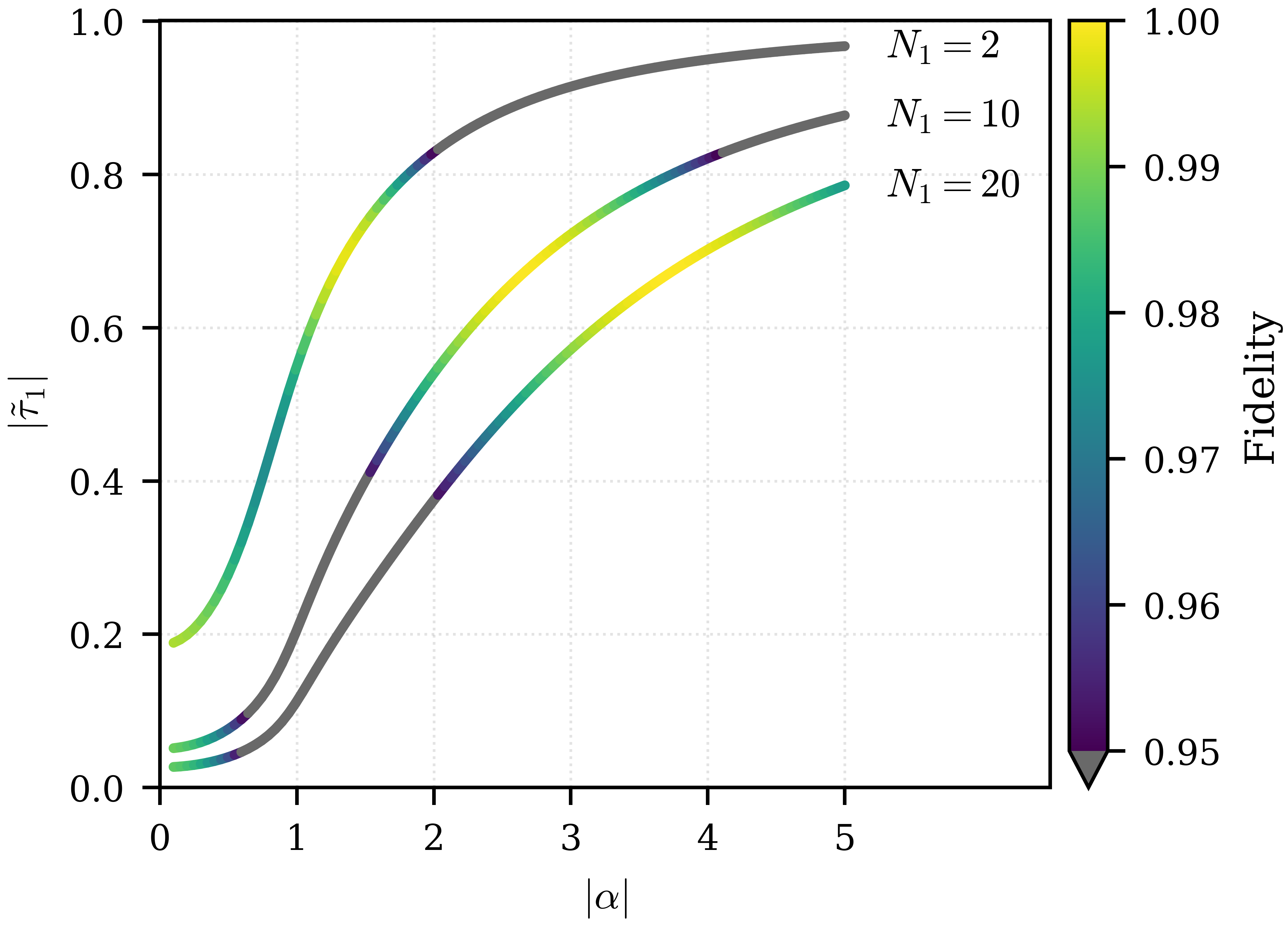}
    \caption{Cat state fidelity ridges for the target-driven view for a variation of even subtracted photon numbers $N_1$ at fixed $\tau_2 = 0.5$ in the nonlinear interferometer scheme. Grey lines describe fidelities below $0.95$.}
    \label{fig:ridgecurvesNI}
\end{figure}

Having established the optimal phase relations, we next examine how the additional squeezing stage modifies the fidelity landscape discussed for plain photon subtraction. For fixed $|\tau_2|$, the fidelity landscape in the $(|\alpha|,|\taut_1|)$-plane retains the characteristic ridge-like structure of the plain photon-subtraction scheme. However, the fidelity along these ridges is now generally nonmonotonic and exhibits a localized peak at an intermediate target amplitude. Thus, increasing $|\alpha|$ along a given ridge can initially improve the fidelity, before it decreases again beyond the optimal region. As $N_1$ increases, the ridge shifts toward larger target amplitudes, while the position and width of its maximum also change, see Fig.~\ref{fig:ridgecurvesNI} for an example with $\tau_2 = 0.5$. Since both the position and height of the localized maxima depend on $|\tau_2|$, this parameter must also be included in the full optimization and the displayed landscape represents only an illustrative two-dimensional cut through the full parameter space.

In principle, a resource-driven perspective analogous to the plain photon-subtraction scheme can also be adopted. However, for \(N_1\geq2\) the fidelity depends on two independent squeezing parameters, such that no unique one-dimensional resource-driven optimum exists without specifying how the available squeezing resources are distributed between the two stages. We therefore focus on the target-driven perspective here and address finite squeezing resources explicitly below.

Performing the full fidelity optimization, the numerical results further show that finite optimal solutions generally satisfy $|\taut_1|>|\tau_2|$, with the effective first-stage squeezing also exceeding that of the corresponding plain photon-subtraction scheme. Heuristically, the first squeezing stage may therefore be viewed as overshooting the squeezing required in the plain scheme, while the second, oppositely oriented squeezing operation partially compensates this excess and reshapes the Fock-state distribution towards the target cat state. This reshaping considerably reduces the subtraction order required for a given target fidelity. For even cat states with $|\alpha|=2$, the plain photon-subtraction scheme requires at least $N_1=8$ to reach $F>0.98$, whereas the nonlinear-interferometer-type scheme reaches this threshold already for $N_1=4$. For $|\alpha|=3$, the corresponding subtraction order is reduced from $N_1=18$ to $N_1=6$. This reduction in the required subtraction order comes, however, at the price of substantially larger squeezing resources. For several of these low-order branches, the global fidelity maximum is in fact approached only asymptotically. For $|\alpha|=3$, this occurs for $2\leq N_1\leq8$, where $|\taut_1|$ and $|\tau_2|$ both approach unity in a correlated manner while the combined output squeezing remains finite. Finite fidelity optima are recovered from $N_1=9$ onward. For $|\alpha|=2$, the corresponding asymptotic behavior occurs for $N_1=2$ and $3$, while the optimum is finite from $N_1=4$ onward. Nevertheless, these asymptotic fidelity optima can already be approximated very closely with finite squeezing resources. For example, for $|\alpha|=3$ and $N_1=6$, the finite realization $|\xi_1|=4.939$, $|\xi_2|=3.135$, and $T_1 = 0.99877$ yields $F\simeq0.98455$, only slightly below the asymptotic value $F\simeq0.98465$.

Beyond this low-order, boundary-dominated regime, the additional squeezing cost becomes progressively less pronounced. For $|\alpha|=3$, when the fidelity-optimal state is realized at the maximal heralding probability, the required first-stage squeezing initially decreases with increasing $N_1$ and approaches the values of the plain photon-subtraction scheme, while the second-stage squeezing becomes progressively weaker. From approximately $N_1\geq18$ onward, the first-stage squeezing follows the slowly increasing trend of the plain photon-subtraction scheme and remains only slightly larger, whereas the optimal second-stage squeezing continues to decrease. Thus, the largest resource overhead of the nonlinear-interferometer-type scheme occurs precisely in the low-$N_1$ regime in which the second squeezing stage provides the strongest reshaping advantage.

\subsection{Heralding probability and fidelity-probability trade-off}
As already discussed at the beginning of this section, the second squeezing operation is a unitary transformation and therefore does not directly modify the probability of detecting a given photon number $N_1$. Thus the probability of heralding a particular event is given by Eq.~\eqref{eq:ProbPS} as in the plain photon-subtraction scheme. 
Nonetheless, an implicit dependence is expected as the fidelity optimization generally yields different optimal values of $|\taut_1|$ in the two schemes. Although one might therefore expect correspondingly different maximal heralding probabilities, $P_{N_1}$ varies only weakly with $|\taut_1|$ over a broad parameter range for sufficiently large $|\taut_1|$ for $N_1\geq2$, see Fig.~\ref{fig:prob_vs_tau}. Thus, the maximal heralding probabilities at fixed $N_1 \geq 2$ remain similar in both schemes.
For $N_1=1$, the situation is slightly different because the fidelity depends only on the combined squeezing $\breve{\tau}_2$. For $|\alpha|=3$, fidelity optimization gives $|\breve{\tau}_{2,\mathrm{opt}}|\simeq0.84713$, yielding the same maximal fidelity as in the plain photon-subtraction scheme. Within the resulting degenerate family of fidelity-optimal realizations, maximizing the heralding probability drives $|\taut_1|$ and $T_1$ towards unity, with the second squeezing adjusted such that $\breve{\tau}_2$ remains fixed. The corresponding heralding probability approaches $P_{N_1=1}\simeq0.19245$, compared with $0.18135$ in the plain photon-subtraction scheme.

\begin{figure}[t!]
    \includegraphics[width=0.99\linewidth]{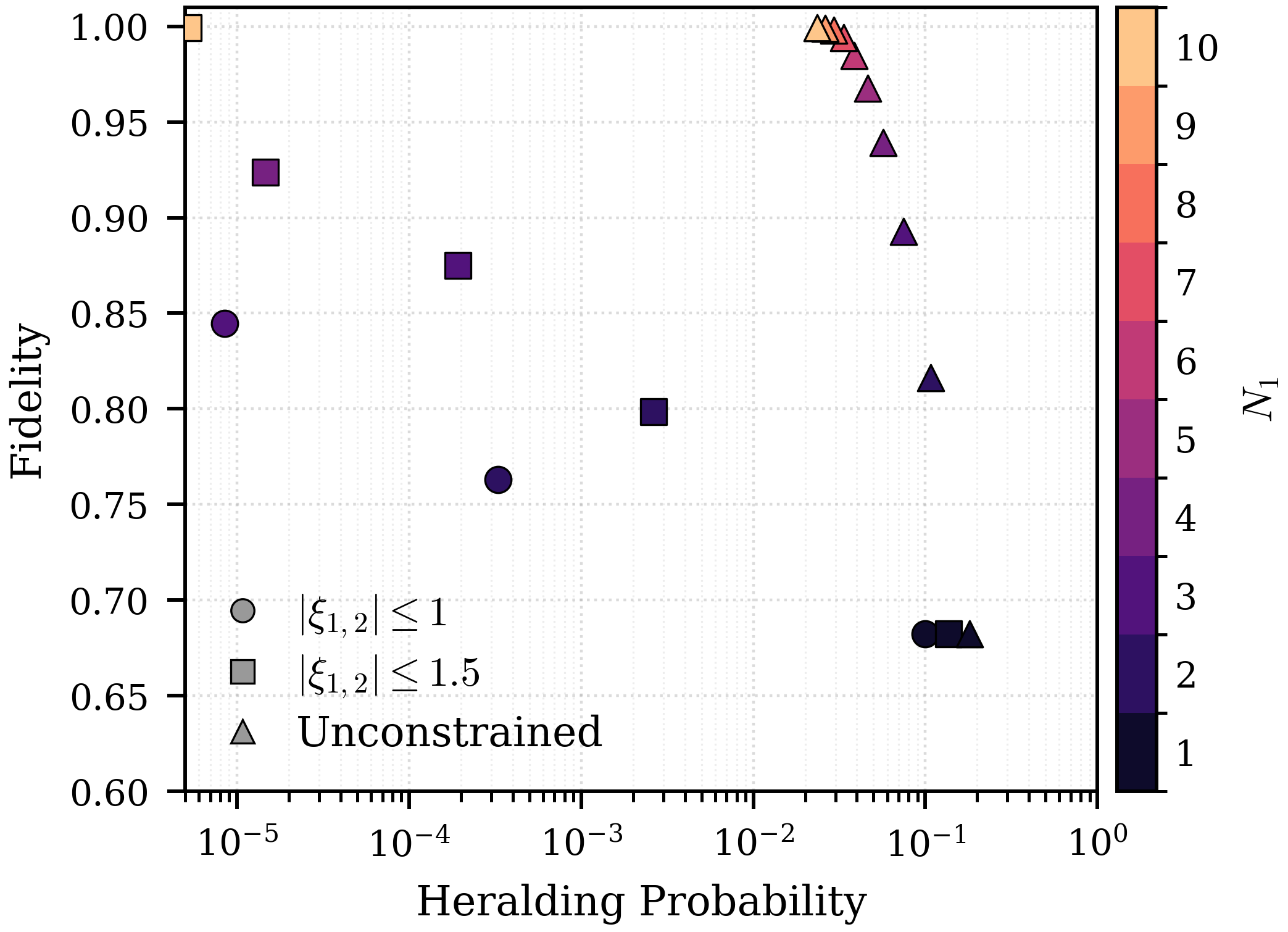}
    \caption{Fidelity over heralding probability for the nonlinear-interferometer scheme for $|\alpha| = 3$ cat states. We compare different maximum squeezing levels $|\xi_{\mathrm{max}}|$ (markers) and different photon subtraction numbers $N_1$ (color encoded).}
\label{fig:fidelity_probability_NI}
\end{figure}

Nevertheless, one achieves higher probabilities for given fidelities in an indirect fashion as reshaping the conditional output state allows substantially higher fidelities to be reached at comparatively small subtraction numbers $N_1$. This behavior is illustrated in Fig.~\ref{fig:fidelity_probability_NI}, where the unconstrained optima reach or asymptotically approach fidelities close to unity already for moderate subtraction numbers while retaining heralding probabilities on the order of $10^{-2}$ to $10^{-1}$. Since lower-order subtraction events generally occur with higher probability, this leads to an improved trade-off between the fidelity and the heralding probability relative to the plain photon-subtraction scheme.

Once the available squeezing is constrained, the achievable fidelity and heralding probability are reduced significantly if the constraint excludes the unconstrained optimum, see Fig.~\ref{fig:fidelity_probability_NI}. This occurs more readily in the nonlinear-interferometer-type scheme because it generally requires stronger squeezing than the plain one. Even for the comparatively high bound $|\xi_{1,\mathrm{max}}|=1.5$, the unconstrained fidelity solutions for $N_1<10$ require effective first-stage squeezing beyond the physically accessible range. The corresponding unconstrained fidelity maxima therefore cannot be reached under this constraint. Approaching the largest attainable effective squeezing then requires $T_1\to1$, which suppresses the heralding probability. Under tighter bounds, the limitations are even more severe, see the $|\xi_{1,\mathrm{max}}| = 1$ example in Fig.~\ref{fig:fidelity_probability_NI}. 
Although the maximal fidelities remain higher than in the plain photon-subtraction scheme, this advantage largely disappears if fidelity is sacrificed to recover a comparable heralding probability of both schemes. For a fixed heralding branch $N_1$ and a given heralding probability defined by the plain photon-subtraction scheme, the optimal second squeezing is then typically weak, being of the order of $0.1$, and the fidelity gain is below one percentage point compared to the plain photon-subtraction scheme. A substantial advantage from the second squeezer therefore requires significantly larger squeezing resources. For the example $|\alpha|=3$, this advantage emerges at $N_1=10$ only for $|\xi_{1,\mathrm{max}}|\gtrsim2.5$, while the required bound increases monotonically for larger $N_1$. For lower subtraction orders, even stronger squeezing is needed for the current example. The precise thresholds also depend on the target amplitude $|\alpha|$.

An advantage nevertheless remains when the comparison is made at a fixed minimum fidelity rather than for a fixed heralding branch. For the example of an even cat state with $|\alpha|=3$ and maximal available squeezing resources of $|\xi_{1,2}|\leq 1.5$, a fidelity threshold of $F_{\mathrm{min}}=0.95$ requires at least an $N_1=10$ photon event in the plain photon-subtraction scheme, with $P_{N_1=10} = 0.0107$, whereas the extended scheme reaches the same threshold already for $N_1=6$, with a comparable heralding probability of $0.0104$. In this case, the first squeezer nearly saturates the bound, $|\xi_1|=1.4965$, while only weak second-stage squeezing, $|\xi_2|=0.3205$, is required. These effects are more pronounced at higher fidelities. For $F_{\mathrm{min}}=0.98$ and the same squeezing constraint, the plain photon-subtraction scheme requires at least $N_1=18$, yielding a heralding probability of $0.0030$, while the extended scheme already succeeds for $N_1=8$ with $P_{N_1=8} = 0.0048$, $|\xi_1|=1.4992$, and $|\xi_2|=0.4071$. Thus, the nonlinear-interferometer scheme can substantially reduce the required photon-number resolution to satisfy a given fidelity threshold.

\subsection{Fidelity-constrained multibranch success probability optimization}
\label{sec:NI-Multibranch}
\begin{figure}[t]
    \centering
    \includegraphics[width=0.99\linewidth]{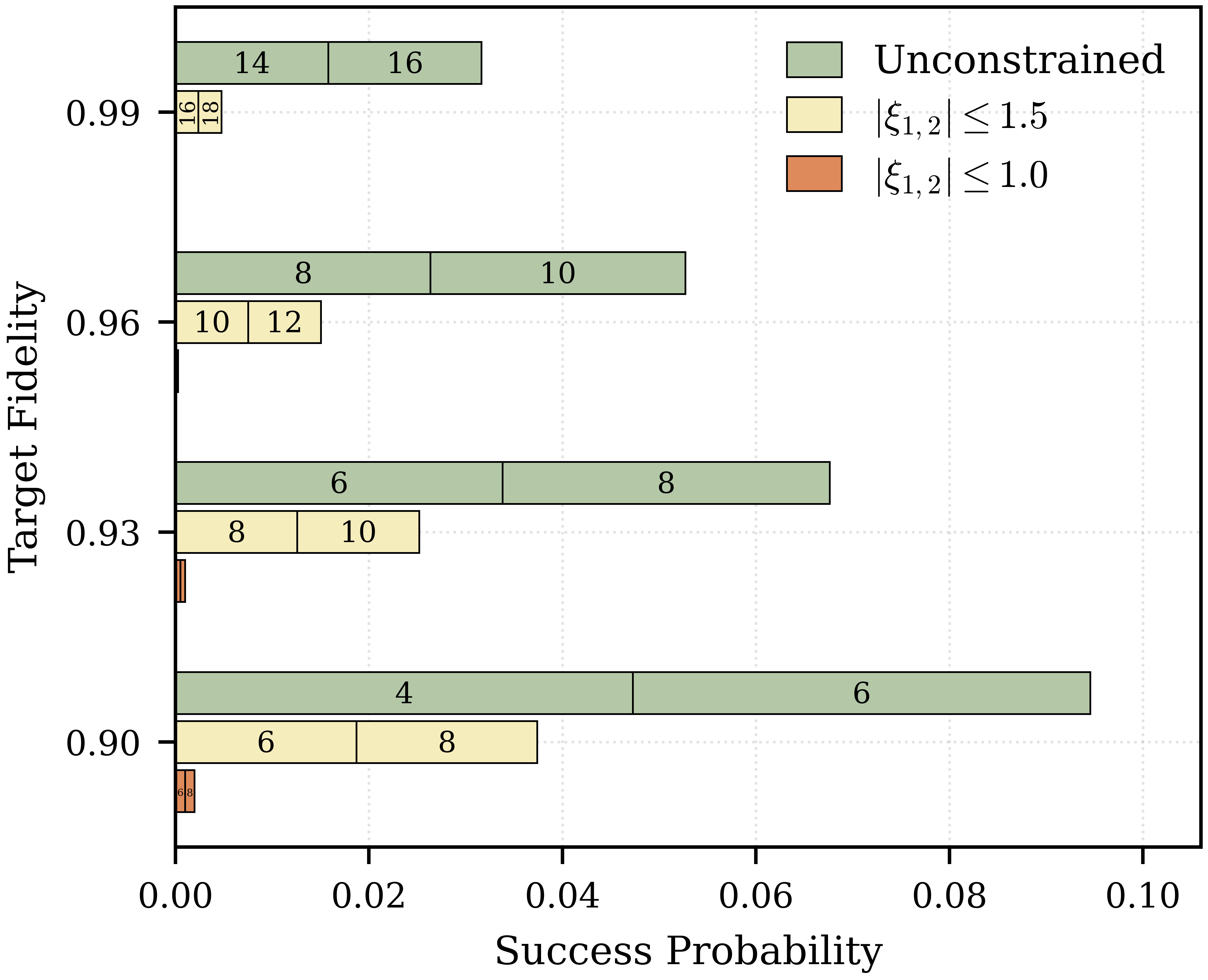}
    \caption{Fidelity--success-probability trade-off for multibranch even cat state generation with amplitude $|\alpha|=3$ in the nonlinear interferometer scheme. The total optimized success probability is plotted as a function of the target fidelity constraint across different hardware limits on the squeezing parameter $|\xi_i|$ in each stage $i$. Discrete stacked bars decompose the cumulative success probability into the absolute contributions of individual heralding modes ($N_1 = 2,4,6,...$). Colors indicate distinct squeezing constraints ($|\xi_i| \le 1.0$, $|\xi_i| \le 1.5$, and the unconstrained parameter space).}
    \label{fig:ni_multibranchOptimization}
\end{figure}

For a proper overall comparison, we again consider the multibranch optimization problem as for the plain photon-subtraction scheme, retaining all photon-number branches with fidelity above $F_{\mathrm{min}}$ and maximizing their total success probability. 
For an $|\alpha|=3$ even cat state, we show illustrative fidelity examples in Fig.~\ref{fig:ni_multibranchOptimization} and further details and examples in Tabs.~\ref{tab:NLI-multibranch-even} and \ref{tab:NLI-multibranch-odd}.

In the absence of squeezing constraints, the nonlinear-interferometer scheme consistently improves the fidelity--success-probability trade-off. Over the considered range ${0.90\leq F_{\mathrm{min}}\leq0.99}$, the optimized total success probability exceeds that of the plain photon-subtraction scheme by factors ranging from approximately $1.13$ to $1.73$, with the largest relative gain occurring at the highest fidelity threshold, see also Fig.~\ref{fig:SuccessProbability-Comparison} for a direct comparison of all three studied schemes. At the same time, the required detected photon numbers are substantially reduced. For example, at $F_{\mathrm{min}}=0.98$, the extended scheme accepts the branches $N_1\in\{10,12\}$, whereas the plain photon-subtraction scheme requires $N_1\in\{50,52,\ldots,60\}$ to maximize the success probability under the same fidelity constraint. Similarly, at $F_{\mathrm{min}}=0.99$, the extended scheme accepts $N_1\in\{14,16\}$, compared with $N_1\in\{74,76,\ldots,84\}$ in the plain photon-subtraction scheme.

As in the plain photon-subtraction analysis, we repeated the unconstrained multibranch optimization for $|\alpha|=2$ and $4$, revealing the same overall scaling with the cat amplitude. For all tested fidelity thresholds, increasing $|\alpha|$ decreases the optimized success probability and shifts the accepted heralding outcomes toward larger photon numbers.

For the nonlinear-interferometer-type scheme, the squeezing requirements of the unconstrained optimization are more subtle as already discussed for the single-branch optimization. While finite global optima are obtained for $F_{\mathrm{min}}=0.91$, $0.92$, $0.98$, and $0.99$, the optimal success probabilities at $F_{\mathrm{min}}=0.90$ and ${0.93\leq F_{\mathrm{min}}\leq0.97}$ are approached only asymptotically for even cat states with $|\alpha| = 3$. In these cases, the optimization drives $|\tilde{\tau}_1|$, $|\tau_2|$, and $T_1$ towards unity and hence both physical squeezing amplitudes towards infinity. As in the single-branch optimization, the two squeezing operations approach this limit in a correlated manner with opposite squeezing directions, such that their large individual contributions partially compensate and the effective squeezing $\breve{\tau}_2$ of the conditional output state remains finite. The corresponding optimized probabilities should therefore be understood as suprema in the infinite-squeezing limit rather than as probabilities attained at a finite parameter point.

Importantly, this asymptotic behavior does not imply that comparable success probabilities require arbitrarily large squeezing. Representative finite-resource parameter sets already approach the limiting probabilities very closely. Using, for example, $|\xi_1|\simeq 3.19$ and $|\xi_2|\simeq 1.37$, a finite realization for $F_{\mathrm{min}}=0.90$ yields $P_{\mathrm{succ}}\simeq 0.09455$, compared with the asymptotic value $0.09474$. For the asymptotic branch set $\{6,8\}$, which is optimal for $0.93 \leq F_{\mathrm{min}} \leq 0.95$, the corresponding finite realization gives $P_{\mathrm{succ}}\simeq0.06762$ instead of $0.06771$ with $|\xi_1|\simeq 3.35$ and $|\xi_2|\simeq 1.53$. Likewise, for $F_{\mathrm{min}}=0.96$ and $0.97$, a finite realization of the optimal branch set $\{8,10\}$ reaches $P_{\mathrm{succ}} \simeq 0.05272$, compared with the asymptotic value $0.05278$, using $|\xi_1|\simeq 3.48$ and $|\xi_2|\simeq 1.66$. Thus, although the exact unconstrained optima can lie on an infinite-squeezing boundary, more than $99.8\%$ of the corresponding limiting success probability can already be obtained with finite squeezing resources.

These representative operating points are chosen to closely approximate the absolute optimum of the extended scheme and therefore still require comparatively large squeezing. Such large squeezing is, however, not necessary to outperform plain photon subtraction. This is already evident from the finite global optimum at $F_{\mathrm{min}}=0.99$, for which the extended scheme yields $P_{\mathrm{succ}}\simeq0.03163$ with $|\xi_1|\simeq2.523$ and $|\xi_2|\simeq0.476$, whereas the unconstrained plain photon-subtraction scheme reaches only $P_{\mathrm{succ}}\simeq0.01829$ and requires $|\xi_1|\simeq2.934$. Hence, the nonlinear-interferometer scheme increases the success probability by a factor of about $1.73$ while requiring a smaller peak squeezing amplitude, although the simple sum of the two squeezing magnitudes, $|\xi_1|+|\xi_2|\simeq2.999$, is slightly larger than the squeezing required in the plain photon-subtraction scheme.

This motivates the complementary resource-oriented question of how much squeezing is required to merely outperform the unconstrained plain photon-subtraction optimum, rather than to attain the absolute optimum of the extended scheme. For example, at $F_{\mathrm{min}}=0.98$, the optimization of the plain photon-subtraction scheme gives $P_{\mathrm{succ}}\simeq0.02625$ and requires $|\xi_1|\simeq2.775$. In the extended scheme, a finite parameter set with only $|\xi_1|=2$ and $|\xi_2| \simeq 0.216$ as well as $T_1 \simeq 0.52851$ already reaches $P_{\mathrm{succ}}\simeq0.03100$ using the branches $\{12,14\}$. The success probability is therefore increased by a factor of about $1.18$, while both the peak squeezing and even the simple sum of the squeezing magnitudes, $|\xi_1|+|\xi_2|\simeq2.216$, remain below the squeezing required by the plain photon-subtraction scheme. Since $|\xi_1|+|\xi_2| < |\xi_{\mathrm{PS}}|$, this operating point also satisfies $|\xi_1|^2+|\xi_2|^2 < |\xi_{\mathrm{PS}}|^2$. Thus, the reduction persists also for the quadratic correlated-resource measure discussed below. The additional squeezer can therefore improve the fidelity--success-probability trade-off without necessarily increasing either the maximum individual squeezing amplitude or these simple measures of the total squeezing requirement.

For the experimentally more relevant bound $|\xi_{1,\mathrm{max}}|=|\xi_{2,\mathrm{max}}|=1.5$, the advantage is smaller but remains significant, particularly at high fidelity. The optimized success probability is comparable to that of the plain photon-subtraction scheme at low thresholds and exceeds it by factors of approximately $1.38$, $1.96$, and $3.87$ at $F_{\mathrm{min}}=0.95$, $0.98$, and $0.99$, respectively. The accepted photon numbers are again reduced, for instance from $\{20,22\}$ to $\{12,14\}$ at $F_{\mathrm{min}}=0.98$ and from $\{30,32\}$ to $\{16,18\}$ at $F_{\mathrm{min}}=0.99$. In these constrained optima, the first squeezer essentially saturates the available bound, while the second squeezer remains comparatively weak with $|\xi_2|\lesssim 0.22$ over the considered thresholds.
Under the tighter bound $|\xi_{1,\mathrm{max}}|=|\xi_{2,\mathrm{max}}|=1$, the extended scheme coincides with the plain photon-subtraction scheme for $0.9 \leq F_{\mathrm{min}}\leq0.93$, as the optimum reduces to $\xi_2=0$. At higher thresholds, a weak but nonzero second squeezing stage leads to a progressively increasing advantage by permitting lower-order heralding branches to satisfy the fidelity constraint. At $F_{\mathrm{min}}=0.98$, for example, the success probability is enhanced by a factor of about $2.31$ with $|\xi_2|\simeq0.124$, while the required photon number decreases from $N_1=18$ to $N_1=12$. At $F_{\mathrm{min}}=0.99$, the enhancement reaches a factor of approximately $15.7$ with $|\xi_2|\simeq0.171$, while the optimal branch shifts from $N_1=28$ to $N_1=14$. The corresponding absolute probabilities increase from approximately $1.57\times10^{-6}$ to $2.46\times10^{-5}$ in the latter case and therefore remain small despite the large relative improvement.

Across the constrained cases, the number of accepted branches remains comparable in both schemes. In the unconstrained case, however, the nonlinear-interferometer scheme requires fewer branches for almost all fidelity thresholds, particularly at high fidelities. The improvement in both regimes therefore arises primarily from shifting the accepted outcomes toward lower photon numbers and increasing their combined heralding probability, rather than from accepting a larger number of branches.

So far, the squeezing constraints have been imposed independently on the two squeezers. In practice, however, the available resources may also be correlated, for example if both stages are pumped by the same laser source with limited total power. Such a scenario could be modeled by a joint constraint on $\xi_1$ and $\xi_2$. A simple phenomenological model could take the form $|\xi_1|^2+|\xi_2|^2\leq \xi_{\mathrm{tot}}^2$,
reflecting the distribution of a limited pump power between two $\chi^{(2)}$ nonlinear stages. The precise relation may differ depending on the nonlinear media, interaction lengths, losses, and pump-coupling efficiencies. Such a constraint introduces a direct trade-off between the two stages, since increasing the squeezing in one stage reduces the resources available to the other. Nevertheless, we do not expect this to alter the present conclusions substantially, since the optimal second-stage squeezing is typically weak and the resource demand is dominated by the first squeezer.

Finally, the odd-parity case exhibits the same overall qualitative behavior as the even-parity case, but with slightly more pronounced parity-dependent variations than for plain photon subtraction. In the unconstrained regime, the optimized success probabilities can differ by factors up to $1.15$ at individual fidelity thresholds, and the occurrence of finite versus asymptotic optima also differs between the two parities. Under finite squeezing constraints, the optimized probabilities are again generally of comparable magnitude, although noticeable differences remain at some thresholds. No systematic advantage of either parity is observed over the full fidelity range.

\section{Alternating Scheme}
\label{sec:AlternatingScheme}
\begin{figure}[t!]
    \centering
    \includegraphics[width=0.75\linewidth]{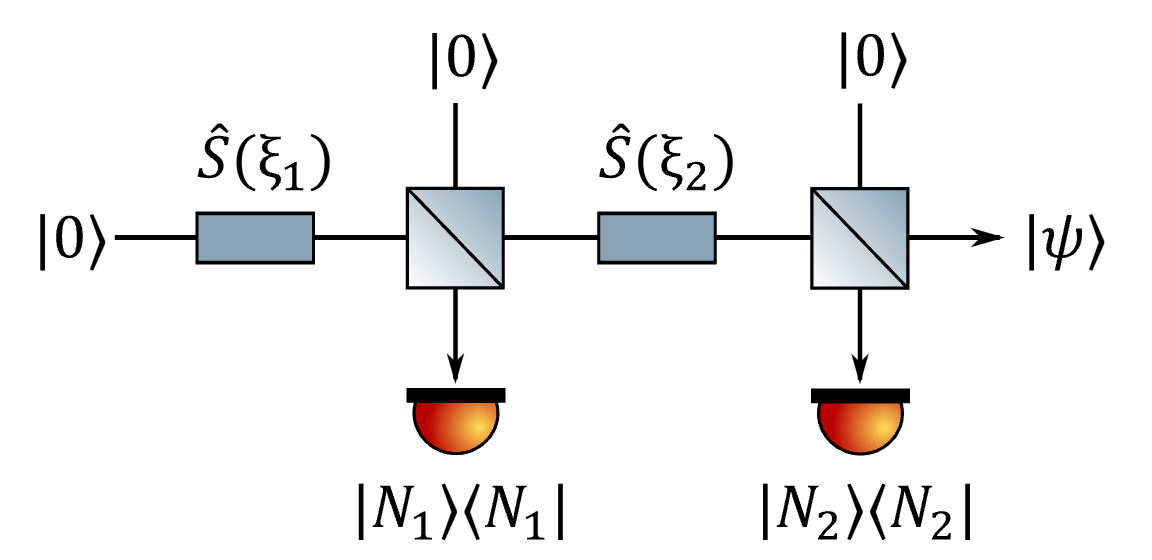}
    \caption{Setup of alternating squeezing and photon subtraction operations with two stages each.}
    \label{fig:setup_double}
\end{figure}

A natural extension of the nonlinear-interferometer-type scheme is to introduce a further beam-splitter-based photon subtraction after the second squeezing operation, see Fig.~\ref{fig:setup_double}. In the following, we investigate whether this additional conditioning stage further improves the fidelity with a target cat state and, subsequently, the success probability to generate it. At the same time, we analyze whether an improved state approximation can be achieved with reduced squeezing resources or less demanding photon-number resolution.

We denote the photon numbers detected at the first and second subtraction stages by $N_1$ and $N_2$, respectively. The second subtraction is implemented by a beam splitter with transmittance $T_2$ and transmission phase $\phi_{\rmT_2}$. In analogy to the first subtraction stage, it attenuates the combined squeezing parameter $\breve{\tau}_2$ introduced in the previous section according to $\tilde{\breve{\tau}}_2 = T_2\E^{2\I\phi_{\rmT_2}}\breve{\tau}_2$. It is also convenient to introduce the correspondingly attenuated parameter associated with the second squeezer alone, $\taut_2 = T_2\E^{2\I\phi_{\rmT_2}}\tau_2$. As for the first subtraction stage, the beam-splitter transmission phase can be absorbed into the remaining optical phases, and we therefore set $\phi_{\rmT_2} = 0$ without loss of generality.

Applying the second photon-subtraction to the state generated by the nonlinear-interferometer-type scheme gives the structural form
\begin{align}
 \ket{\psi_{N_1,N_2}} &\sim \!\! \sum_{k=0}^{\min(N_1,N_2)} \!\!\!\! c_k \,
 \He_{N_1-k}\!\bigg(\sqrt{\tfrac{1}{2}(\tilde{\breve{\tau}}_2-\taut_2)} \, \hat{a}^{\dagger}\bigg) \notag \\
 &\qquad\qquad \times
 \He_{N_2-k}\!\bigg(\sqrt{\tfrac{1}{2}\tilde{\breve{\tau}}_2}\, \hat{a}^{\dagger}\bigg)
 \hat{S}(\tilde{\breve{\xi}}_2)\ket{0}.
\label{eq:alternating-state-product}
\end{align}
The coefficients $c_k = c_k(\tilde{\breve{\tau}}_2,\taut_2,N_1,N_2)$ contain combinatorial factors and squeezing-parameter-dependent prefactors. Their structure follows directly by commuting the $N_2$ annihilation operators through the non-Gaussian Hermite polynomial and the Gaussian exponential. Equation~\eqref{eq:alternating-state-product} permits a simple heuristic interpretation. Out of the $N_2$ photons subtracted at the second stage, $k$ annihilation operators act on the non-Gaussian part generated by the first subtraction event and reduce the order of its Hermite polynomial from $N_1$ to $N_1-k$. The remaining $N_2-k$ annihilation operators act on the Gaussian part and generate the second Hermite polynomial of order $N_2-k$. Since the photons detected at the second stage cannot be assigned to either contribution, the coherent sum over all possibilities $k=0,\ldots,\min(N_1,N_2)$ must be retained.

For the subsequent analysis, it is more convenient to expand the sum over products of Hermite polynomials into a sum of single Hermite polynomials. 
This can be derived directly from the generating function of the Hermite polynomials. The two differently scaled arguments are first expressed in terms of the common argument associated with the final effective squeezing. Multiplication of the corresponding generating functions then combines each product into Hermite polynomials whose orders differ by two. After inserting the coefficients $c_k$ and collecting terms of equal order, the normalized conditional state can finally be written as
\begin{widetext}
\begin{align}
 \ket{\psi_{N_1,N_2}}
 &= \frac{1}{\mathcal{N}_{N_1,N_2}}
 \sum_{k=0}^{\lfloor \frac{N_1}{2}\rfloor} \frac{\gamma_2^k}{k!(N_1-2k)!} 
 \He_{N_1+N_2-2k}\!\bigg( \sqrt{\tfrac{1}{2}\tilde{\breve{\tau}}_2}\,\hat{a}^{\dagger}\bigg)
 \hat{S}(\tilde{\breve{\xi}}_2)\ket{0}, 
 \notag \\
 \mathcal{N}_{N_1,N_2}^{2} &= 
 \sum_{j=0}^{\lfloor \frac{N_1+N_2}{2}\rfloor}
 \frac{ (2|\tilde{\breve{\tau}}_2|)^{N_1+N_2-2j}}{(N_1+N_2-2j)!} \;\;
 \bigg|\sum_{l=0}^{\min(j,\lfloor \frac{N_1}{2}\rfloor)} \!\! 
 \frac{(-1)^{l} (N_1+N_2-2l)!}{(N_1-2l)!l!(j-l)!}
 \frac{\gamma_2^l}{(1-|\tilde{\breve{\tau}}_2|^2)^{\frac{N_1+N_2}{2}-l}}
 \bigg|^2.
\label{eq:alternating-state}
\end{align}
\end{widetext}
where we introduced the abbreviation
\begin{align}
 \gamma_2 = \frac{\tau_2}{\tau_2-\breve{\tau}_2}
 = \frac{\taut_2/\tilde{\breve{\tau}}_2}{\taut_2/\tilde{\breve{\tau}}_2-1},
\label{eq:gamma-alternating}
\end{align}
depending only on the ratio $\taut_2/\tilde{\breve{\tau}}_2$. 
This formulation makes explicit that the normalized conditional state depends only on two complex parameters, namely $\tilde{\breve{\tau}}_2$ and the ratio $\taut_2/\tilde{\breve{\tau}}_2$ encoded in $\gamma_2$. The separate dependence on the beam-splitter parameters and the input squeezing parameters reappears when reconstructing a physical implementation and calculating the corresponding heralding probability. As before, a continuous family of experimental parameter choices may therefore generate the same normalized conditional state while exhibiting different heralding probabilities.

Several particular cases can be identified directly at the state level. First, for $N_1=0$ and $N_1=1$, the sum in Eq.~\eqref{eq:alternating-state} consists of its $k=0$ contribution only. Consequently,
\begin{align}
 \ket{\psi_{0,N_2}}
 &\sim \He_{N_2}\!\Bigg(
 \sqrt{\frac{\tilde{\breve{\tau}}_2}{2}}\hat{a}^{\dagger}\Bigg)
 \hat{S}(\tilde{\breve{\xi}}_2)\ket{0}, \\
 \ket{\psi_{1,N_2}}
 &\sim \He_{N_2+1}\!\Bigg(
 \sqrt{\frac{\tilde{\breve{\tau}}_2}{2}}\hat{a}^{\dagger}\Bigg)
 \hat{S}(\tilde{\breve{\xi}}_2)\ket{0}.
\end{align}
These states belong to the same families as plain $N_2$- and $(N_2+1)$-photon subtraction, respectively. The parameter $\gamma_2$ is therefore redundant for these two branches. Second, for $N_2=0$, Eq.~\eqref{eq:alternating-state} recovers the nonlinear-interferometer-type state discussed in the previous section, see Eq.~\eqref{eq:psiNPSNI}. Third, setting $\tau_2 = 0$ gives $\gamma_2 = 0$, such that only the $k=0$ contribution remains and the state reduces to plain $(N_1+N_2)$-photon subtraction with effective squeezing parameter $\tilde{\breve{\tau}}_2 = T_2\taut_1 = T_2 T_1 \tau_1$. 
The latter reduction also clarifies the relation to cascaded photon-subtraction schemes considered in Refs.~\cite{Kuts2022,Podoshvedov2023}. If the intermediate squeezing operation is absent, all partitions $(N_1,N_2)$ with the same total detected photon number $N=N_1+N_2$ prepare the same normalized conditional state. However, the corresponding heralding probabilities are branch dependent in general. For two sequential subtraction beam splitters, summing these probabilities over all partitions reproduces the heralding probability of plain $N$-photon subtraction, $ \sum_{N_1+N_2=N}P_{N_1,N_2} = P_{N}$, where $P_{N}$ is the probability given in Eq.~\eqref{eq:ProbPS} with $T_1$ replaced by the effective total transmittance $T_{\mathrm{eff}}=T_1T_2$, and $P_{N_1,N_2}$ denotes the joint probability of heralding $N_1$
and $N_2$ photons at the first and second subtraction stages, respectively. Hence, the probability penalty associated with selecting a single partition disappears if all branches with the same total count can be accepted. We return to this point when discussing the success probability. 
Fourth, a further equivalence occurs between the $N_1=2$ and $N_1=3$ state families at fixed total photon number $N=N_1+N_2$. Suppressing their common argument and squeezing operator, their non-Gaussian polynomial parts read $\frac{1}{2}\He_N+\gamma_2\He_{N-2}$ and $\frac{1}{6}\He_N+\gamma_2\He_{N-2}$, respectively. Both polynomial families can be related by the rescaling $\gamma_2\to\gamma_2/3$, under which the $N_1=3$ polynomial becomes identical to the $N_1=2$ polynomial up to an overall factor of three. Thus, in the absence of restrictions on the physically accessible parameter domain, the heralding patterns $(2,N-2)$ and $(3,N-3)$ describe the same normalized state family.

Except for these particular reductions, changing $N_1$ or $N_2$ modifies both the involved Hermite orders and their fixed relative coefficients. The corresponding state families are therefore generally not nested. Consequently, no general monotonicity of the fidelity with either subtraction number should be expected solely from the state structure.

\subsection{Fidelity with cat states}
The conditional state in Eq.~\eqref{eq:alternating-state} has parity $(-1)^{N_1+N_2}$. It is therefore compared with an even cat state for even $N_1+N_2$ and with an odd cat state for odd $N_1+N_2$. Evaluating the overlap with the corresponding cat state gives
\begin{align}
 F &= \frac{1\pm(-1)^{N_1+N_2}}{\E^{|\alpha|^2}\pm\E^{-|\alpha|^2}}
 \frac{\sqrt{1-|\tilde{\breve{\tau}}_2|^2}}{\mathcal{N}_{N_1,N_2}^{2}}
 \E^{-\Re(\tilde{\breve{\tau}}_2\alpha^{*2})}
 \notag \\
 &\quad \times
 \Bigg|\sum_{k=0}^{\lfloor \frac{N_1}{2}\rfloor}
 \frac{\gamma_2^k}{k!(N_1-2k)!}
 \He_{N_1+N_2-2k}\!\Bigg(
 \sqrt{\frac{\tilde{\breve{\tau}}_2}{2}}\alpha^*
 \Bigg)\Bigg|^2.
\label{eq:fidelityS-PS-S-PS}
\end{align}
The prefactor $1\pm(-1)^{N_1+N_2}$ again enforces the parity selection rule. Since the output state is fully characterized by $\tilde{\breve{\tau}}_2$ and $\gamma_2$ for fixed $N_1$ and $N_2$, the fidelity depends on the experimental parameters only through these two complex variables, which remain to be optimized. Additional variables are required only when the physical input squeezing parameters and the beam-splitter transmittances are reconstructed for the probability optimization similar to the previous cases.

The particular state reductions discussed above immediately carry over to the fidelity. The branches $(0,N_2)$ and $(1,N_2)$ reproduce the fidelities of plain $N_2$- and $(N_2+1)$-photon subtraction. We also obtain plain $(N_1+N_2)$-photon subtraction for $\tau_2=0$, whereas Eq.~\eqref{eq:fidelityS-PS-S-PS} reduces to the fidelity of the nonlinear-interferometer-type scheme for $N_2=0$. At fixed total photon number $N=N_1+N_2$, the branches $(2,N-2)$ and $(3,N-3)$ attain the same set of fidelities under the rescaling $\gamma_2 \to \gamma_2/3$, provided that the full parameter range is physically accessible. As these are identities between normalized state families, their heralding probabilities and required physical parameters need not coincide.

To obtain further insight into the fidelity in general and the phase optimization in particular, we consider the large-$|\alpha|$ expansion of the fidelity in analogy to the previous section. Its leading-order contribution reads
\begin{align}
 F &\approx
 \frac{1\pm(-1)^N}{\E^{|\alpha|^2}} \frac{\sqrt{1-|\tilde{\breve{\tau}}_2|^2}}{\mathcal{N}_{N_1,N_2}^{2}}
 \frac{(2|\tilde{\breve{\tau}}_2|)^N|\alpha|^{2N}}{(N_1!)^2}
 \E^{-\Re(\tilde{\breve{\tau}}_2\alpha^{*2})}.
\label{eq:alternating-large-alpha-fidelity}
\end{align}
The dependence on the target-state phase is only contained in the exponential factor at leading order, being maximized for
\begin{align}
 \arg(\tilde{\breve{\tau}}_2)-2\varphi_\alpha=\pi.
\label{eq:alternating-optimal-q-phase}
\end{align}
The remaining phase of $\gamma_2$ enters in this approximation through the normalization factor $\mathcal{N}_{N_1,N_2}$. Thus, minimizing $\mathcal{N}_{N_1,N_2}$ with respect to $\arg(\gamma_2)$ maximizes the fidelity. This optimization gives
\begin{align}
 \arg(\gamma_2)=0
\label{eq:alternating-optimal-gamma-phase}
\end{align}
for generic unrestricted solutions. This can be heuristically inferred from the explicit form of $\mathcal{N}_{N_1,N_2}$ given in Eq.~\eqref{eq:alternating-state}. For each fixed $j$, the inner sum has the structure $(-1)^l d_{jl} \E^{\I l \arg(\gamma_2)}$ with positive coefficients $d_{jl}$. Thus, for vanishing $\arg(\gamma_2)$, all contributions are collinear with alternating signs, allowing for destructive interference. 
Hence, the ratio $\tau_2/\breve{\tau}_2$ is real, i.e., its phase is either zero or $\pi$. These two possibilities distinguish which of the two physical squeezers shares the effective phase of $\tilde{\breve{\tau}}_2$. Choosing $\alpha$ real and $\tilde{\breve{\tau}}_2<0$, one finds that $0<\gamma_2 < 1$ corresponds to the first squeezer sharing the phase of $\tilde{\breve{\tau}}_2$, while the second squeezer differs by $\pi$. For $\gamma_2 > 1$, the second squeezer shares the effective phase and the first differs by $\pi$. 
Thus, the generic unrestricted optimum is characterized by two orthogonal squeezing axes, with one squeezing parameter satisfying $\varphi_i-2\varphi_\alpha=\pi$, i.e., being aligned with the cat state orientation in phase space, and the other $\varphi_j-2\varphi_\alpha=0$. Which squeezing stage assumes either relation is determined by the optimized value of $|\gamma_2|$.

Similar to the previous case, the large-$|\alpha|$ analysis provides a good indication of the optimal phase structure beyond the asymptotic regime. 
In particular, the phase structure in Eqs.~\eqref{eq:alternating-optimal-q-phase} and \eqref{eq:alternating-optimal-gamma-phase} is confirmed by direct numerical optimization for all tested generic branches across different values of $|\alpha|$, in particular for $|\alpha| = 3$.

Numerically, we also observe a characteristic crossover between the two alignment possibilities of the squeezers with the target cat state depending on the measured photon numbers. Increasing $N_2$ at fixed $N_1$ generally produces a crossover from the first to the second squeezer, while increasing $N_1$ at fixed $N_2$ consistently produces the reverse trend. For the example $|\alpha| = 3$, we find that all tested branches with $N_2>N_1$ favor the second squeezer, whereas branches with $N_1>N_2$ generally favor the first squeezer when the difference $N_1-N_2$ is sufficiently large.
However, for many branches favoring the first-squeezer alignment, the single-branch fidelity optimum in the unrestricted parameter space is approached only asymptotically in the infinite-squeezing limit. Close to the diagonal $N_1=N_2$, the alignment is branch and resource dependent.

A systematic scan over all partitions with $N_1+N_2\leq20$ finds the same phase structure as in the unrestricted squeezing analysis for squeezing constraints down to $|\xi_i|\leq1.5$. There are, however, two relevant qualifications. First, for $N_1=0$ and $N_1=1$, the state and its fidelity are independent of $\gamma_2$, such that the relative phase of the physical squeezers is not fixed by the fidelity. The combined squeezing parameter $\breve{\tau}_2$ can generally be realized by a continuous family of choices for $\taut_1$ and $\tau_2$ which generate the same normalized conditional state but can differ in their required squeezing resources and heralding probabilities. In particular, if the individual squeezing magnitudes are restricted below the value required to produce the optimal effective squeezing with one stage, aligned squeezers can add coherently according to $ |\breve{\tau}_2| = \frac{|\taut_1|+|\tau_2|}{1+|\taut_1\tau_2|} = \tanh(|\tilde{\xi}_1|+|\xi_2|)$. This aligned configuration is selected when the optimal effective squeezing cannot be reached by less constructive phase arrangements, or when the required squeezing resources are explicitly minimized. Second, sufficiently strong squeezing constraints can also change the phase optimum of generic branches. For $(N_1,N_2)=(2,1)$, $|\alpha|=3$, and $|\xi_i|\leq1$, direct complex-phase optimization yields an intermediate relative phase rather than the orthogonal-squeezer arrangement. However, the resulting fidelity improvement is small, from $0.83741$ to $0.83970$, and does not change the globally optimal $N=N_1+N_2$ fixed-total branch, which remains $(3,0)$ (see discussion below) with orthogonal squeezers and a fidelity $F_{3,0} = 0.84638$. Under the stronger constraint $|\xi_i|\leq0.5$, the deviation can become more pronounced. As an example, for $N_1=N_2=3$, $|\alpha|=3$, and $|\xi_i|\leq0.5$, the numerical fidelity optimum is approached by aligned rather than orthogonal squeezers. Thus, the phase structure found in the unrestricted and moderately constrained regimes is robust over the tested range but not universal under arbitrary resource constraints.

Furthermore, we observe a hierarchy among the fidelities at fixed total detected photon number $N=N_1+N_2$. For sufficiently large $N$, the endpoint branch in which all photons are detected at the first subtraction stage, $(N_1,N_2)=(N,0)$, yields the largest fidelity among all partitions. In particular, the nonlinear-interferometer-type endpoint always outperforms the plain-photon-subtraction endpoint $(0,N)$ in the unrestricted regime. For example, for the target amplitude $|\alpha|=3$, the complete numerical scan over all partitions up to $N=20$ confirms that $(N,0)$ gives the largest fidelity for $5\leq N\leq20$ as well as $N=2$. The low-order subtractions $N=3$ and $N=4$ constitute exceptions, where the largest fidelities are obtained for $(2,1)$ and $(2,2)$, respectively.

Under the squeezing constraint $|\xi_i|\leq1.5$, the endpoint pattern is recovered for $6\leq N\leq20$ and $N=2$. At $N=5$, however, the constraint prevents the $(5,0)$ branch from reaching its unrestricted optimum, such that the $(2,3)$ branch attains the largest constrained fidelity. This exception therefore results from the boundary of the physically reachable parameter domain. For $6\leq N <10$ and $N=2$, the constraint still reduces the fidelity of the $(N,0)$ branch relative to its unrestricted value, but this branch nevertheless remains optimal among all partitions. For $N \geq 10$, the constraint no longer changes the fidelity maximum of the optimal endpoint branch $(N,0)$ for all tested examples.
Tightening the constraint to $|\xi_i|\leq1$ changes the low-order behavior more substantially. The endpoint $(N,0)$ then provides the largest fidelity for $N=3$ and $8 \leq N \leq 20$, while the branches $(2,N-2)$ are optimal for $4\leq N \leq 7$. For $N=2$, the branches $(0,2)$ and $(1,1)$ belong to the same plain two-photon-subtracted state family being optimal under this constraint compared to the nonlinear-interferometer-type endpoint $(2,0)$. From $N=12$ onward, the squeezing constraint no longer changes the maximal achievable fidelity of the optimal $(N,0)$ endpoint. The constrained endpoint optima for $N=3$ and $8\leq N \leq 11$, as well as the $(2,2)$ optimum for $N=4$, lie on the boundary of the physically reachable state domain, such that their maximal fidelities have to be understood as limiting values when a strictly positive heralding probability is required. Although the fidelity advantage of $(N,0)$ over the nearest competing partitions decreases with increasing $N$, no further crossover is observed up to $N=20$.

Finally, the onset of this endpoint hierarchy depends on the target amplitude. For $|\alpha|=2$, only $N=3$ remains exceptional, with $(2,1)$ slightly outperforming $(3,0)$, whereas $(N,0)$ is optimal for $4\leq N\leq20$. For $|\alpha|=1$, the endpoint is likewise optimal throughout $4\leq N \leq 20$, while at $N=3$ the branches $(2,1)$ and $(3,0)$ attain the same maximal fidelity, consistent with their state-family equivalence discussed above. These branch orderings remain unchanged under the squeezing constraints $|\xi_i| \leq 1.5$ and $|\xi_i| \leq 1$, although the fidelity differences become extremely small as the optimal fidelities approach unity.

\subsection{Heralding probability and fidelity-probability trade-off}
For the present calculation, it is convenient to factorize the joint probability of detecting the heralding pattern $(N_1,N_2)$ as
\begin{align}
 P_{N_1,N_2} = P_{N_1}\,P_{N_2|N_1},
\label{eq:alternating-joint-probability}
\end{align}
where $P_{N_1}$ is the probability of the first subtraction event given by Eq.~\eqref{eq:ProbPS} and $P_{N_2|N_1}$ is the conditional probability of detecting $N_2$ photons at the second stage after the first outcome has occurred. The latter reads,
\begin{align}
 P_{N_2|N_1}
 &= \frac{N_1!}{N_2!} \mathcal{N}_{N_1,N_2}^{2}
 \sqrt{\frac{1-\frac{|\tilde{\breve{\tau}}_2|^2}{T_2^2}}{1-|\tilde{\breve{\tau}}_2|^2}}
 \left( \frac{1-T_2}{T_2}\frac{|\tilde{\breve{\tau}}_2|}{2} \right)^{N_2} \notag\\
 &\quad\times
 \frac{\left(1-\frac{|\tilde{\breve{\tau}}_2|^2}{T_2^2}\right)^{N_1}}{\left(\frac{2|\tilde{\breve{\tau}}_2|}{T_2}\right)^{N_1}
 \Fhyp\!\!\left(\frac{1-N_1}{2},-\frac{N_1}{2},1,\frac{1}{|\taut_1|^2}\right)}.
\label{eq:ProbAlternatingConditional}
\end{align}

Similar to the previous schemes, we first optimize the fidelity and subsequently keep the resulting state fixed while varying the beam-splitter transmittances to optimize the heralding probability. From the fidelity optimization, we directly obtain $\tilde{\breve{\tau}}_2$ and the attenuated parameter of the second squeezer $\taut_2$ encoded in $\gamma_2$. At this point, it might be tempting to exploit the factorization in Eq.~\eqref{eq:alternating-joint-probability} and optimize the two probability factors independently, as $P_{N_1}$ depends explicitly only on the transmittance of the first beam splitter $T_1$ while it can be shown that $P_{N_2|N_1}$ depends only on $T_2$. Indeed, inserting
\begin{align}
 \taut_1 = \frac{T_2(\tilde{\breve{\tau}}_2-\taut_2)}{T_2^2-\tilde{\breve{\tau}}_2\taut_2^*}
\label{eq:alternating-taut1-reconstruction}
\end{align}
into Eq.~\eqref{eq:ProbAlternatingConditional}, the conditional probability $P_{N_2|N_1}$ becomes a function of $T_2$ only once $\tilde{\breve{\tau}}_2$ and $\taut_2$ are fixed. Having $T_2$, we are able to reconstruct the experimental input parameter of the second squeezer via $\tau_2=\frac{\taut_2}{T_2}$ and the attenuated parameter $\taut_1$ of the first squeezer. We can finally determine the optimal choice for $T_1$ from $P_{N_1}$. However, the two optimizations are not independent because the first-stage probability $P_{N_1}$ also depends indirectly on $T_2$ through the reconstructed parameter $\taut_1(T_2)$. Consequently, maximizing $P_{N_2|N_1}$ with respect to $T_2$ first and subsequently maximizing $P_{N_1}$ with respect to $T_1$ does not generally maximize their product. Nevertheless, this sequential procedure often provides a good approximation when the optimized first-stage probability $P_{N_1,\mathrm{opt}}\big(\taut_1(T_2),T_1\big)$ varies only weakly over the relevant range of $T_2$, as occurs for many unrestricted solutions close to the large-squeezing plateau discussed below. It can, however, fail substantially near squeezing or reachability boundaries, where small changes in $T_2$ can strongly modify $P_{N_1}$.

The joint optimization can nevertheless be reduced to a single-variable numerical optimization over $T_2$. The optimal value of $T_1$ can still be determined analytically from $P_{N_1}$ for fixed $T_2$. In the absence of a squeezing constraint, this leads to the same quadratic relation as in the previous schemes, with solution given in Eq.~\eqref{eq:ToptSucProbPS}, evaluated at $\taut_1=\taut_1(T_2)$. Inserting the relation for $\taut_1(T_2)$ given by Eq.~\eqref{eq:alternating-taut1-reconstruction} also into the conditional probability part, the resulting reduced objective is
\begin{align}
 P_{N_1,N_2}(T_2) = P_{N_1}\!\big(\taut_1(T_2),T_{1,\mathrm{opt}}(T_2) \big)\, P_{N_2|N_1}(T_2).
\label{eq:alternating-reduced-probability}
\end{align}
The optimal $T_2$ is obtained by maximizing this product rather than the conditional probability alone. 
In contrast to the simple optimization of $T_1$ in the previous schemes, there is no analogous quadratic expression available for $T_2$ in general. In addition to the explicit powers of $T_2$ in Eq.~\eqref{eq:ProbAlternatingConditional}, it enters the hypergeometric-function argument and the optimized first-stage probability through $\taut_1(T_2)$. The remaining optimization over $T_2$ therefore has to be performed numerically.

There are, however, several particular cases in which this general reconstruction and optimization procedure either simplifies or requires additional qualification. These are related to particular branches that we already identified above at the level of the conditional state and its fidelity. 
A simple analytic exception occurs for $N_1=0$ for which the hypergeometric function in Eq.~\eqref{eq:ProbAlternatingConditional} equals unity. The first subtraction stage can be effectively removed by choosing $T_1=1$, such that $P_0=1$, and the setup reduces to plain $N_2$-photon subtraction with effective squeezing parameter $\tilde{\breve{\tau}}_2$. For $N_2\geq1$, the optimal second beam-splitter transmittance is therefore given by the quadratic solution given in direct analogy to Eq.~\eqref{eq:ToptSucProbPS}, $T_{2,\mathrm{opt}} = \frac{ \sqrt{|\tilde{\breve{\tau}}_2|^4+4N_2(N_2+1)|\tilde{\breve{\tau}}_2|^2} - |\tilde{\breve{\tau}}_2|^2}{2N_2}$. Since the fidelity is independent of $\gamma_2$ for $N_1=0$, the decomposition of the effective squeezing between the two physical squeezers remains free and may be chosen during the probability or resource optimization.

For $N_1=1$, the hypergeometric function also reduces to unity and the normalized state belongs to the class of plain $(N_2+1)$-photon-subtracted states. Nevertheless, the selected branch requires detections at both stages, and $P_{N_1=1}$ depends indirectly on $T_2$ through $\taut_1(T_2)$. Hence, the optimization of $T_2$ does not generally reduce to the quadratic plain-photon-subtraction result. As for $N_1=0$, $\gamma_2$ is not fixed by the fidelity and remains available for the probability or resource optimization. Another endpoint simplification occurs for $N_2=0$. Since only a zero-photon outcome is required at the second stage, its optimal beam splitter is transparent, $T_2=1$, and $P_{0|N_1}=1$. The joint probability then reduces to that of the nonlinear-interferometer-type scheme. Finally, for $\tau_2=0$, the conditional state becomes a plain $(N_1+N_2)$-photon-subtracted state. Nevertheless, the probability of a particular partition $(N_1,N_2)$ remains branch dependent. Only after summing over all partitions with fixed $N=N_1+N_2$ does one recover the plain photon-subtraction probability with effective total transmittance $T_{\mathrm{eff}}=T_1T_2$, as discussed above. Thus, $\tau_2=0$ provides a state-level simplification but not another general quadratic solution for the probability of an individual branch.

The reconstruction formulas for input squeezing parameters also make transparent how finite squeezing resources restrict the probability optimization. Imposing the input constraint $|\xi_i|\leq\xi_{\mathrm{max}}$, the admissible transmittances must satisfy $T_2 > |\tilde{\breve{\tau}}_2|$, $T_2 \geq \frac{|\taut_2|}{\tau_{\mathrm{max}}}$, $T_1 \geq \frac{|\taut_1(T_2)|}{\tau_{\mathrm{max}}}$, together with $T_i\leq1$, where $T_i<1$ is required whenever a nonzero photon number is detected at the corresponding stage. For each fixed $T_2$, the unconstrained analytic optimum for $T_1$ given in Eq.~\eqref{eq:ToptSucProbPS} is evaluated at $\taut_1=\taut_1(T_2)$. If this value lies below the allowed lower bound, the constrained optimum is attained at $T_1=\frac{|\taut_1(T_2)|}{\tau_{\mathrm{max}}}$, such that the first input squeezer saturates its resource bound. Similarly, the optimization over $T_2$ frequently favors the smallest admissible value. Whenever this boundary is set by $T_2=\frac{|\taut_2|}{\tau_{\mathrm{max}}}$, the second input squeezer saturates its bound. Physically, decreasing $T_2$ increases the fraction of light directed towards the second detector, while keeping $\taut_2$ fixed simultaneously requires a larger input squeezing, $|\tau_2|=\frac{|\taut_2|}{T_2}$.

Without a finite squeezing constraint, the tendency to favor smaller transmittances frequently drives one or both input parameters towards $|\tau_i|\to1$, corresponding to $|\xi_i|\to\infty$. The very large squeezing values obtained in an unrestricted numerical optimization should therefore be interpreted as representatives of an asymptotic boundary rather than as well-defined finite-resource optima. Imposing a finite squeezing bound regularizes this behavior and commonly leads to one or both squeezers saturating the available value. Saturating the input squeezing constraint alone does not imply a vanishing probability, since the corresponding beam splitter may still have $T_i<1$.

A different situation arises if the fidelity optimum itself lies on the boundary of the physically reachable state domain. For example, if it requires $|\taut_2|=\tau_{\mathrm{max}}$, its realization would demand $T_2=1$ and $|\tau_2|=\tau_{\mathrm{max}}$. For $N_2>0$, the second-stage probability then vanishes because no light is reflected towards the detector. Similarly, $|\taut_1|=\tau_{\mathrm{max}}$ requires $T_1=1$, leading to $P_{N_1>0}=0$. Such fidelity maxima are therefore only limiting values when a strictly positive heralding probability is required. A practical parameter configuration has to be chosen slightly inside the reachable domain, resulting in a small reduction of the fidelity. More generally, these cases are better characterized by optimizing the probability under a fidelity constraint than by attaching a probability to the exact fidelity supremum.

The endpoint reductions discussed for the fidelity optimization also organize the probability hierarchy at fixed total photon number $N=N_1+N_2$. Every interior partition with $N_1,N_2>0$ requires two nontrivial detection events, such that both factors in Eq.~\eqref{eq:alternating-joint-probability} are smaller than unity. This tends to suppress the joint probability relative to the endpoint schemes, for which one conditioning factor can be made unity. Accordingly, in the numerical scans over all partitions for $|\alpha|=3$ and total photon numbers up to $N=20$, the largest branchwise probability at the respective fidelity optimum is always found at one of the two endpoints. For $2\leq N\leq8$, the plain-photon-subtraction endpoint $(0,N)$ provides the largest probability, whereas the nonlinear-interferometer-type endpoint $(N,0)$ becomes slightly more probable for $9\leq N\leq20$. This probability hierarchy must, however, be considered together with the corresponding fidelities. The plain-photon-subtraction endpoint $(0,N)$ generally provides the largest probability but also the smallest fidelity at fixed $N$, whereas interior branches can improve the fidelity at the cost of a reduced heralding probability.

Albeit the functional form of the first-stage probability remains the same as in the previous schemes, its optimized numerical value is not generally independent of $N_2$. Changing $N_2$ changes the fidelity-optimal values of $\tilde{\breve{\tau}}_2$ and $\taut_2$, and thereby the reconstructed value of the attenuated first-squeezer parameter $\taut_1(T_2)$. Nevertheless, we observe nearly constant values of $P_{N_1}$ at fixed $N_1$ in the unrestricted data. This behavior follows from the fact that $P_{N_1}$ varies only slowly with respect to $\taut_1$ after optimization over $T_1$ and for sufficiently large values of $\taut_1$ as discussed for the previous schemes, see Fig.~\ref{fig:prob_vs_tau}. This near constancy is therefore not a general property of the alternating scheme. 
Away from the large-squeezing plateau, and in particular under finite squeezing constraints, both $\taut_1$ and $P_{N_1}$ generally vary with $N_2$.

By contrast, the conditional second-stage factor $P_{N_2|N_1}$ exhibits an overall decrease with increasing $N_2$, reflecting the suppression of higher photon-count outcomes. For example, in the unrestricted $|\alpha|=3$ data for $N_1=2$, it decreases from $0.09096$ for $N_2=1$ to $0.00557$ for $N_2=10$. This behavior is not strictly monotonic for every $N_1$, since the fidelity-optimal state parameters and the optimal transmittance change between branches. Consequently, similar to the fidelity, the joint probability does not generally vary monotonically with either $N_1$ at fixed $N_2$ or $N_2$ at fixed $N_1$. Nevertheless, the occurrence of two nonzero heralding events leads to an overall suppression of interior branches as the detected photon numbers increase.

\begin{figure}[t]
 \centering
 \includegraphics[width=0.99\linewidth]{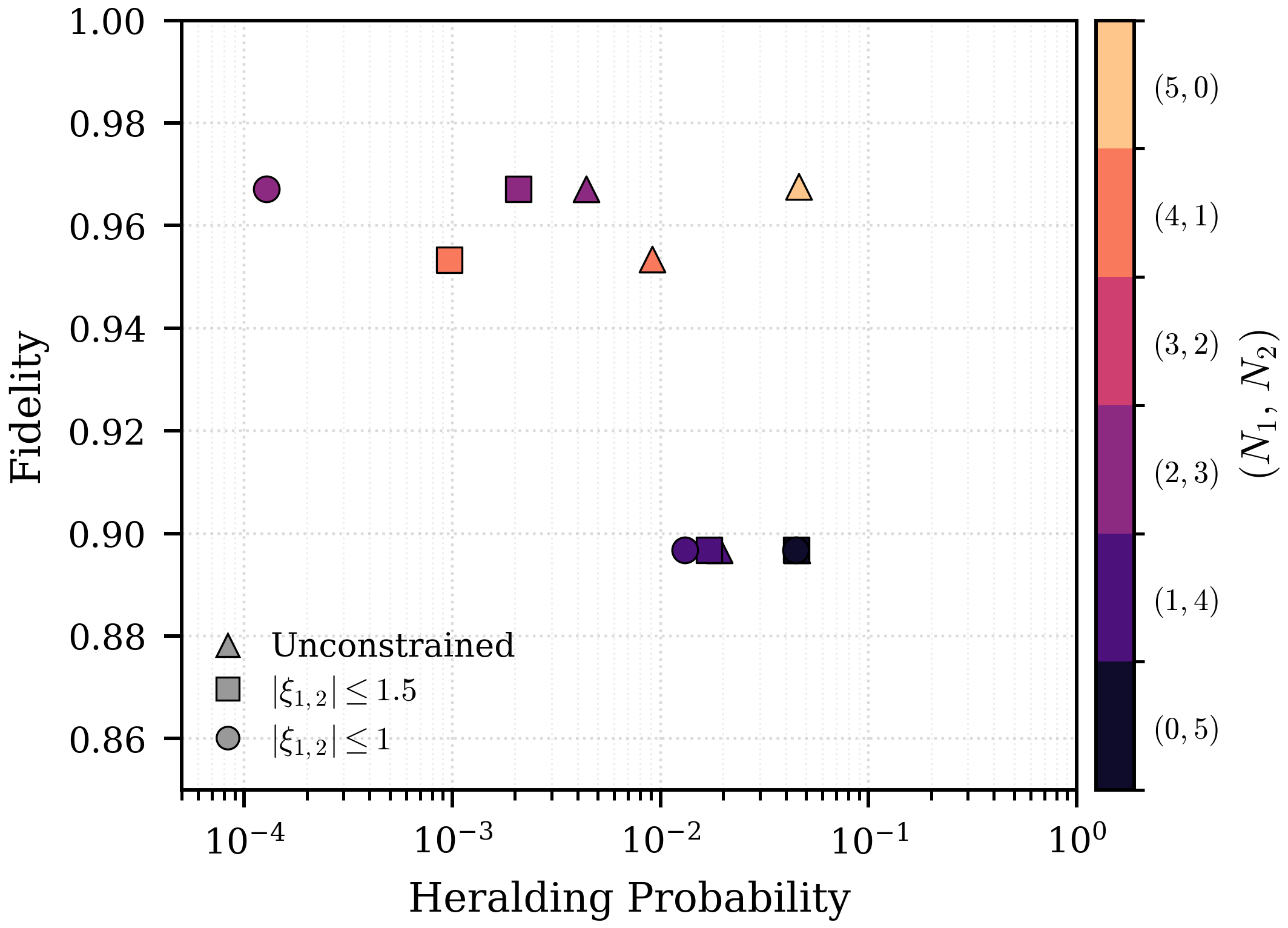}
 \caption{Fidelity over heralding probability for the alternating scheme for $|\alpha| = 3$ cat states. We compare different maximum squeezing levels $|\xi_{\mathrm{max}}|$ (markers) and different photon subtraction number combinations $N_1,N_2$ (color encoded). All branches satisfy $N_1+N_2=5$. For the constrained cases, the plotted points are obtained with $T_{1,2}<0.99$. Points with heralding probabilities below the displayed range are omitted.}
 \label{fig:fid_vs_prob_alt}
\end{figure}

The case $N=5$ also provides a representative example of how a squeezing constraint can change the relative performance of the three schemes. The corresponding fidelity--heralding-probability pairs for the different partitions with $N_1+N_2=5$ are shown in Fig.~\ref{fig:fid_vs_prob_alt}. In the unrestricted regime, the nonlinear-interferometer-type endpoint $(5,0)$ is favorable. Its maximal fidelity and corresponding optimized heralding probability approach $F_{5,0}\simeq0.96763$ and $P_{5,0}\simeq0.0461$, respectively, in the large-squeezing limit. Plain photon subtraction, corresponding to $(0,5)$, reaches the substantially lower fidelity $F_{0,5}=0.89682$, although with a comparable heralding probability $P_{0,5}\simeq0.0452$. The alternating branch $(2,3)$ already reaches a similarly high fidelity as the $(5,0)$ branch, $F_{2,3}=0.96708$, but with the considerably smaller probability $P_{2,3}=0.0044$. Among the remaining partitions, $(4,1)$ reaches $F_{4,1}=0.95346$ with $P_{4,1}=0.00912$, while $(1,4)$ reproduces the fidelity of plain photon subtraction with a lower probability $P_{1,4}=0.01925$. For $(3,2)$, the unrestricted fidelity maximum $F_{3,2}=0.92670$ is attained only at vanishing heralding probability and therefore not displayed in Fig.~\ref{fig:fid_vs_prob_alt}. Thus, without a squeezing constraint, the additional subtraction stage provides no advantage for this example.

The situation changes once the available squeezing is restricted. For $|\xi_i|\leq1.5$, the maximal fidelity of the $(5,0)$ endpoint is reduced to $F_{5,0}=0.95655$, whereas the alternating branch $(2,3)$ retains $F_{2,3}=0.96708$ and therefore becomes fidelity-optimal among the considered partitions. By contrast, the plain photon-subtraction endpoint $(0,5)$ remains essentially unaffected, with $F_{0,5}=0.89682$ and $P_{0,5}\simeq0.0452$. Tightening the constraint to $|\xi_i|\leq1$ further reduces the $(5,0)$ fidelity to $F_{5,0}=0.93289$, while the $(2,3)$ branch again retains $F_{2,3}=0.96708$. The plain photon-subtraction branch likewise retains its unconstrained value $F_{0,5}=0.89682$ and its heralding probability changes only marginally to $P_{0,5}\simeq 0.0445$. Thus, the three $(0,5)$ points almost overlap in Fig.~\ref{fig:fid_vs_prob_alt} and the difference is barely visible by eye. The fidelity of the $(1,4)$ branch also remains unchanged under the considered squeezing constraints, while their heralding probabilities decrease more noticeably from $0.01925$ to $0.01710$ to $0.01307$. The $(4,1)$ branch retains nearly its unrestricted fidelity for $|\xi_i|\leq1.5$ ($F_{4,1}=0.95338$), but falls to $F_{4,1}=0.91585$ at $|\xi_i|\leq1$, with a considerable decrease in the heralding probability $P_{4,1}\simeq1.61\times10^{-8}$ being not displayed in the figure. The $(3,2)$ branch likewise has negligible probabilities, $P_{3,2}\leq4.4\times10^{-7}$, in the constrained cases.

Thus, an alternating branch provides the largest fidelity under the finite squeezing constraints, but this advantage comes at the cost of a strongly reduced heralding probability. For $(2,3)$, the probability decreases from $P_{2,3}=0.0044$ in the unrestricted boundary limit to $0.00207$ for $|\xi_i|\leq1.5$ and $0.000128$ for $|\xi_i|\leq1$, with both input squeezers saturating the respective bounds in the constrained cases. Hence, although the squeezing constraints do not change the fidelity-optimal normalized state of this branch, they strongly restrict the range of transmittances available for optimizing its heralding probability. By contrast, the exact constrained fidelity maxima of the $(5,0)$ endpoint lie on a reachability boundary and therefore have vanishing heralding probability. A finite-probability realization must consequently move away from these maxima at the cost of a reduced fidelity. 
For the finite transmittances $T_{1,2}<0.99$ considered in Fig.~\ref{fig:fid_vs_prob_alt}, the constrained $(5,0)$ branch reaches slightly lower fidelities of $0.95475$ and $0.93028$, with corresponding heralding probabilities of $1.05\times10^{-6}$ and $5.88\times10^{-9}$, respectively. These points, together with all $(3,2)$ points and the $(4,1)$ point for $|\xi_i|\leq1$, lie below the displayed probability range.

These results show that the second subtraction stage is most useful when the available peak squeezing or the detectable photon number is limited. It provides additional freedom to reshape the non-Gaussian polynomial and can thereby reach fidelities that are inaccessible to plain photon subtraction and the nonlinear-interferometer-type endpoint under the same individual squeezing bound.  
This advantage is accompanied by additional experimental complexity, as two nonzero heralding outcomes are required, both squeezing stages are frequently operated close to their respective maximal available strengths, and the joint probability can be orders of magnitude smaller than for plain photon subtraction.

\subsection{Fidelity-constrained multibranch success probability optimization}
\begin{figure}[t]
    \centering
    \includegraphics[width=0.99\linewidth]{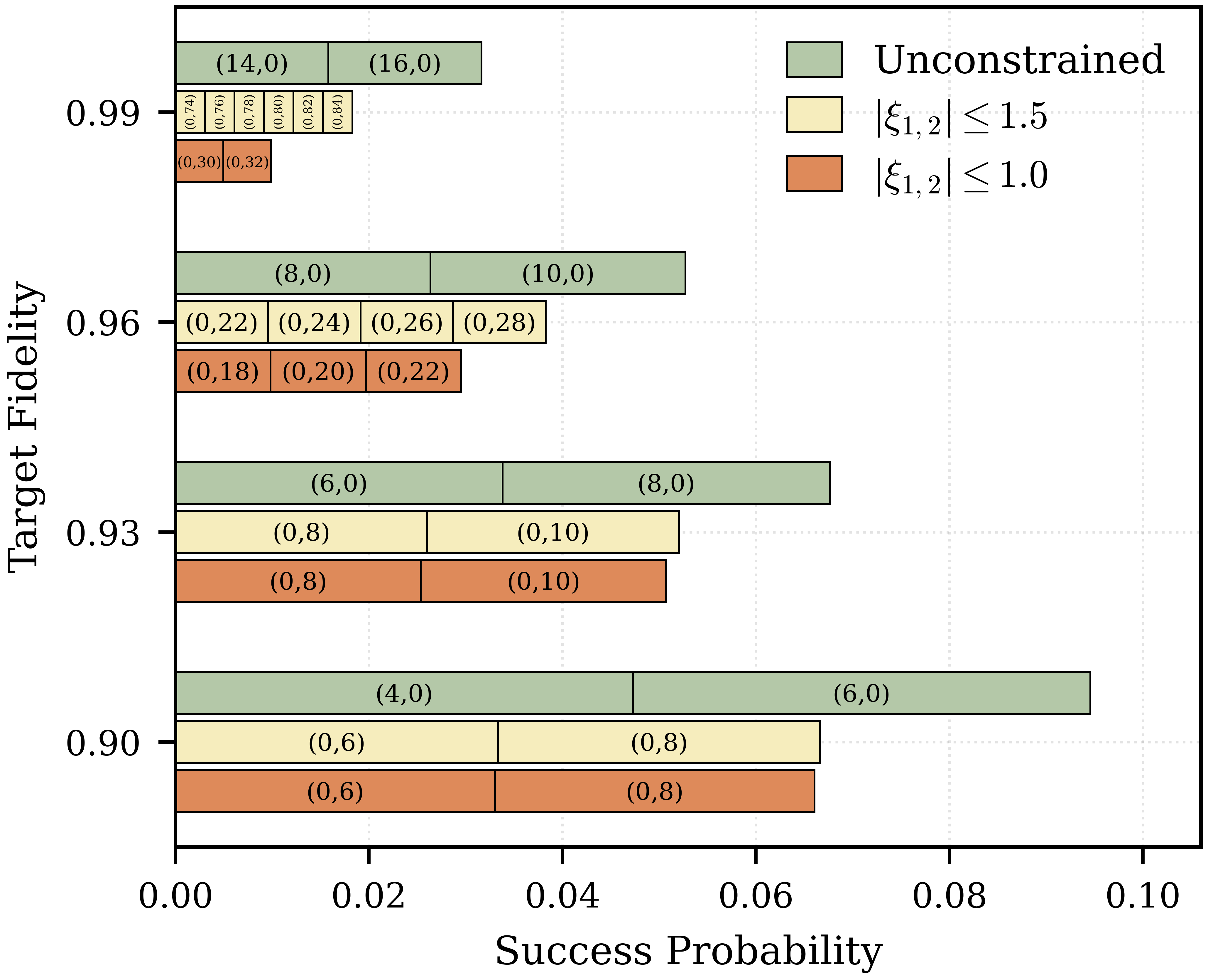}
    \caption{Fidelity--success-probability trade-off for multibranch even cat state generation with amplitude $|\alpha|=3$. The total optimized success probability is plotted as a function of the target fidelity constraint across different hardware limits on the squeezing parameters $|\xi_{1,2}|$. Discrete stacked bars decompose the cumulative success probability into the absolute contributions of individual heralding modes ($(N_1,N_2) = (0,2),(2,2),...$). Colors indicate distinct squeezing constraints ($|\xi_1| \le 1.0$, $|\xi_1| \le 1.5$, and the unconstrained parameter space).}
    \label{fig:alt_multibranchOptimization}
\end{figure}

Finally, we consider the discontinuous multibranch optimization problem, analogous to the investigations in Secs.~\ref{sec:PhotonSubtraction} and \ref{sec:SqSuSq}. Generalizing the concept of the total probability of accepted outcomes from a single-subtraction stage to the case of two subtraction stages, we have
\begin{align}
 P_{\mathrm{succ}}(F_{\mathrm{min}}) = \max_{\substack{0<T_1,T_2\leq1, \\ |\tilde{\breve{\tau}}_2|<T_2, \, |\taut_2|<T_2, \\ \taut_1(T_2)|<T_1}} \sum_{(N_1,N_2)\in\mathcal{S}(F_{\mathrm{min}})} \!\!\!\! P_{N_1,N_2},
\label{eq:alternating-multibranch-probability}
\end{align}
with $\mathcal{S}(F_{\mathrm{min}}) = \left\{(N_1,N_2)\,\middle|\,F_{N_1,N_2}\geq F_{\mathrm{min}}\right\}$. The common experimental parameters are optimized such that the success probability of all accepted branches is maximized. The resulting optimized trade-offs for even and odd target cat states with $|\alpha|=3$ are summarized in Tabs.~\ref{tab:alternating_multibranch_even} and \ref{tab:alternating_multibranch_odd}, respectively. Representative examples are illustrated in Fig.~\ref{fig:alt_multibranchOptimization}.

In the absence of squeezing constraints, the numerical optimization selects the nonlinear-interferometer-type endpoint $N_2=0$ for every considered fidelity threshold. The second subtraction beam splitter is therefore transparent, $T_2=1$, and the optimized success probabilities and accepted branches coincide exactly with those of the nonlinear-interferometer-type scheme discussed in the previous section.  
Consequently, the finite versus asymptotic character of the unconstrained solutions is inherited directly from the nonlinear-interferometer-type optimization. Thus, although interior alternating branches can achieve higher fidelities than either endpoint at certain fixed total photon numbers, the corresponding gain is not sufficient to compensate the probability penalty associated with a second nontrivial heralding event in the unrestricted multibranch optimization.

The situation changes qualitatively once the squeezing in each stage is restricted to $|\xi_i|\leq1.5$. For all considered fidelity thresholds, the global optimum moves to the opposite endpoint, $N_1=0$, with $T_1=1$. The first subtraction stage is therefore inactive, while the two squeezing operations can be combined coherently before the remaining photon-subtraction event. In this configuration, the conditional state belongs to the plain photon-subtraction family, but the required effective squeezing can be distributed between the two physical squeezing stages. Since the individual bound $|\xi_i|\leq1.5$ permits a combined squeezing amplitude of up to $|\breve{\xi}_2|=3$, the entire unconstrained plain photon-subtraction optimum over the considered range $0.90\leq F_{\mathrm{min}}\leq0.99$ remains accessible. Accordingly, the optimized success probabilities and branch sets coincide with those of unconstrained plain photon subtraction. The physical decomposition between the two squeezers is not unique in this case. The equal distribution listed in Tab.~\ref{tab:alternating_multibranch_even} represents one convenient choice that minimizes the maximum individual squeezing amplitude.

Compared with the nonlinear-interferometer-type scheme under the same individual squeezing bound, this endpoint switch substantially increases the success probability. For the even target state, the enhancement factors are approximately $2.42$, $3.13$, and $3.84$ at $F_{\mathrm{min}}=0.95$, $0.98$, and $0.99$, respectively. At the two highest thresholds, for example, the alternating setup reaches $P_{\mathrm{succ}}\simeq0.02625$ and $0.01829$, compared with $0.00837$ and $0.00476$ for the constrained nonlinear-interferometer-type scheme. The corresponding accepted outcomes, however, are shifted to substantially larger photon numbers as required by plain photon subtraction, see Tab.~\ref{tab:alternating_multibranch_even}. Hence, the gain in heralding probability is accompanied by a considerably stronger requirement on the photon-number resolution of the second detector.

Under the tighter bound $|\xi_i|\leq1$, the same $N_1=0$ endpoint remains optimal throughout the considered fidelity range. In contrast to the $|\xi_i|\leq1.5$ case, however, the unrestricted plain optimum can no longer be reached because the required effective squeezing exceeds the largest accessible combined value for $F_\mathrm{min} \geq 0.9$. The probability optimization therefore drives both physical squeezers to the boundary, $|\xi_1|=|\xi_2|=1$, with aligned squeezing directions, giving the maximal combined squeezing $|\breve{\xi}_2|=2$. Despite this restriction, the resulting success probabilities remain comparatively large. At $F_{\mathrm{min}}=0.98$ and $0.99$, for example, we obtain $P_{\mathrm{succ}}\simeq0.01823$ and $0.00990$ with the branch sets $\{(0,20),(0,22)\}$ and $\{(0,30),(0,32)\}$, respectively. These values are more than two orders of magnitude larger than for the nonlinear-interferometer-type scheme under the same per-stage squeezing bound, although they remain below the corresponding unconstrained plain photon-subtraction probabilities.

The occurrence of different optimal endpoints in the constrained and unconstrained regimes naturally raises the question how the transition between them takes place as the available squeezing is increased. In particular, it remains to be determined whether the global optimum switches directly from the plain-photon-subtraction-type endpoint to the nonlinear-interferometer-type endpoint, or whether an intermediate squeezing range exists in which genuinely alternating patterns with $N_1,N_2>0$ become favorable. To investigate these points, we vary the common squeezing bound $|\xi_i|\leq\xi_{\mathrm{max}}$ continuously. Numerically, we observe a crossover that depends weakly on the imposed fidelity threshold through the corresponding optimal branch sets. For the even $|\alpha|=3$ cat state and $0.90\leq F_{\mathrm{min}}\leq0.99$, the threshold-dependent crossover values are confined to the comparatively narrow range $1.84\lesssim\xi_{\mathrm{max}}\lesssim1.94$. For each fidelity threshold, the plain-photon-subtraction-type endpoint $N_1=0$ is optimal below its respective crossover, whereas the nonlinear-interferometer-type endpoint $N_2=0$ becomes favorable above it. For example, at $F_{\mathrm{min}}=0.98$ the crossover occurs at approximately $\xi_{\mathrm{max}}\simeq1.88$, accompanied by a change from the high-photon-number plain-photon-subtraction branch set $\{(0,50),(0,52),\ldots,(0,60)\}$ to the much lower-order nonlinear-interferometer branch set $\{(12,0),(14,0)\}$. Importantly, within the numerical resolution of our scan we do not observe an intermediate interval in which genuinely alternating patterns with $N_1,N_2>0$ become globally optimal. Such interior branches can satisfy the fidelity constraint and may provide higher fidelities at fixed total photon number, but the gain does not compensate the probability penalty associated with two nontrivial heralding events.

A different resource comparison is obtained if the squeezing resources of the two stages are constrained jointly rather than independently. As a simple model, we consider the quadratic constraint ${|\xi_1|^2+|\xi_2|^2\leq\xi_{\mathrm{tot}}^2}$. For the plain-photon-subtraction-type endpoint $N_1=0$, the aligned squeezing operations can add coherently, such that the largest accessible combined squeezing is $|\breve{\xi}_2|=\sqrt{2}\,\xi_{\mathrm{tot}}$, attained for an equal distribution between the two stages. By contrast, the nonlinear-interferometer-type endpoint generally favors a strongly asymmetric allocation of the available resource, with most of the squeezing assigned to the first stage and only a comparatively weak second squeezing operation, see Sec.~\ref{sec:NI-Multibranch}. The joint constraint therefore favors the nonlinear-interferometer-type endpoint relative to an independent per-stage bound. For example, for the even $|\alpha|=3$ cat state and $F_{\mathrm{min}}=0.98$, the direct crossover occurs at approximately $\xi_{\mathrm{tot}}\simeq1.883$. At this point, the plain-photon-subtraction-type solution with $\mathcal{S}=\{(0,42),(0,44),\ldots,(0,50)\}$ and the nonlinear-interferometer-type solution with $\mathcal{S}=\{(12,0),(14,0)\}$ both yield $P_{\mathrm{succ}}\simeq0.02608$, while the latter requires approximately $|\xi_1|\simeq1.871$ and $|\xi_2|\simeq0.216$. For comparison, reproducing the unrestricted plain-photon-subtraction optimum at the same fidelity requires ${\xi_{\mathrm{tot}}\gtrsim2.775/\sqrt{2}\simeq1.96}$. A numerical scan around the endpoint crossover, allowing the relative squeezing phase and both beam-splitter transmittances to vary, does not reveal an intermediate range in which genuinely alternating patterns with $N_1,N_2>0$ become globally optimal. Instead, the optimization again approaches one of the two endpoint families. Thus, the correlated resource constraint shifts the crossover in favor of asymmetric nonlinear-interferometer-type solutions, but does not alter the endpoint-dominated structure observed for the independent squeezing constraints within the tested parameter range.

A different picture emerges when the photon-number resolution is limited in addition to the available peak squeezing. Imposing a bound $N_i\leq N_{\mathrm{det,max}}$ on the photon number resolved by each detector can make genuinely alternating branches globally favorable in the multibranch optimization. For example, for an even $|\alpha|=3$ cat state with $F_{\mathrm{min}}=0.98$, $|\xi_i|\leq1$, and $N_{\mathrm{max}}=8$, the nonlinear-interferometer-type scheme reaches only $P_{\mathrm{succ}}\simeq3.1\times10^{-7}$, whereas the alternating scheme yields $P_{\mathrm{succ}}\simeq6.6\times10^{-5}$ by accepting the branches $\{(2,6),(7,7)\}$, corresponding to an enhancement by more than two orders of magnitude. This advantage is not solely due to distributing a larger total photon number over two detectors. Restricting additionally to $N_1+N_2\leq8$ retains the branch $(2,6)$ and still gives $P_{\mathrm{succ}}\simeq6.1\times10^{-5}$. Hence, when both the peak squeezing and the accessible photon-number range are limited, the second subtraction stage can provide a genuine multibranch advantage over the nonlinear-interferometer-type endpoint, while requiring similarly low or even lower photon-number resolution and a smaller maximum squeezing amplitude.

Finally, the odd-parity case exhibits the same overall structure as the even-parity case. The unconstrained optimum again reduces to the nonlinear-interferometer-type endpoint, whereas for the two finite squeezing bounds considered explicitly above, $|\xi_i|\leq 1.5$ and ${|\xi_i|\leq 1}$, the optima are obtained for the plain-photon-subtraction-type endpoint $N_1=0$ with aligned squeezing operations. For $|\xi_i|\leq1.5$, the unconstrained plain odd-parity optimum can likewise be reproduced, while for $|\xi_i|\leq1$ both squeezers saturate the available bound. As in the even-parity case, continuously varying the common squeezing constraint reveals a direct crossover between the two endpoint families, without an intermediate range in which genuinely alternating patterns with $N_1,N_2>0$ become globally optimal. The corresponding crossover values are again confined to a comparatively narrow range of squeezing strengths, $1.85\lesssim\xi_{\mathrm{max}}\lesssim1.95$ over $0.90\leq F_{\mathrm{min}}\leq0.99$, depending only weakly on the imposed fidelity threshold. The optimized success probabilities show moderate parity-dependent variations at individual fidelity thresholds, but no systematic advantage of either parity is observed over the full fidelity range.

\section{Robustness of multibranch optima}
\label{sec:Robustness}

\begin{figure*}[t]
 \centering
 \includegraphics[width=0.99\linewidth]{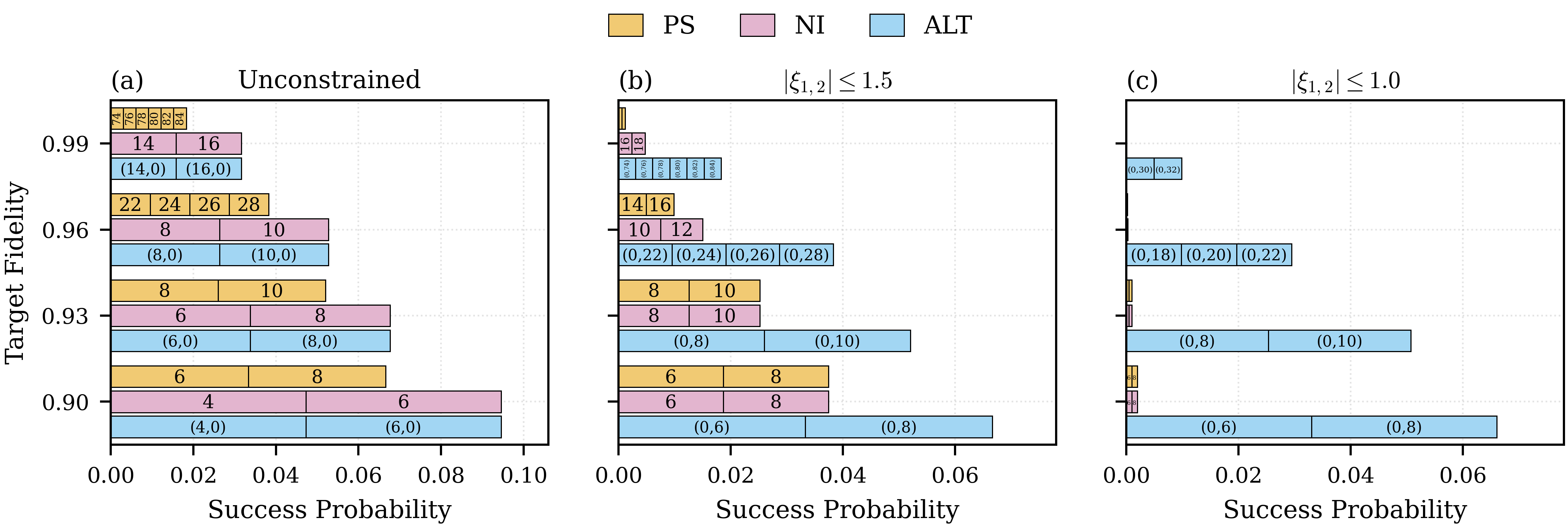}
 \caption{Fidelity--success-probability trade-off for multibranch even cat state generation with amplitude $|\alpha|=3$ with unconstrained squeezing (left panel), squeezing constrained to $|\xi|\leq 1.5$ (middle panel), squeezing constrained to $|\xi|\leq 1.0$ (right panel). The total optimized success probability is plotted as a function of the target fidelity constraint across the three schemes. Discrete stacked bars decompose the total success probability into contributions from individual accepted heralding outcomes, specified by $N_1$ for PS and NI and by $(N_1,N_2)$ for ALT. Colors indicate the three schemes: plain photon subtraction (PS) in yellow, nonlinear interferometer (NI) in magenta, and the alternating scheme (ALT) in blue.}
 \label{fig:SuccessProbability-Comparison}
\end{figure*}

Beyond identifying optimal fidelity-probability trade-offs, it is important to assess how robust the corresponding parameter settings are under realistic experimental imperfections. In particular, experimentally useful solutions should not rely on excessively fine tuning of the optical parameters.
Secs. \ref{sec:PhotonSubtraction}, \ref{sec:SqSuSq}, and \ref{sec:AlternatingScheme} established the individual fidelity-probability trade-offs of the three considered schemes. To facilitate a direct visual comparison under matched squeezing constraints, Fig.~\ref{fig:SuccessProbability-Comparison} summarizes their optimized success probabilities for fidelity thresholds $0.9\leq F_{\mathrm{min}}\leq0.99$.
In the unconstrained case (left panel of Fig.~\ref{fig:SuccessProbability-Comparison}), the nonlinear interferometer's second squeezing operation yields moderately higher success probabilities than plain photon subtraction, by up to a factor of $1.73$ (Sec.~\ref{sec:SqSuSq}), while the alternating scheme's optimum reduces to $N_2=0$ and therefore coincides exactly with the nonlinear-interferometer result. Under the constraint $|\xi_{1,2}|\leq 1.5$ (middle panel of Fig.~\ref{fig:SuccessProbability-Comparison}), the alternating scheme starts to outperform plain subtraction and nonlinear interferometer. The alternating scheme's $N_1=0$ branch draws on two individually-bounded squeezers to reach the combined effective squeezing needed to match the unconstrained plain-photon-subtraction optimum for the considered fidelity range (Sec.~\ref{sec:AlternatingScheme}). This advantage sharpens considerably under the tighter bound $|\xi_{1,2}|\leq 1.0$, see right panel of Fig.~\ref{fig:SuccessProbability-Comparison}, where the alternating scheme outperforms plain photon subtraction and the nonlinear interferometer by more than three or two orders of magnitude, respectively, at the highest tested fidelity of $0.99$.

\subsection{Sensitivity of branch fidelities}
\begin{figure*}[t!]
    \centering
    \includegraphics[width=0.99\linewidth]{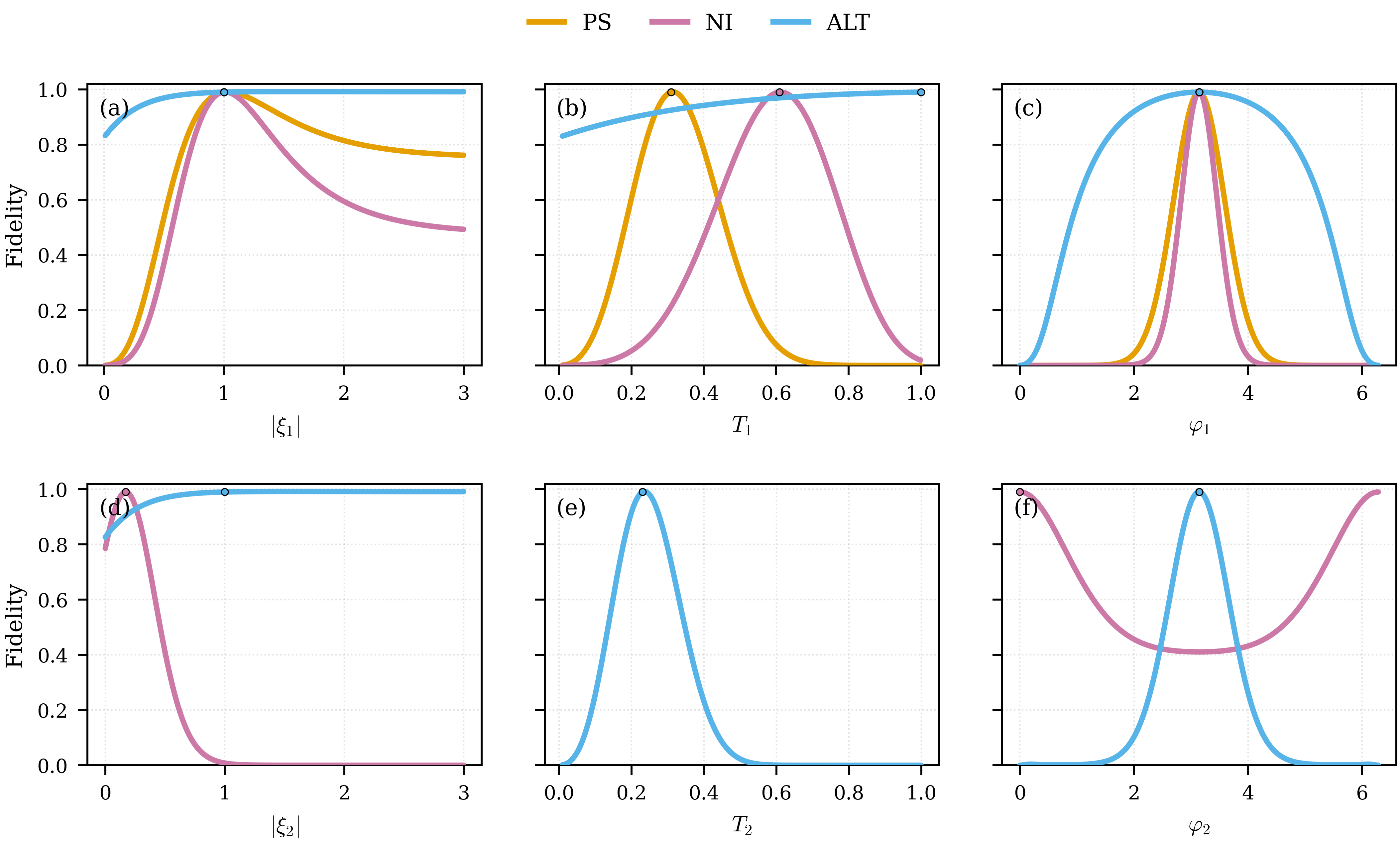}
    \caption{Sensitivity of the output state fidelity to individual experimental parameter perturbations around the fidelity-optimal operating point for an $|\alpha|=3$ cat state at threshold fidelity $0.99$ under squeezing bound $|\xi_{1,2}|\leq 1$, comparing plain photon subtraction (PS, $N_1=28$, yellow curves), the nonlinear-interferometer-type scheme (NI, $N_1=14$, magenta curves), and the alternating two-stage scheme (ALT, $N_1=0, N_2=30$, blue curves). In each panel, one parameter is varied while all others are held fixed at the respective scheme's optimum, marked by the colored circles: (a) first-squeezer amplitude $|\xi_{1}|$; (b) first beam-splitter transmittance $T_1$; (c) first-squeezer phase $\varphi_1$; (d) second-squeezer amplitude  $|\xi_{2}|$; (e) second beam-splitter transmittance $T_2$; (f) second-squeezer phase $\varphi_2$. The squeezing phases are specified relative to twice the phase of the cat-state parameter. PS is omitted from panels (d)–(f), as it involves no second squeezing stage; NI is additionally omitted from panel (e), as it employs no second beam-splitter.}
    \label{fig:Fidelity-sensitivity}
\end{figure*}

These gains, however, all assume access to the precise optimal parameter values identified above. From an experimental standpoint, the robustness of the fidelity to deviations from these optima is therefore just as important as the optimum itself. Since finite squeezing resources are likely to remain an important constraint for near-term implementations, we focus the following robustness analysis on the case $|\xi_{1,2}|\leq1.0$, which also exhibits the largest performance differences between the three schemes. Specifically, we fix all parameters to the previously determined multibranch optima for a cat state with $|\alpha|=3$ and $F_{\mathrm{min}}=0.99$, and then vary each experimental parameter independently. We first focus on the robustness of the branch fidelities against independent variations of the experimental parameters. We consider $N_1=28$ for plain photon subtraction, $N_1=14$ for the nonlinear-interferometer scheme, and $(N_1,N_2)=(0,30)$ for the alternating scheme. In the first two cases, the selected branch is the sole contribution to the optimized success probability, while in the alternating scheme we select $(N_1,N_2)=(0,30)$ as a representative example from the set of accepted branches at this fidelity threshold. We obtain similar results for the $(0,32)$ branch. Beyond the peak fidelity itself, the width of each curve in Fig.~\ref{fig:Fidelity-sensitivity} around its optimum indicates how precisely the corresponding parameter must be controlled experimentally.

The results indicate an asymmetry between the three schemes. The alternating scheme is remarkably robust in terms of miscalibrated squeezing amplitudes on either stage. This broad robustness of the alternating scheme can be understood from the particular branches favored under the considered squeezing constraints. For $(N_1,N_2)=(0,N)$, the normalized conditional state depends only on the combined effective squeezing $\tilde{\breve{\tau}}_2$. At the optimum, the two squeezers are aligned and therefore add constructively, $|\tilde{\breve{\tau}}_2| = T_2\tanh\big(|\xit_1| + |\xi_2|\big)$ with $\xit_1 = \xi_1$ as $T_1=1$ for the optimal parameter setting. Consequently, variations of the individual squeezing amplitudes affect the fidelity only through this single combined parameter. For increasing squeezing, the hyperbolic tangent rapidly approaches unity, while $T_2$ provides an upper bound for this effective squeezing parameter, $|\tilde{\breve{\tau}}_2|<T_2$. Since the fidelity-optimal value of $|\tilde{\breve{\tau}}_2|$ already is close to this bound, increasing either squeezing amplitude merely drives the effective parameter slightly closer to $T_2$. The resulting state therefore changes only weakly, which explains the broad high-squeezing plateau observed in Fig.~\ref{fig:Fidelity-sensitivity}. A related saturation occurs for plain photon subtraction and for variations of the first squeezing amplitude in the nonlinear-interferometer
scheme. In both cases, the first beam splitter bounds the corresponding effective squeezing according to $|\taut_1|=T_1\tanh|\xi_1|$, such that the conditional state approaches a finite limiting state for $|\xi_1|\rightarrow\infty$. However, the resulting plateaus occur at substantially lower fidelities than in the alternating scheme. In plain photon subtraction, increasing $|\xi_1|$ drives the single effective squeezing parameter away from its fidelity-optimal value, whereas in the nonlinear-interferometer scheme it additionally disturbs the optimal balance between the two squeezing operations. For the second squeezing amplitude of the nonlinear-interferometer scheme no analogous finite high-fidelity plateau occurs. In the absence of a subsequent attenuating beam splitter, increasing $|\xi_2|$ drives the combined output squeezing towards the infinitely squeezed limit and the fidelity with the fixed target cat state rapidly vanishes. In the alternating scheme, by contrast, the second subtraction beam splitter bounds the final effective squeezing by $T_2$, and this bound lies close to its fidelity-optimal value for the branches considered here.

Towards lower $|\xi_1|$, plain photon subtraction and the nonlinear-interferometer scheme both exhibit a rapid fidelity loss, which is slightly more pronounced for the nonlinear-interferometer scheme. In plain photon subtraction, reducing $|\xi_1|$ weakens the only available squeezing, while in the nonlinear-interferometer scheme it additionally moves the system away from the balance between the two squeezing operations required at the optimum. The alternating $(0,N)$ branch behaves differently. Even for $|\xi_1|\to0$, the second squeezer still provides a finite effective squeezing parameter, such that the fidelity does not vanish as in the other two schemes due to the symmetric splitting of squeezing resources.

The same effective-squeezing picture also explains the comparatively weak sensitivity of the alternating scheme to variations of $T_1$ and $\varphi_1$. Reducing $T_1$ attenuates the contribution of the first squeezer according to $\taut_1=T_1\tau_1$ and therefore lowers the squeezing available from the first stage. For the relevant branches of type $(N_1,N_2)=(0,N)$, however, the second squeezer still contributes to the combined effective parameter $\tilde{\breve{\tau}}_2$, such that the total squeezing does not vanish even when the first-stage contribution is strongly reduced. This makes the fidelity comparatively insensitive to moderate deviations of $T_1$. In plain photon subtraction, by contrast, the first squeezer is the only source of squeezing, so variations of $T_1$ directly modify the effective squeezing that determines the conditional state. In the nonlinear-interferometer scheme, the fidelity depends separately on the first and second squeezing parameters, and its optimum relies on a specific balance between their contributions. Deviations of $T_1$ therefore disturb this balance usually more strongly than in the alternating $(0,N)$ branches.

The $\varphi_1$ sensitivity can be understood similarly, although its interpretation differs between the schemes. In plain photon subtraction, only a single relative phase enters, such that deviations of $\varphi_1$ simply rotate the effective squeezing axis away from the phase chosen for the target cat state. Equivalently, this may be viewed as preparing a cat state with a correspondingly rotated phase rather than as an intrinsic loss of state quality. 
Both the nonlinear-interferometer and alternating schemes depend on two relative phases. For the alternating $(0,N)$ branches, however, the two squeezing operations can be combined into a single effective squeezing parameter. Thus, varying $\varphi_1$ rotates the effective squeezing axis, which can again be interpreted as merely changing the phase orientation of the generated cat state. At the same time, it changes the relative phase between the squeezers, thereby weakening their constructive combination. In the nonlinear-interferometer scheme, by contrast, photon subtraction separates the two squeezing operations, such that their phase dependence cannot be reduced to a single effective squeezing parameter. Variations of $\varphi_1$ therefore affect the conditional state more intricately, consistent with the narrower fidelity peak observed for the nonlinear-interferometer scheme in Fig.~\ref{fig:Fidelity-sensitivity}.
Finally, a particularly direct illustration of the phase structure derived in Sec. \ref{sec:AlternatingScheme} appears in Fig. \ref{fig:Fidelity-sensitivity}. The nonlinear interferometer and alternating schemes require opposite phase conventions for the second squeezer, numerically confirming the relative-phase argument obtained analytically from the asymptotic analyses above.

The dependence of the alternating $(0,N)$ branch on the second-stage parameters $T_2$ and $\varphi_2$ closely resembles the dependence of plain photon subtraction on $T_1$ and $\varphi_1$, respectively. This reflects the fact that the $(0,N)$ conditional state belongs to the same family as plain $N$-photon subtraction, with $\tilde{\breve{\tau}}_2$ replacing the effective squeezing parameter $\taut_1$ of the plain photon-subtraction scheme. Accordingly, $T_2$ controls its attenuation through the final subtraction stage, while $\varphi_2$ controls the complex phase of the combined effective squeezing. However, the analogy for $\varphi_2$ is only qualitative, since varying this phase also changes the relative phase between the two physical squeezers.

\subsection{Heralding probabilities and multibranch stability}
So far, this robustness analysis has focused on the fidelity of individual, fixed heralding branches under parameter deviations. Two further aspects are relevant in practice: how the associated heralding probability itself responds to the same deviations, and whether the set of accepted branches remains stable as parameters drift. Since the practical generation rate depends not only on the fidelity but also on how often a given outcome is actually heralded, we additionally examine the sensitivity of the branch-wise probability $P_{N_1}$ (plain photon subtraction, nonlinear interferometer) or $P_{N_1,N_2}$ (alternating scheme) to the same parameter deviations considered above. This distinction matters experimentally. A branch whose fidelity is robust but whose probability collapses under miscalibration offers little practical advantage.

\begin{figure*}[t!]
    \centering
    \includegraphics[width=0.99\linewidth]{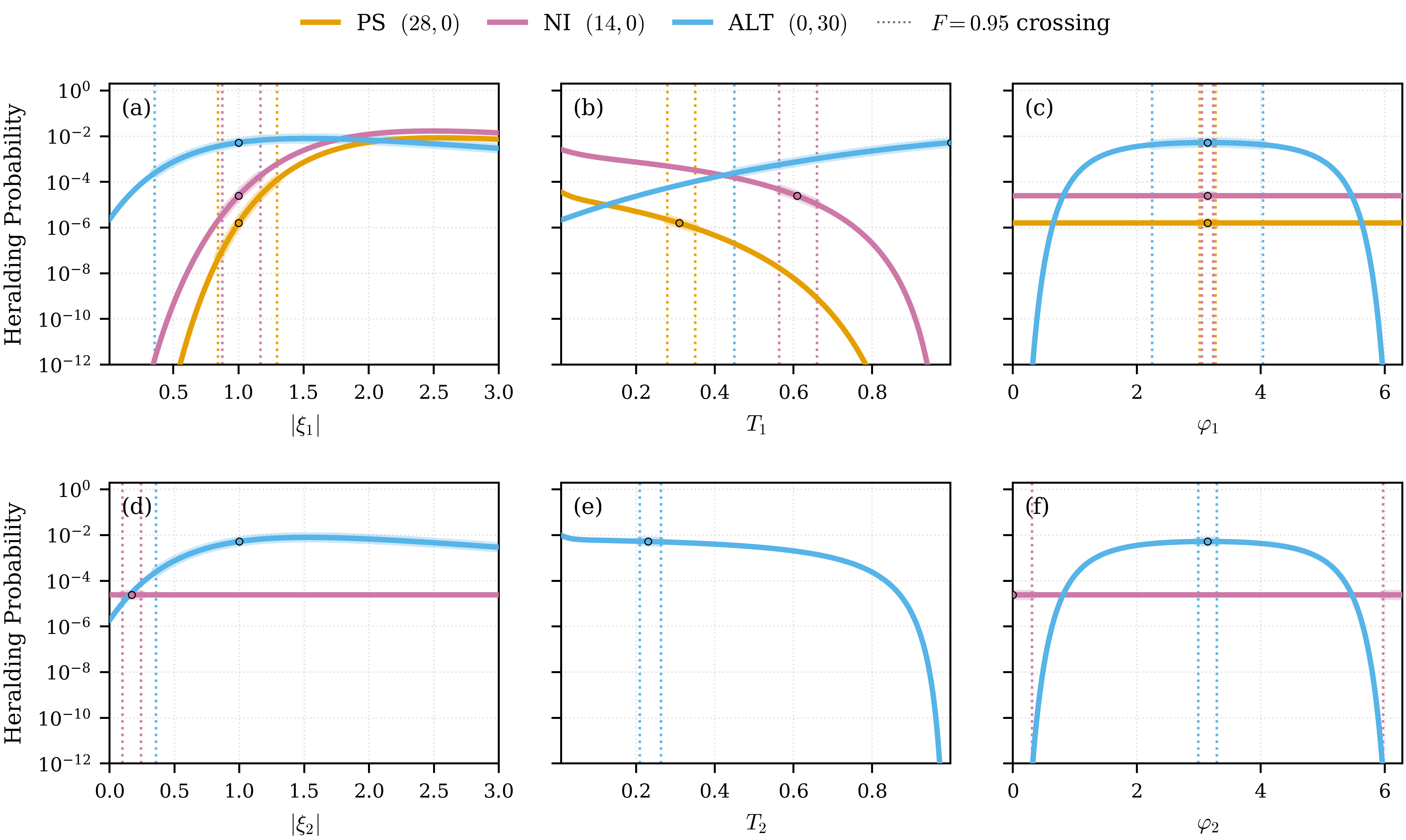}
    \caption{Heralding probability of the same three branches as in Fig.~\ref{fig:Fidelity-sensitivity} under independent variations of each experimental parameter around its optimum, shown on a logarithmic scale. Panels (a)–(f) correspond to the same parameters as in Fig. \ref{fig:Fidelity-sensitivity}. Colored circles mark each scheme's optimal operating point. Dotted vertical lines indicate where the corresponding branch's fidelity crosses $F=0.95$, delimiting the parameter range over which that branch remains accepted under this representative fidelity threshold. PS is omitted from panels (d)–(f) and NI from panel (e), for the same reasons as in Fig. \ref{fig:Fidelity-sensitivity}.}
    \label{fig:Probability-sensitivity}
\end{figure*}

The branch-wise heralding probabilities show a related but not identical robustness pattern compared to the fidelity sensitivity on parameter variations. For plain photon subtraction and the nonlinear-interferometer scheme, the branch-wise heralding probability is fixed entirely by the first subtraction stage and depends only on $T_1$ and the modulus $|\taut_1|$, but not on the squeezing phases. In the nonlinear-interferometer scheme, the subsequent squeezing operation is unitary and therefore does not affect the probability of the preceding heralding event, as seen from the constant dependence on $|\xi_2|$ and $\varphi_2$ in Fig.~\ref{fig:Probability-sensitivity}. Their strongest probability variations instead occur with $|\xi_1|$ and $T_1$, where reduced squeezing or increased beam-splitter transmission strongly suppresses the probability of detecting the comparatively large photon numbers considered here. For the alternating $(0,N)$ branch, the behavior again follows from its reduction to an effective plain $N$-photon-subtraction process at the second stage. Consequently, the probability is comparatively insensitive to moderate variations of the individual squeezing amplitudes, while $T_2$ produces the expected direct trade-off between transmission and heralding probability. Variations of either squeezing phase modify the combined effective squeezing and therefore also the heralding probability, leading to the broad phase dependence observed around the optimum.

A further complication arises specifically in the multibranch scenario, where several heralding outcomes are accepted simultaneously whenever their individual fidelity exceeds a specified threshold. Because this acceptance criterion is evaluated separately for each branch, a deviation in the shared experimental parameters can push individual branches across the threshold, causing them to drop out of, or newly enter, the accepted set. This represents a qualitatively different behavior from the smooth fidelity degradation discussed above. Rather than varying continuously, the total success probability can change abruptly as the accepted set changes. When a branch drops below the specified threshold, its entire probability contribution is lost, whereas a newly accepted branch adds its contribution at once. The parameter range over which the accepted set remains unchanged is therefore directly related to the fidelity robustness of the individual branches.

\begin{figure*}[t!]
    \centering
    \includegraphics[width=0.99\linewidth]{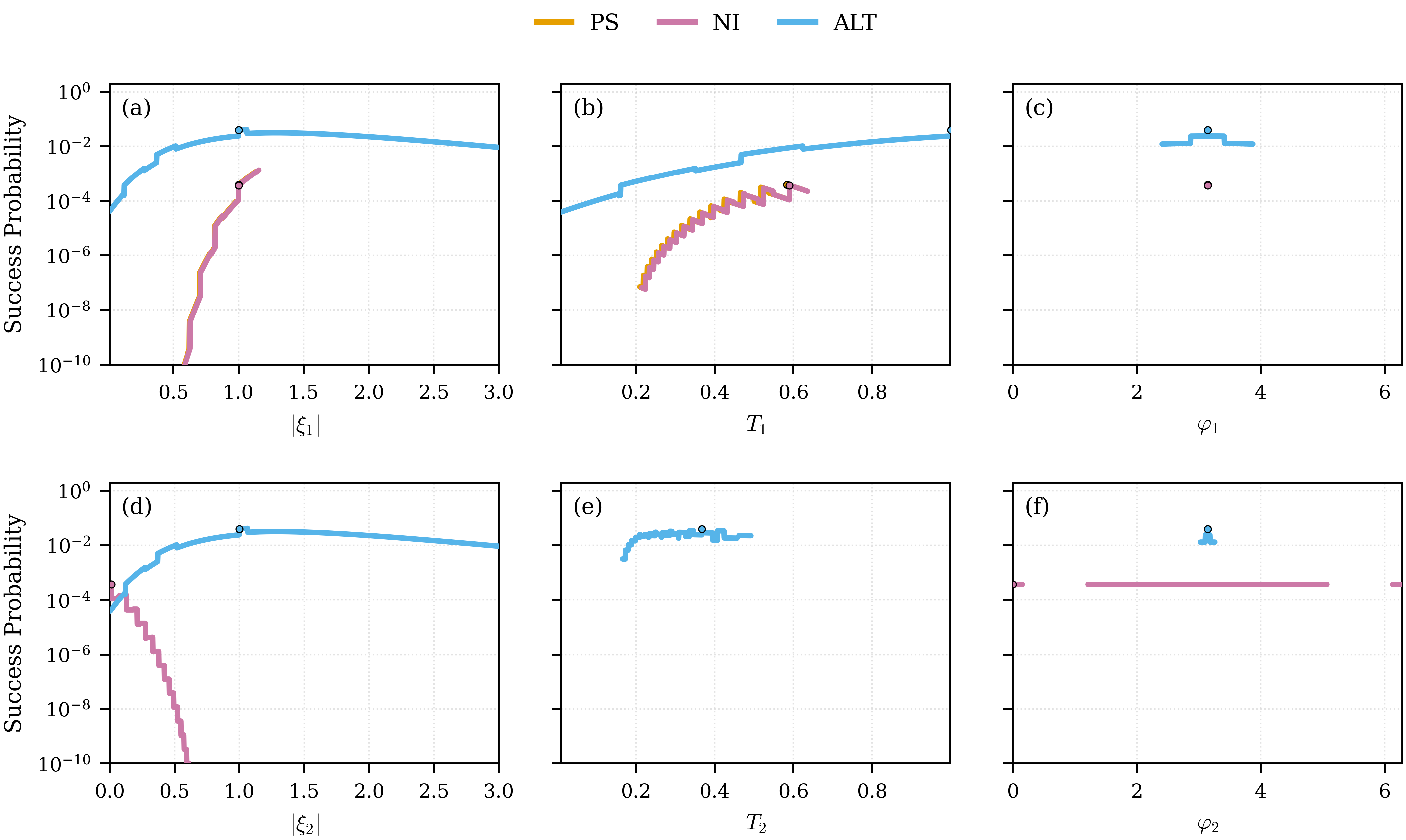}
    \caption{Multi-branch success probability around the fidelity-optimal operating point for an $|\alpha| = 3$ cat state at threshold fidelity $0.95$ under squeezing bound $|\xi_{1,2}| \leq 1$  under independent variations of each experimental parameter around its optimum, shown on a logarithmic scale. Panels (a)–(f) correspond to the same parameters as in Fig. \ref{fig:Fidelity-sensitivity}. Colored circles mark each scheme's optimal operating point. PS is omitted from panels (d)–(f) and NI from panel (e), for the same reasons as in Fig. \ref{fig:Fidelity-sensitivity}.}
    \label{fig:Multibranch-sensitivity}
\end{figure*}

This discontinuous behavior is illustrated in Fig.~\ref{fig:Multibranch-sensitivity}, where we use the less restrictive threshold $F_{\mathrm{min}}=0.95$ rather than $0.99$ considered above. At $F_{\mathrm{min}}=0.99$, the fidelity margins are sufficiently narrow for $\varphi_1$, $\varphi_2$, and $T_2$, such that varying one of these parameters quickly removes the contributing branches, making the changes of the accepted set less informative. At $F_{\mathrm{min}}=0.95$, the step-like changes in the success probability directly reveal branches entering and leaving the accepted set. Plain photon subtraction and the nonlinear-interferometer scheme show almost identical behavior, consistent with the very weak second squeezing required by the latter at this threshold, $|\xi_2|\simeq0.014$.

\subsection{Loss in the heralding arms and reoptimization}
Finally, we consider the effect of imperfect photon-number detection in the heralding arms. We restrict this analysis to the alternating scheme under the squeezing constraint $|\xi_{1,2}|\leq1$, which provides the largest success probabilities among the considered schemes in this resource regime and was found to be comparatively robust against several parameter variations. Moreover, the relevant optimum belongs to the $(0,N)$ endpoint family and is therefore equivalent at the state level to plain photon subtraction with the combined effective squeezing parameter $\tilde{\breve{\tau}}_2$. Accordingly, the qualitative conclusions of the following loss analysis also apply to the plain photon subtraction scheme, albeit at the reduced effective squeezing accessible with only a single squeezing stage.

We model losses in the heralding arm by a loss channel of transmittance $\eta$ preceding an otherwise ideal photon-number-resolving detector, such that $1-\eta$ denotes the effective loss in the heralding arm.
For a registered photon number $N$, the corresponding conditional state is then no longer pure, but becomes the mixture
\begin{align}
 \rho_{0,N}^{(\eta)} &= \frac{1}{P_{0,N}^{(\eta)}}
 \sum_{k=N}^{\infty} \binom{k}{N}\eta^N(1-\eta)^{k-N} P_{0,k} \ket{\psi_{0,k}}\!\bra{\psi_{0,k}},
\label{eq:lossy-heralded-state}
\end{align}
with probability
\begin{align}
 P_{0,N}^{(\eta)} &= \sum_{k=N}^{\infty} \binom{k}{N}\eta^N(1-\eta)^{k-N}P_{0,k}.
\label{eq:lossy-heralding-probability}
\end{align}
Accordingly, its fidelity with the target cat state is the probability-weighted average of the ideal branch fidelities $F_{0,k}$,
\begin{align}
 F_{0,N}^{(\eta)} &= \frac{1}{P_{0,N}^{(\eta)}} \sum_{k=N}^{\infty} \binom{k}{N}\eta^N(1-\eta)^{k-N} P_{0,k} F_{0,k}.
\label{eq:lossy-heralding-fidelity}
\end{align}

For the alternating optimum at all considered $F_{\mathrm{min}}$ and $|\xi_{1,2}|\leq1$, the first subtraction stage is inactive, $N_1=0$ and $T_1=1$, such that only loss in the second heralding arm is relevant. Keeping the lossless optimal parameters fixed shows that loss affects the heralding probability and the conditional fidelity in qualitatively different ways. The probability of registering the nominally accepted outcomes changes only moderately with decreasing $\eta$. A registered $N$-photon event no longer originates exclusively from an ideal $N$-photon outcome as events with $k>N$ photons can also be registered as $N$ if $k-N$ photons are lost. These additional contributions partly compensate the reduced probability of correctly detecting the original $N$-photon event. The conditional fidelity, however, decreases rapidly, as a registered outcome of fixed parity acquires contributions from ideal branches of both parities. Components with parity opposite to that of the target cat state have vanishing overlap with the target and therefore strongly reduce the fidelity of the resulting mixture. The originally accepted branches can therefore cross below the fidelity threshold already for modest heralding-arm loss. While lower-order heralding events are generally less affected than higher-order ones, the same qualitative loss-induced fidelity degradation persists. For the fidelity thresholds considered in this work $0.9 \leq F_{\mathrm{min}} \leq 0.99$, already at $\eta=0.98$ none of the originally accepted outcomes remains above threshold.

If the heralding arm losses are known in advance, part of this degradation can in principle be compensated by reoptimizing the system parameters. First, we restrict this reoptimization to the $N_1=0$ endpoint family rather than performing a full lossy optimization over arbitrary $(N_1,N_2)$ branches. This reoptimization can restore the imposed fidelity constraint over the considered loss range, but only at the cost of a substantial reduction in the heralding probability. For instance, for $F_{\mathrm{min}}=0.95$, already at $\eta=0.98$ the optimum switches to the single registered branch $(0,12)$ with $T_2\simeq0.918$ and $P_{\mathrm{succ}}\simeq 4.9\times10^{-16}$. For $\eta=0.96$, the same branch remains optimal while $T_2$ increases to approximately $0.957$ and the success probability decreases further to $P_{\mathrm{succ}}\simeq 9.5\times10^{-20}$. Thus, although reoptimization can formally recover the required fidelity, the corresponding generation rate becomes negligible.

The strong reduction of the optimized success probability is a direct consequence of the parameter adjustment required to preserve the fidelity. The optimization shifts the second beam-splitter transmittance towards unity. This reduces the probability of generating higher actual photon numbers in the heralding arm and therefore suppresses their admixture into a fixed registered outcome after loss. At the same time, however, less light is directed towards the heralding detector, such that the probability of the desired nonzero photon-number event is strongly reduced. The recovery of the fidelity threshold is therefore achieved by trading heralding rate for reduced contamination of the conditional state.

This trade-off becomes less severe when the fidelity threshold is relaxed. For $F_{\mathrm{min}}=0.90$, the lossless optimum accepts the branches $(0,6)$ and $(0,8)$ with ${P_{\mathrm{succ}}\simeq0.066}$. At $\eta=0.98$, the reoptimized solution retains only $(0,8)$, with $T_2\simeq 0.801$ and $P_{\mathrm{succ}}\simeq 2.9\times 10^{-6}$. For lower efficiencies, the optimum switches to $(0,6)$ and progressively drives the second beam splitter towards transparency. At $\eta=0.96$, $0.94$, $0.92$, and $0.90$, we obtain $T_2\simeq 0.955$, $0.969$, $0.977$, and $0.981$, respectively, while the corresponding success probabilities decrease to approximately $1.1\times 10^{-8}$, $8.6\times 10^{-10}$, $1.4\times 10^{-10}$, and $3.1\times 10^{-11}$. Thus, lowering the fidelity threshold substantially improves the tolerance against heralding-arm loss as expected, but does not remove the strong reduction in generation rate.

The preceding endpoint reoptimization shows that the loss-induced fidelity degradation can be compensated only by strongly suppressing the heralding rate. This raises the question whether the additional state-shaping freedom of genuinely alternating heralding patterns can provide a more favorable trade-off once loss is included directly in the optimization. To address this question, we repeat the lossy optimization over generic heralding patterns $(N_1,N_2)$, while retaining the squeezing constraint $|\xi_{1,2}|\leq1$. Rather than repeating this analysis for all fidelity thresholds, which would go beyond the scope of the present work, we focus on the two representative cases $F_{\mathrm{min}}=0.90$ and $0.95$. Allowing arbitrary branches substantially improves the achievable success probabilities compared with the restriction to the $N_1=0$ endpoint family. For $F_{\mathrm{min}}=0.90$, the optimum at $\eta=0.98$ is found at the nonlinear-interferometer-type endpoint $(6,0)$ with $P_{\mathrm{succ}}\simeq 1.5\times 10^{-4}$. For lower efficiencies, the optimum moves to the genuinely alternating branch $(2,4)$, yielding $P_{\mathrm{succ}}\simeq 2.2\times 10^{-5}$, $2.4\times 10^{-6}$, $4.0\times 10^{-7}$, and $9.6\times 10^{-8}$ for $\eta=0.96$, $0.94$, $0.92$, and $0.90$, respectively. A similar behavior is obtained for $F_{\mathrm{min}}=0.95$. At $\eta=0.98$, the optimum is again located at the $N_2=0$ endpoint, now for the branch $(8,0)$, with $P_{\mathrm{succ}}\simeq 2.2\times 10^{-7}$. For lower efficiencies, the optimum switches to $(2,4)$, with success probabilities of approximately $2.1\times 10^{-9}$, $1.6\times 10^{-10}$, $2.5\times 10^{-11}$, and $5.8\times 10^{-12}$ for $\eta=0.96$, $0.94$, $0.92$, and $0.90$, respectively. Thus, the additional freedom of the alternating scheme can mitigate the loss-induced rate penalty by several orders of magnitude compared with reoptimization within the $N_1=0$ family, although the absolute success probability still decreases rapidly as the heralding-arm efficiency is reduced.

The change of the optimal heralding pattern reflects the balance between fidelity margin and heralding probability. For weak loss, an $N_2=0$ endpoint can be favorable because it avoids a second nontrivial lossy heralding event altogether. As the loss increases further, a genuinely alternating branch can become preferable, since the additional squeezing and subtraction stage provide more freedom to reshape the conditional state and thereby increase its tolerance against the admixture of neighboring photon-number sectors. For the parameters considered here, this leads to the branch $(2,4)$ over most of the investigated loss range.

Although both heralding arms will generally be affected by loss in an experiment, considering losses in each arm separately helps to clarify their influence on the optimized architecture. Since the lossless constrained optimum belongs to the $N_1=0$ endpoint family, loss in the first heralding arm does not affect this solution. The first subtraction stage is inactive in this case and no nonzero detection event is required there. By contrast, loss in the second heralding arm directly acts on the active heralding stage and therefore affects the lossless optimum. In the limiting case where only the second heralding arm is lossy, reoptimization can instead favor the opposite endpoint $N_2=0$, thereby avoiding the lossy detector altogether. In this case, the scheme reduces to the nonlinear-interferometer-type configuration and retains its corresponding lossless success probability. Thus, the distribution of loss between the two heralding arms influences which subtraction stages are favored by the reoptimization. Even for genuinely alternating branches, where both subtraction stages remain active, losses in the two heralding arms can have different effects. In particular, for the representative branch $(2,4)$, loss in the second arm is more detrimental than loss in the first, consistent with the larger detected photon number at the second stage and the correspondingly stronger mixing of higher ideal photon-number sectors into a fixed registered outcome.

\section{conclusion}
\label{sec:Conclusion}

\subsection{Summary}
In this work, we have investigated how additional squeezing and photon-subtraction stages can improve the generation of optical Schr\"odinger cat states starting from photon-subtracted squeezed vacuum. Plain photon subtraction provides the simplest configuration, requiring only a single squeezing operation and one PNR heralding stage. In principle, arbitrarily high fidelities can be approached by increasing the photon-subtraction order and appropriately adjusting the squeezing. For larger cat-state amplitudes, however, this quickly translates into demanding photon-number resolution and rapidly decreasing heralding probabilities. Rather than optimizing only a single heralding outcome, we also considered a fidelity-constrained multibranch strategy. For a common experimental setting, all photon-number outcomes whose conditional states satisfy a prescribed fidelity threshold are accepted simultaneously, and their probabilities are summed to define the total success probability. This allows different photon-subtraction orders to contribute to the generation of the same target state and provides a more complete measure of the experimentally usable generation rate. Under restricted squeezing, however, fewer branches can satisfy the fidelity requirement simultaneously, reducing the corresponding multibranch gain.

Adding a second squeezing operation after the photon-subtraction stage provides a qualitatively different degree of freedom. The nonlinear-interferometer-type configuration can reshape the non-Gaussian photon-number distribution after the heralding event without directly changing its probability. As a consequence, a given target fidelity can generally be reached with substantially fewer subtracted photons than for the plain photon subtraction scheme. For example, for an $|\alpha|=3$ even cat state, fidelities above $0.98$ are already accessible with $N_1=6$, compared with $N_1=18$ for plain photon subtraction. This reduction of the required subtraction order improves the fidelity-probability trade-off. The main improvement therefore arises from the increase in fidelity achievable at low $N_1$, rather than from an enhanced probability for an individual heralding event. In the unconstrained multibranch optimization, the total success probability exceeds that of plain photon subtraction by factors between approximately $1.13$ and $1.73$ over the considered fidelity range. At the same time, the accepted photon numbers are strongly reduced, e.g., from $N_1\in\{50,52,\ldots,60\}$ to $N_1\in\{10,12\}$ at $F_{\mathrm{min}}=0.98$. The main cost of this improvement is the increased squeezing requirement. In particular, some of the low-order fidelity optima are approached only asymptotically for large individual squeezing amplitudes, although their success probabilities can already be approximated closely with finite resources.

The alternating scheme introduces a second photon-subtraction stage and thereby extends the accessible non-Gaussian state family beyond the nonlinear-interferometer-type configuration. At fixed total detected photon number, the nonlinear-interferometer-type endpoint provides the largest fidelity in the unrestricted regime for most cases, while genuinely alternating branches can become favorable for low subtraction orders or when the available squeezing is constrained. Their additional degree of freedom becomes particularly relevant when both the available squeezing and the photon-number resolution are limited, since the total subtraction order can be distributed between two detectors while the intermediate squeezing reshapes the resulting non-Gaussian state.

For the lossless multibranch success-probability optimization without an explicit bound on the detected photon number, the two endpoint families play a dominant role. In the unrestricted squeezing regime, the optimum reduces to the nonlinear-interferometer-type endpoint $N_2=0$. Under independent per-stage squeezing constraints, the optimum can instead move towards $N_1=0$, where both squeezing operations can be combined before a single nonzero subtraction event. As the available squeezing is varied, we observe a direct, threshold-dependent crossover between these two endpoint families. Below the crossover, the plain-photon-subtraction-type endpoint $N_1=0$ is favored, whereas above it, including the unrestricted limit, the nonlinear-interferometer-type endpoint $N_2=0$ becomes optimal. Within the numerical resolution of our scan, no intermediate squeezing regime is found in which genuinely alternating branches with $N_1,N_2>0$ become globally optimal. For $|\xi_i|\leq1.5$, the $N_1=0$ endpoint already provides sufficient combined squeezing to reproduce the unrestricted plain-photon-subtraction optimum throughout the considered fidelity range. For the tighter bound $|\xi_i|\leq1$, both squeezers are driven towards their maximal values and the resulting success probabilities remain substantially larger than those of either plain photon subtraction or the constrained nonlinear-interferometer-type configuration. This advantage comes with an increased requirement on photon-number resolution, reflecting a trade-off between the available squeezing resources and the photon-subtraction orders required to reach a given fidelity. Correspondingly, once the maximum photon number resolved by each detector is restricted as well, genuinely alternating branches can again become favorable and may outperform the endpoint configurations by more than two orders of magnitude. 

The comparison therefore does not identify a single configuration that is preferable independently of the available resources. Plain photon subtraction minimizes the optical complexity, the nonlinear-interferometer-type scheme provides an efficient route towards high fidelities with comparatively small detected photon numbers, and the alternating configuration offers the largest flexibility when the available resources are constrained.

The analysis of experimental imperfections emphasizes the importance of the heralding detectors. Loss in a PNR heralding arm mixes different actual photon-number sectors into the same registered outcome and can therefore rapidly reduce the conditional fidelity. Reoptimizing an endpoint configuration can recover a prescribed fidelity only by strongly suppressing the heralding rate. Allowing genuinely alternating patterns provides additional freedom to compensate this admixture and can improve the lossy success probability by several orders of magnitude compared with such endpoint reoptimization, although the absolute generation rate still decreases rapidly with increasing heralding-arm loss.

Finally, the considered schemes also admit a complementary interpretation in terms of Gaussian-state preparation followed by non-Gaussian conditioning. Since Gaussian operations acting on the retained mode commute with measurements on modes that are no longer involved in the subsequent evolution, the nonlinear-interferometer-type configuration can equivalently be regarded as the preparation of a pure nondisplaced two-mode Gaussian state followed by a PNR projection on one mode. By the Bloch-Messiah decomposition, the same Gaussian state can alternatively be prepared from two squeezed-vacuum inputs mixed on a passive two-mode interferometer, up to local phase rotations. This corresponds to a realization of generalized photon subtraction as investigated in Ref.~\cite{Takase2021}. Equivalently, Gaussian processing acting on the detected mode can also be absorbed into the measurement basis. A squeezer preceding PNR detection is equivalent to projection onto a squeezed Fock state. The same conditional state family may therefore also be viewed as arising from a squeezed vacuum mixed with vacuum and conditioned on such a squeezed-Fock-state measurement. Although these representations are equivalent at the level of the accessible conditional states in the unrestricted Gaussian setting, they need not remain resource equivalent once constraints are imposed on the available squeezing operations or on the measurement implementation. Likewise, the two-stage alternating scheme can be viewed as preparing a structured three-mode Gaussian state followed by PNR conditioning on two modes. This establishes a direct connection between the present sequential description and GBS-like Gaussian-PNR state-engineering architectures~\cite{Su:2019vdv}.

\subsection{Comparison to the state of the art}
Table~\ref{tab:cat_generation} places the schemes studied in this work alongside representative theoretical proposals and experimental demonstrations of optical Schr\"odinger cat state generation. Since the considered approaches differ in their input resources as well as their fidelity and heralding definitions, the reported values provide representative benchmarks rather than a direct performance ranking. For a common target amplitude of $|\alpha|=3$, all three schemes studied here can reach a fidelity threshold of $F_{\mathrm{min}}=0.99$ using only squeezed vacuum inputs, beam splitters, and PNR detection, without requiring an engineered non-Gaussian ancilla such as the squeezed few-photon superposition state used in Ref.~\cite{luo2025}.

Within this restricted resource class, generalized photon subtraction as investigated in Ref.~\cite{Takase2021} provides a particularly relevant reference. Its two-mode Gaussian state realization is closely related to the nonlinear-interferometer-type configuration considered here, although the two implementations need not require equivalent squeezing resources. Moreover, the reported heralding probabilities in Ref.~\cite{Takase2021} refer to individual measurement outcomes approximating squeezed cat states, whereas our results concern fidelity-constrained multibranch optimization for nonsqueezed cat states. The corresponding probabilities therefore cannot be compared directly. Generalized photon subtraction has also recently been demonstrated experimentally in the picosecond pulsed regime, with photon-number-resolved heralding up to $N=4$ photons~\cite{endo2026}. The reported fidelity of $0.996$ at an effective cat-state amplitude of $\alpha_{\mathrm{eff}}=1.69$, however, is computed against a loss-included theoretical model rather than an ideal cat state, unlike the fidelities used throughout this work and in Refs.~\cite{Sychev2017,caron2026}, which quantify closeness to an ideal (possibly squeezed) cat state, either directly for the lossless theoretical states considered here or after correcting the measured state for specified experimental losses. The reported fidelity of $0.996$, therefore reflects agreement between the experiment and its own loss-included model rather than proximity to an ideal cat state, and should not be read as directly comparable to those other values, even though the demonstrated heralding order itself marks substantial experimental progress towards the large-amplitude, high-fidelity regime.

Other routes to larger cat amplitudes illustrate the different fidelity-resource trade-offs, including cat breeding at $|\alpha|\approx1.85$ with a fidelity of $77\%$ after efficiency correction ~\cite{Sychev2017} and coherent-state superpositions of free-space optical pulses reaching $|\alpha|\approx2.47$ at a fidelity of only about $53\%$~\cite{caron2026}. These demonstrations underline the experimental challenge of generating large-amplitude cat states with high fidelity, while the results presented here identify the squeezing and photon-number resolution requirements for approaching this regime.

\begin{table*}[htbp]
\centering
\setlength{\tabcolsep}{5pt}
\renewcommand{\arraystretch}{1.35}

\newcommand{\CatLeft}[2]{%
    \parbox[t]{#1}{\raggedright #2}%
}
\newcommand{\CatCenter}[2]{%
    \parbox[t]{#1}{\centering #2}%
}

\begin{tabular}{@{}lcccc@{\hspace{14pt}}ll@{}}
\toprule

\CatLeft{2.0cm}{\textbf{Source}} &
\CatCenter{1.2cm}{$\boldsymbol{|\alpha|}$} &
\CatCenter{1.8cm}{\textbf{Fidelity}} &
\CatCenter{2.1cm}{\textbf{Success Prob.}} &
\CatCenter{1.2cm}{\textbf{Th./Exp.}} &
\CatLeft{3.0cm}{\textbf{Resources}} &
\CatLeft{3.9cm}{\textbf{Scheme}} \\

\midrule

\CatLeft{2.0cm}{Takase (2021)\cite{Takase2021}} &
\CatCenter{1.2cm}{$3.0$} &
\CatCenter{1.8cm}{$0.997$} &
\CatCenter{2.1cm}{$\sim10^{-2}$} &
\CatCenter{1.2cm}{Th.} &
\CatLeft{3.0cm}{Two-mode squeezing interference; PNR, squeezing $1.15$} &
\CatLeft{3.9cm}{Generalized photon subtraction (proposal)} \\

\CatLeft{2.0cm}{Podoshvedov (2023)\cite{Podoshvedov2023}} &
\CatCenter{1.2cm}{$2.5$} &
\CatCenter{1.8cm}{$0.97$} &
\CatCenter{2.1cm}{$10^{-7}$} &
\CatCenter{1.2cm}{Th.} &
\CatLeft{3.0cm}{$>10$ dB squeezing; PNR $10$} &
\CatLeft{3.9cm}{Sequential photon subtraction} \\

\CatLeft{2.0cm}{Luo (2025)\cite{luo2025}} &
\CatCenter{1.2cm}{$2.24$} &
\CatCenter{1.8cm}{$>0.99$} &
\CatCenter{2.1cm}{$\ge0.5$} &
\CatCenter{1.2cm}{Th.} &
\CatLeft{3.0cm}{$10$ dB squeezing; two-photon Fock ancilla} &
\CatLeft{3.9cm}{Photon subtraction from squeezed few-photon superposition states} \\

\CatLeft{2.0cm}{Sychev (2017)\cite{Sychev2017}} &
\CatCenter{1.2cm}{$1.85$} &
\CatCenter{1.8cm}{$0.77$} &
\CatCenter{2.1cm}{$\sim0.2$ (only amplification)} &
\CatCenter{1.2cm}{Exp.} &
\CatLeft{3.0cm}{Two odd cats with amplitude $1.15$} &
\CatLeft{3.9cm}{Cat ``breeding''} \\

\CatLeft{2.0cm}{Endo (2026)\cite{endo2026}} &
\CatCenter{1.2cm}{$1.69$} &
\CatCenter{1.8cm}{$0.996^{a}$} &
\CatCenter{2.1cm}{n/a} &
\CatCenter{1.2cm}{Exp.} &
\CatLeft{3.0cm}{Squeezed vacuum; PNR} &
\CatLeft{3.9cm}{Generalized photon subtraction (four photons)} \\

\CatLeft{2.0cm}{Caron (2026)\cite{caron2026}} &
\CatCenter{1.2cm}{$2.47$} &
\CatCenter{1.8cm}{$0.53$} &
\CatCenter{2.1cm}{n/a} &
\CatCenter{1.2cm}{Exp.} &
\CatLeft{3.0cm}{Fock states $|1\rangle,|2\rangle$} &
\CatLeft{3.9cm}{Fock state mixing (coherent state superposition)} \\

\CatLeft{2.0cm}{This manuscript} &
\CatCenter{1.2cm}{$3.0$} &
\CatCenter{1.8cm}{$0.99$} &
\CatCenter{2.1cm}{$1.6\times10^{-6}$} &
\CatCenter{1.2cm}{Th.} &
\CatLeft{3.0cm}{Squeezing $1.0$,\\ PNR $28$} &
\CatLeft{3.9cm}{Photon subtraction} \\

\CatLeft{2.0cm}{This manuscript} &
\CatCenter{1.2cm}{$3.0$} &
\CatCenter{1.8cm}{$0.99$} &
\CatCenter{2.1cm}{$2\times10^{-5}$} &
\CatCenter{1.2cm}{Th.} &
\CatLeft{3.0cm}{Squeezing $1.0$,\\ PNR $14$} &
\CatLeft{3.9cm}{Nonlinear interferometer} \\

\CatLeft{2.0cm}{This manuscript} &
\CatCenter{1.2cm}{$3.0$} &
\CatCenter{1.8cm}{$0.99$} &
\CatCenter{2.1cm}{$0.0099$} &
\CatCenter{1.2cm}{Th.} &
\CatLeft{3.0cm}{Squeezing $1.0$,\\ PNR $32$} &
\CatLeft{3.9cm}{Alternating squeezing and photon subtraction ($N_1=0$ endpoint)} \\

\bottomrule
\end{tabular}
\caption{Comparison of optical (squeezed) Schr\"odinger cat state generation schemes. Fidelity is evaluated against an ideal cat state for Refs.~\cite{Podoshvedov2023,luo2025} and for the schemes studied in this work, and against a squeezed cat state for Refs.~\cite{Takase2021,Sychev2017, endo2026, caron2026}. The reported fidelities and success probabilities follow the conventions of the respective references and are not directly comparable across all entries. For the schemes studied here, the probabilities correspond to the optimized lossless multibranch success probability at $F_{\mathrm{min}}=0.99$ under $|\xi_i|\leq1$.\\
\footnotesize $^{a}$Fidelity between the reconstructed experimental state and a loss included simulated target, not a measure of closeness to an ideal cat state; the density matrix itself is reconstructed without loss correction. This differs from the fidelities reported in this work and in Refs.~\cite{Sychev2017,caron2026}, which quantify closeness to an ideal cat state, either directly for the lossless theoretical states of this work or after correcting the measured state for known losses in Refs.~\cite{Sychev2017,caron2026}.
}
\label{tab:cat_generation}
\end{table*}

\subsection{Outlook}
The alternating construction considered in this work naturally suggests an extension to more than two squeezing and photon-subtraction stages. Importantly, the structure obtained in Sec.~\ref{sec:AlternatingScheme} is well suited for such an iterative treatment. After two subtraction stages, the conditional state can be written as a finite superposition of single Hermite polynomials acting on one effective squeezed vacuum,
\begin{align*}
 \ket{\psi_{N_1,N_2}} \sim
\sum_{k=0}^{\lfloor \frac{N_1}{2}\rfloor} C_k^{(2)}  \He_{N_1+N_2-2k}\!\bigg(\sqrt{\tfrac{\tilde{\breve{\tau}}_2}{2}}\, \hat a^\dagger \bigg)
 \hat S\Big(\tilde{\breve{\xi}}_2\Big)\ket{0},
\end{align*}
where the coefficients $C_k^{(2)}$ are given by Eq.~\eqref{eq:alternating-state}. This representation provides a convenient starting point for adding a third squeezing and subtraction stage.

Applying a third squeezing operation $\hat S(\xi_3)$, the squeezer can first be commuted through each Hermite polynomial. At the same time, it combines with the effective squeezing of the preceding stages according to
\begin{align}
 \breve{\tau}_3 = \frac{\tilde{\breve{\tau}}_2+\tau_3}{1+\tilde{\breve{\tau}}_2\tau_3^*}.
\end{align}
The subsequent beam splitter attenuates the combined squeezing parameter to $\tilde{\breve{\tau}}_3 = T_3\breve{\tau}_3$, $\taut_3=T_3\tau_3$,
and $\tilde{\breve{\xi}}_3$ is defined, in direct analogy to the preceding stages, through $\E^{\I\tilde{\breve{\varphi}}_3}\tanh|\tilde{\breve{\xi}}_3| = \tilde{\breve{\tau}}_3$. Conditioning on an \mbox{$N_3$-photon} detection event then requires commuting the corresponding annihilation operators through each Hermite component and the Gaussian part. As for the second subtraction stage, this initially generates products of Hermite polynomials. Using the same Hermite-polynomial multiplication and rescaling identities as for the two-stage problem, these products can again be collected into a  sum over single Hermite polynomials. The resulting three-stage conditional state can therefore be written as
\begin{align}
 \ket{\psi_{N_1,N_2,N_3}} &= \frac{1}{\mathcal{N}_{N_1,N_2,N_3}}
 \sum_{r=0}^{\lfloor \frac{(N_1+N_2)}{2}\rfloor}  C_r^{(3)}
\\
 &\quad\times
 \He_{N_1+N_2+N_3-2r}\!\bigg( \sqrt{\tfrac{\tilde{\breve{\tau}}_3}{2}}\,\hat a^\dagger \bigg)
 \hat S\Big(\tilde{\breve{\xi}}_3\Big)\ket{0},  \notag
\end{align}
where
\begin{align*}
 C_r^{(3)} &= \frac{1}{(N_1+N_2-2r)!}
 \!\! \sum_{k=0}^{\min(r,\lfloor \frac{N_1}{2} \rfloor)} \!\! \frac{(N_1+N_2-2k)!}{k!(N_1-2k)!(r-k)!}
 \notag\\
 &\quad\times
 \left[ (1+\tilde{\breve{\tau}}_2\tau_3^*) \frac{\breve{\tau}_3}{\breve{\tau}_3-\tau_3} \gamma_2 \right]^k
 \left( \frac{\tau_3}{\tau_3-\breve{\tau}_3} \right)^{r-k},
\end{align*}
and 
\begin{align*}
 &\mathcal{N}_{N_1,N_2,N_3}^{2} =
 \sum_{j=0}^{\lfloor \frac{N_1+N_2+N_3}{2}\rfloor}
 \frac{(2|\tilde{\breve{\tau}}_3|)^{N_1+N_2+N_3-2j}}{(N_1+N_2+N_3-2j)!} \notag\\
 &\quad\times
 \left| \sum_{r=0}^{\min\left(j,\lfloor \frac{N_1+N_2}{2}\rfloor\right)} \frac{(-1)^r (N_1+N_2+N_3-2r)!C_r^{(3)}}{(j-r)!\left(1-|\tilde{\breve{\tau}}_3|^2\right)^{\frac{N_1+N_2+N_3}{2}-r}} \right|^2.
\end{align*}
Thus, the third stage does not introduce a qualitatively different functional form. Instead, it changes the coefficients of the Hermite-polynomial superposition, increases its maximal order by $N_3$, and replaces the Gaussian part by the new effective squeezing $\tilde{\breve{\xi}}_3$.

The same steps can be repeated iteratively for further squeezing and subtraction stages. Before the $j$th subtraction, the newly applied squeezing operation combines with the effective squeezing produced by all preceding stages according to
\begin{align}
 \breve{\tau}_j &= \frac{\tilde{\breve{\tau}}_{j-1}+\tau_j}{1+\tilde{\breve{\tau}}_{j-1}\tau_j^*},
 \qquad
 \tilde{\breve{\tau}}_j = T_j\breve{\tau}_j,
\end{align}
with $\E^{\I\tilde{\breve{\varphi}}_j}\tanh|\tilde{\breve{\xi}}_j| = \tilde{\breve{\tau}}_j$. 
After conditioning on the corresponding photon-number outcome and collecting the resulting Hermite-polynomial products, the state after an arbitrary finite number $j$ of stages retains the general form
\begin{align}
 \ket{\psi_{N_1,\ldots,N_j}} &= \frac{1}{\mathcal{N}_{N_1,\ldots,N_j}}
 \sum_{r=0}^{\left\lfloor
 \frac{N_1+\cdots+N_{j-1}}{2} \right\rfloor} C_r^{(j)} \\
 &\quad\times \He_{N_1+\cdots+N_j-2r}\!\bigg( \sqrt{\tfrac{\tilde{\breve{\tau}}_j}{2}}\, \hat a^\dagger \bigg)
 \hat S\Big(\tilde{\breve{\xi}}_j\Big)\ket{0}.  \notag
\end{align}
A useful structural consequence is that the increasing number of stages adds complexity primarily to the coefficients $C_r^{(j)}$, but not to the functional basis of the conditional state. After every additional squeezing-plus-subtraction stage, the Gaussian contribution can again be absorbed into the single effective squeezing parameter $\tilde{\breve{\tau}}_j$, while the non-Gaussian part remains a finite superposition of Hermite polynomials with one common argument and orders differing by two. Additional stages therefore enlarge the number and tunability of the coefficients in this superposition without requiring a qualitatively new class of basis functions.

Already for three subtraction stages, the endpoint families $(N,0,0)$, $(0,N,0)$, and $(0,0,N)$ correspond to distinct distributions of the available squeezing operations around a single nonzero heralding event. The first two configurations belong to the nonlinear-interferometer-type family. For $(N,0,0)$, one squeezer precedes the active subtraction while the two subsequent squeezing operations can be combined into an effective
post-subtraction squeezer. For $(0,N,0)$, two squeezing operations can instead be combined before the active subtraction, while the third remains available afterwards. By contrast, for $(0,0,N)$ all squeezing operations precede the only nonzero subtraction event and can be combined into a single effective input squeezing, such that the conditional state belongs to the plain photon-subtraction family. Under independent per-stage squeezing constraints, the two nonlinear-interferometer-type endpoints are nevertheless not resource equivalent, since they allow different amounts of the available squeezing to be combined before and after the active subtraction event. This suggests the conjecture that finite per-stage squeezing constraints may shift the optimal nonlinear-interferometer-type realization from $(N,0,0)$ to $(0,N,0)$ as the available squeezing resources are varied.

More generally, distributing a fixed total detected photon number over several subtraction stages with intermediate squeezing enlarges the set of accessible Hermite-polynomial coefficients while reducing the photon number that must be resolved by an individual detector. Additional squeezing operations, whether placed between or after the nonzero subtraction events, provide further unitary degrees of freedom for reshaping the conditional state without necessarily introducing another nonzero heralding requirement. It is therefore conceivable that additional stages may allow a given target fidelity to be reached with a smaller maximum resolved photon number, or even with fewer detected photons in total, owing to the larger flexibility provided by the combined action of photon subtraction and squeezing.

The same freedom may also be relevant in the presence of heralding loss. Each additional nonzero heralding event introduces another loss-sensitive conditioning step and therefore generally reduces the joint probability. At the same time, however, the larger number of squeezing parameters and photon-number partitions provides additional freedom to increase the fidelity margin against the admixture of neighboring photon-number sectors. The two-stage loss analysis already indicates that the optimum need not remain on a single-subtraction endpoint once detector loss is included. Extending this optimization to three or more stages would therefore allow one to investigate whether the increased flexibility in combining squeezing and photon subtraction can outweigh the probability penalty associated with further heralding events, and whether configurations with only one or two active subtraction stages embedded in a longer squeezing sequence are preferable to fully alternating patterns.

Finally, the state-engineering framework developed here is not restricted to cat states. The Hermite-polynomial coefficients generated by alternating squeezing and PNR conditioning provide a tunable non-Gaussian Fock-space structure that could also be optimized for other bosonic resource states. Possible targets include finite-energy GKP states~\cite{Gottesman:2000di, Su:2019vdv, Eaton:2019eyl, Takase:2024fia} and binomial code states~\cite{Michael:2016eaq, Nehra:2021hxw}, for which different photon-number amplitudes and relative phases are required. Exploring to what extent the tunable Hermite-polynomial structure generated by alternating squeezing and photon-number conditioning can approximate such bosonic resource states, and what fidelity--success-probability trade-offs can be achieved, constitutes another natural direction for future work.

\section*{Acknowledgements}
We thank Gil Zimmermann and Gregor Sauer for insightful discussions. We acknowledge funding by the Deutsche Forschungsgemeinschaft (DFG, German Research Foundation) -- Project-ID 398816777 -- SFB 1375 (projects A7 and C6). JC and FS are members of the Max Planck School of Photonics, supported by the Dieter Schwarz Foundation, the BMFTR, and the Max Planck Society.

\clearpage

\begin{widetext}
\appendix*
\section{Numerical Results for Multibranch Fidelity--Success-Probability Optimization}
\phantomsection\label{app:numerical-results}

\begin{table*}[h]
\centering
\setlength{\tabcolsep}{5pt}

\begin{tabular}{clcccc}
\multicolumn{6}{c}{\textbf{Even cat, plain photon subtraction}}\\[1mm]
\toprule
$F_{\min}$ & $\;\; P_{\mathrm{succ}}$ & $|\tilde{\tau}_1|$ & $T_1$ & $|\xi_1|$ & $\mathcal{S}$ \\
\midrule
\multicolumn{6}{c}{\textit{Unconstrained squeezing}} \\
\midrule
0.90 & 0.06663 & 0.55391 & 0.57096 & 2.0946 & $\{6,8\}$ \\
0.91 & 0.06383 & 0.37139 & 0.37917 & 2.2846 & $\{12,14,16,18\}$ \\
0.92 & 0.05934 & 0.42389 & 0.43395 & 2.2229 & $\{10,12,14\}$ \\
0.93 & 0.05203 & 0.49160 & 0.50509 & 2.1512 & $\{8,10\}$ \\
0.94 & 0.04689 & 0.19928 & 0.20151 & 2.5957 & $\{30,32,\ldots,42\}$ \\
0.95 & 0.04456 & 0.35583 & 0.36292 & 2.3094 & $\{14,16,18\}$ \\
0.96 & 0.03827 & 0.26310 & 0.26696 & 2.4612 & $\{22,24,26,28\}$ \\
0.97 & 0.03334 & 0.19855 & 0.20076 & 2.5984 & $\{32,34,36,38,40\}$ \\
0.98 & 0.02625 & 0.13997 & 0.14106 & 2.7761 & $\{50,52,\ldots,60\}$ \\
0.99 & 0.01829 & 0.10187 & 0.10245 & 2.9322 & $\{74,76,\ldots,84\}$ \\
\midrule
\multicolumn{6}{c}{\textit{Squeezing constraint $|\xi_1|\leq 1.5$}} \\
\midrule
0.90 & 0.03740 & 0.54939 & 0.60697 & 1.4999 & $\{6,8\}$ \\
0.91 & 0.02638 & 0.47398 & 0.52365 & 1.5000 & $\{8,10\}$ \\
0.92 & 0.02587 & 0.48107 & 0.53149 & 1.4999 & $\{8,10\}$ \\
0.93 & 0.02520 & 0.49019 & 0.54156 & 1.5000 & $\{8,10\}$ \\
0.94 & 0.01823 & 0.43427 & 0.47978 & 1.5000 & $\{10,12\}$ \\
0.95 & 0.01335 & 0.39309 & 0.43429 & 1.4999 & $\{12,14\}$ \\
0.96 & 0.00988 & 0.36144 & 0.39932 & 1.4999 & $\{14,16\}$ \\
0.97 & 0.00735 & 0.33711 & 0.37244 & 1.4999 & $\{16,18\}$ \\
0.98 & 0.00426 & 0.29452 & 0.32539 & 1.4999 & $\{20,22\}$ \\
0.99 & 0.00123 & 0.22220 & 0.24549 & 1.4999 & $\{30,32\}$ \\
\midrule
\multicolumn{6}{c}{ \textit{Squeezing constraint $|\xi_1|\leq1$} } \\
\midrule
0.90 & 0.00199 & 0.54939 & 0.72137 & 1.0000 & $\{6,8\}$ \\
0.91 & 0.00140 & 0.56118 & 0.73685 & 1.0000 & $\{6\}$ \\
0.92 & 0.00114 & 0.48107 & 0.63167 & 1.0000 & $\{8,10\}$ \\
0.93 & 0.00101 & 0.49019 & 0.64364 & 1.0000 & $\{8,10\}$ \\
0.94 & 0.00059 & 0.43427 & 0.57022 & 1.0000 & $\{10,12\}$ \\
0.95 & 0.00039 & 0.44487 & 0.58414 & 1.0000 & $\{10\}$ \\
0.96 & 0.00021 & 0.40308 & 0.52926 & 1.0000 & $\{12\}$ \\
0.97 & 0.00011 & 0.37316 & 0.48998 & 1.0000 & $\{14\}$ \\
0.98 & 0.00003 & 0.32111 & 0.42163 & 1.0000 & $\{18\}$ \\
0.99 & 0.0000016 & 0.23597 & 0.30984 & 1.0000 & $\{28\}$ \\
\bottomrule
\end{tabular}

\caption{Optimized multibranch fidelity--success-probability trade-off for generating an even cat state with $|\alpha|=3$ using plain photon subtraction. For each minimum fidelity $F_{\mathrm{min}}$, $P_{\mathrm{succ}}$ denotes the total probability of the accepted photon-number outcomes $\mathcal{S}$. The listed parameters are conservatively rounded such that all branches in $\mathcal{S}$ satisfy $F_{N_1}\geq F_{\mathrm{min}}$ when evaluated using the displayed values.}
\label{tab:PS-multibranch-even}
\end{table*}

\begin{table*}[h]
\centering
\setlength{\tabcolsep}{5pt}

\begin{tabular}{clcccc}
\multicolumn{6}{c}{\textbf{Odd cat, plain photon subtraction}}\\[1mm]
\toprule
$F_{\min}$ & $\;\;P_{\mathrm{succ}}$ & $|\tilde{\tau}_1|$ & $T_1$ & $|\xi_1|$ & $\mathcal{S}$ \\
\midrule
\multicolumn{6}{c}{\textit{Unconstrained squeezing}} \\
\midrule
0.90 & 0.06844 & 0.38741 & 0.39590 & 2.2623 & $\{11,13,15,17\}$ \\
0.91 & 0.06472 & 0.44457 & 0.45568 & 2.1974 & $\{9,11,13\}$ \\
0.92 & 0.05722 & 0.29826 & 0.30326 & 2.3950 & $\{17,19,21,23,25\}$ \\
0.93 & 0.05478 & 0.40434 & 0.41351 & 2.2454 & $\{11,13,15\}$ \\
0.94 & 0.04801 & 0.26334 & 0.26723 & 2.4578 & $\{21,23,25,27,29\}$ \\
0.95 & 0.04348 & 0.28855 & 0.29321 & 2.4135 & $\{19,21,23,25\}$ \\
0.96 & 0.03871 & 0.22351 & 0.22631 & 2.5396 & $\{27,29,31,33,35\}$ \\
0.97 & 0.03280 & 0.16886 & 0.17046 & 2.6785 & $\{39,41,43,45,47,49\}$ \\
0.98 & 0.02669 & 0.16561 & 0.16715 & 2.6878 & $\{41,43,45,47,49\}$ \\
0.99 & 0.01851 & 0.12099 & 0.12181 & 2.8453 & $\{61,63,65,67,69\}$ \\
\midrule
\multicolumn{6}{c}{ \textit{Squeezing constraint $|\xi_1|\leq1.5$} } \\
\midrule
0.90 & 0.03167 & 0.50405 & 0.55688 & 1.4999 & $\{7,9\}$ \\
0.91 & 0.03105 & 0.51144 & 0.56504 & 1.5000 & $\{7,9\}$ \\
0.92 & 0.02211 & 0.44837 & 0.49536 & 1.5000 & $\{9,11\}$ \\
0.93 & 0.02168 & 0.45547 & 0.50320 & 1.5000 & $\{9,11\}$ \\
0.94 & 0.02109 & 0.46484 & 0.51356 & 1.4999 & $\{9,11\}$ \\
0.95 & 0.01537 & 0.41662 & 0.46028 & 1.5000 & $\{11,13\}$ \\
0.96 & 0.01131 & 0.38069 & 0.42059 & 1.4999 & $\{13,15\}$ \\
0.97 & 0.00836 & 0.35370 & 0.39077 & 1.4999 & $\{15,17\}$ \\
0.98 & 0.00377 & 0.28343 & 0.31314 & 1.4999 & $\{21,23\}$ \\
0.99 & 0.00109 & 0.21609 & 0.23874 & 1.4999 & $\{31,33\}$ \\
\midrule
\multicolumn{6}{c}{ \textit{Squeezing constraint $|\xi_1|\leq1$} } \\
\midrule
0.90 & 0.00170 & 0.50405 & 0.66184 & 1.0000 & $\{7,9\}$ \\
0.91 & 0.00154 & 0.51144 & 0.67154 & 1.0000 & $\{7,9\}$ \\
0.92 & 0.00110 & 0.52101 & 0.68411 & 1.0000 & $\{7\}$ \\
0.93 & 0.00082 & 0.45547 & 0.59805 & 1.0000 & $\{9,11\}$ \\
0.94 & 0.00073 & 0.46484 & 0.61036 & 1.0000 & $\{9,11\}$ \\
0.95 & 0.00040 & 0.41662 & 0.54704 & 1.0000 & $\{11,13\}$ \\
0.96 & 0.00025 & 0.43203 & 0.56728 & 1.0000 & $\{11\}$ \\
0.97 & 0.00011 & 0.35370 & 0.46443 & 1.0000 & $\{15,17\}$ \\
0.98 & 0.000023 & 0.30684 & 0.40290 & 1.0000 & $\{19\}$ \\
0.99 & 0.000002 & 0.24429 & 0.32077 & 1.0000 & $\{27\}$ \\
\midrule
\end{tabular}

\caption{Optimized multibranch fidelity--success-probability trade-off for generating an odd cat state with $|\alpha|=3$ using plain photon subtraction. For each minimum fidelity $F_{\mathrm{min}}$, $P_{\mathrm{succ}}$ denotes the total probability of the accepted photon-number outcomes $\mathcal{S}$. The listed parameters are conservatively rounded such that all branches in $\mathcal{S}$ satisfy $F_{N_1}\geq F_{\mathrm{min}}$ when evaluated using the displayed values.}
\label{tab:PS-multibranch-odd}
\end{table*}

\begin{table*}[h]
\centering
\setlength{\tabcolsep}{5pt}

\begin{tabular}{clcccccc}
\multicolumn{8}{c}{\textbf{Even cat, nonlinear interferometer}}\\[1mm]
\toprule
$F_{\min}$ & \multicolumn{1}{c}{$P_{\mathrm{succ}}$} & $|\tilde{\tau}_1|$ & $|\tau_2|$ & $T_1$ & $|\xi_1|$ & $|\xi_2|$ & $\mathcal{S}$ \\
\midrule
\multicolumn{8}{c}{ \textit{Unconstrained squeezing} } \\
\midrule
0.90 & $0.09474^{\dagger}$ & $\to 1$ & $\to 1$ & $\to 1$ & $\to\infty$ & $\to\infty$ & $\{4,6\}$ \\
\quad & 0.09455 & 0.96403 & 0.87778 & 0.96730 & 3.1901 & 1.3660 & $\{4,6\}$ \\[1mm]
0.91 & 0.07215 & 0.97110 & 0.94873 & 0.97247 & 3.6287 & 1.8189 & $\{8,10,12\}$ \\[1mm]
0.92 & 0.07178 & 0.77155 & 0.58602 & 0.78021 & 2.5942 & 0.6716 & $\{8,10,12\}$ \\[1mm]
0.93 & $0.06771^{\dagger}$ & $\to 1$ & $\to 1$ & $\to 1$ & $\to\infty$ & $\to\infty$ & $\{6,8\}$ \\
\quad & 0.06762 & 0.96403 & 0.91094 & 0.96639 & 3.3537 & 1.5330 & $\{6,8\}$ \\[1mm]
0.94 & $0.06771^{\dagger}$ & $\to 1$ & $\to 1$ & $\to 1$ & $\to\infty$ & $\to\infty$ & $\{6,8\}$ \\
\quad & 0.06762 & 0.96403 & 0.91094 & 0.96639 & 3.3537 & 1.5330 & $\{6,8\}$ \\[1mm]
0.95 & $0.06771^{\dagger}$ & $\to 1$ & $\to 1$ & $\to 1$ & $\to\infty$ & $\to\infty$ & $\{6,8\}$ \\
\quad & 0.06762 & 0.96403 & 0.91094 & 0.96639 & 3.3537 & 1.5330 & $\{6,8\}$ \\[1mm]
0.96 & $0.05278^{\dagger}$ & $\to 1$ & $\to 1$ & $\to 1$ & $\to\infty$ & $\to\infty$ & $\{8,10\}$ \\
\quad & 0.05272 & 0.96403 & 0.92985 & 0.96588 & 3.4755 & 1.6573 & $\{8,10\}$ \\[1mm]
0.97 & $0.05278^{\dagger}$ & $\to 1$ & $\to 1$ & $\to 1$ & $\to\infty$ & $\to\infty$ & $\{8,10\}$ \\
\quad & 0.05272 & 0.96403 & 0.92985 & 0.96588 & 3.4755 & 1.6573 & $\{8,10\}$ \\[1mm]
0.98 & 0.04313 & 0.86121 & 0.77208 & 0.86646 & 2.8981 & 1.0254 & $\{10,12\}$ \\[1mm]
0.99 & 0.03163 & 0.60293 & 0.44264 & 0.61074 & 2.5230 & 0.4755 & $\{14,16\}$ \\
\midrule
\multicolumn{8}{c}{ \textit{Squeezing constraint $|\xi_1|,|\xi_2|\leq1.5$} } \\
\midrule
0.90 & 0.03740 & 0.54939 & 0.00000 & 0.60697 & 1.4999 & 0.0000 & $\{6,8\}$ \\
0.91 & 0.03406 & 0.57936 & 0.05459 & 0.64008 & 1.4999 & 0.0546 & $\{6,8\}$ \\
0.92 & 0.02854 & 0.62385 & 0.14577 & 0.68923 & 1.4999 & 0.1468 & $\{6,8\}$ \\
0.93 & 0.02520 & 0.49019 & 0.00000 & 0.54156 & 1.5000 & 0.0000 & $\{8,10\}$ \\
0.94 & 0.02238 & 0.52689 & 0.07113 & 0.58211 & 1.4999 & 0.0712 & $\{8,10\}$ \\
0.95 & 0.01838 & 0.57512 & 0.16667 & 0.63539 & 1.5000 & 0.1682 & $\{8,10\}$ \\
0.96 & 0.01504 & 0.49145 & 0.09811 & 0.54295 & 1.5000 & 0.0984 & $\{10,12\}$ \\
0.97 & 0.01159 & 0.54946 & 0.21104 & 0.60704 & 1.5000 & 0.2142 & $\{10,12\}$ \\
0.98 & 0.00837 & 0.50944 & 0.21260 & 0.56283 & 1.5000 & 0.2159 & $\{12,14\}$ \\
0.99 & 0.00475 & 0.43941 & 0.20419 & 0.48546 & 1.5000 & 0.2071 & $\{16,18\}$ \\
\midrule
\multicolumn{8}{c}{ \textit{Squeezing constraint $|\xi_1|,|\xi_2|\leq1$} } \\
\midrule
0.90 & 0.00198 & 0.54939 & 0.00000 & 0.72137 & 1.0000 & 0.0000 & $\{6,8\}$ \\
0.91 & 0.00140 & 0.56118 & 0.00000 & 0.73685 & 1.0000 & 0.0000 & $\{6\}$ \\
0.92 & 0.00114 & 0.48107 & 0.00000 & 0.63167 & 1.0000 & 0.0000 & $\{8,10\}$ \\
0.93 & 0.00101 & 0.49019 & 0.00000 & 0.64364 & 1.0000 & 0.0000 & $\{8,10\}$ \\
0.94 & 0.00062 & 0.50845 & 0.00693 & 0.66762 & 1.0000 & 0.0069 & $\{8\}$ \\
0.95 & 0.00048 & 0.44942 & 0.01441 & 0.59011 & 1.0000 & 0.0144 & $\{10,12\}$ \\
0.96 & 0.00027 & 0.40429 & 0.00395 & 0.53085 & 1.0000 & 0.0039 & $\{12,14\}$ \\
0.97 & 0.00014 & 0.44607 & 0.08894 & 0.58571 & 1.0000 & 0.0891 & $\{12,14\}$ \\
0.98 & 0.00007 & 0.47572 & 0.12301 & 0.62464 & 1.0000 & 0.1236 & $\{12\}$ \\
0.99 & 0.00002 & 0.46397 & 0.16911 & 0.60921 & 1.0000 & 0.1707 & $\{14\}$ \\
\bottomrule
\end{tabular}

\caption{Optimized multibranch fidelity--success-probability trade-off for generating an even cat state with $|\alpha|=3$ using the nonlinear-interferometer-type scheme. For each minimum fidelity $F_{\min}$, $P_{\mathrm{succ}}$ denotes the total probability of the accepted photon-number outcomes $\mathcal{S}$. A dagger denotes an asymptotic optimum approached in the infinite-squeezing limit. For each such entry, the immediately following row with an empty $F_{\min}$ field gives an approximative representative finite squeezing parameter realization using $|\tilde{\xi}_1|=2$, which attains more than $99.8\%$ of the corresponding asymptotic success probability while satisfying the stated fidelity constraint.}
\label{tab:NLI-multibranch-even}
\end{table*}

\begin{table*}[h]
\centering
\setlength{\tabcolsep}{5pt}

\begin{tabular}{clcccccc}
\multicolumn{8}{c}{\textbf{Odd cat, nonlinear interferometer}}\\[1mm]
\toprule
$F_{\min}$ & \multicolumn{1}{c}{$P_{\mathrm{succ}}$} & $|\tilde{\tau}_1|$ & $|\tau_2|$ & $T_1$ & $|\xi_1|$ & $|\xi_2|$ & $\mathcal{S}$ \\
\midrule
\multicolumn{8}{c}{ \textit{Unconstrained squeezing} } \\
\midrule
0.90 & 0.08017 & 0.91192 & 0.82679 & 0.91629 & 3.0182 & 1.1779 & $\{7,9,11\}$ \\[1mm]
0.91 & $0.07893^{\dagger}$ & $\to 1$ & $\to 1$ & $\to 1$ & $\to\infty$ & $\to\infty$ & $\{5,7\}$ \\
\quad & 0.07881 & 0.96403 & 0.89699 & 0.96677 & 3.2789 & 1.4566 & $\{5,7\}$ \\[1mm]
0.92 & $0.07893^{\dagger}$ & $\to 1$ & $\to 1$ & $\to 1$ & $\to\infty$ & $\to\infty$ & $\{5,7\}$ \\
\quad & 0.07881 & 0.96403 & 0.89699 & 0.96677 & 3.2789 & 1.4566 & $\{5,7\}$ \\[1mm]
0.93 & $0.07893^{\dagger}$ & $\to 1$ & $\to 1$ & $\to 1$ & $\to\infty$ & $\to\infty$ & $\{5,7\}$ \\
\quad & 0.07881 & 0.96403 & 0.89699 & 0.96677 & 3.2789 & 1.4566 & $\{5,7\}$ \\[1mm]
0.94 & $0.05931^{\dagger}$ & $\to 1$ & $\to 1$ & $\to 1$ & $\to\infty$ & $\to\infty$ & $\{7,9\}$ \\
\quad  & 0.05924 & 0.96403 & 0.92152 & 0.96610 & 3.4189 & 1.5990 & $\{7,9\}$ \\[1mm]
0.95 & $0.05931^{\dagger}$ & $\to 1$ & $\to 1$ & $\to 1$ & $\to\infty$ & $\to\infty$ & $\{7,9\}$ \\
\quad & 0.05924 & 0.96403 & 0.92152 & 0.96610 & 3.4189 & 1.5990 & $\{7,9\}$ \\[1mm]
0.96 & $0.05931^{\dagger}$ & $\to 1$ & $\to 1$ & $\to 1$ & $\to\infty$ & $\to\infty$ & $\{7,9\}$ \\
\quad & 0.05924 & 0.96403 & 0.92152 & 0.96610 & 3.4189 & 1.5990 & $\{7,9\}$ \\[1mm]
0.97 & $0.04755^{\dagger}$ & $\to 1$ & $\to 1$ & $\to 1$ & $\to\infty$ & $\to\infty$ & $\{9,11\}$ \\
\quad & 0.04751 & 0.96403 & 0.93656 & 0.96570 & 3.5262 & 1.7093 & $\{9,11\}$ \\[1mm]
0.98 & 0.03958 & 0.85309 & 0.77599 & 0.85815 & 2.9118 & 1.0352 & $\{11,13\}$ \\
0.99 & $0.03337^{\dagger}$ & $\to 1$ & $\to 1$ & $\to 1$ & $\to\infty$ & $\to\infty$ & $\{7\}$ \\
\quad & 0.03333 & 0.96403 & 0.91249 & 0.96634 & 3.3641 & 1.5422 & $\{7\}$ \\
\midrule
\multicolumn{8}{c}{ \textit{Squeezing constraint $|\xi_1|,|\xi_2|\leq1.5$} } \\
\midrule
0.90 & 0.03167 & 0.50405 & 0.00000 & 0.55688 & 1.4999 & 0.0000 & $\{7,9\}$ \\
0.91 & 0.03105 & 0.51143 & 0.00000 & 0.56503 & 1.4999 & 0.0000 & $\{7,9\}$ \\
0.92 & 0.02996 & 0.52384 & 0.00812 & 0.57874 & 1.4999 & 0.0081 & $\{7,9\}$ \\
0.93 & 0.02616 & 0.56437 & 0.08897 & 0.62352 & 1.4999 & 0.0892 & $\{7,9\}$ \\
0.94 & 0.02109 & 0.46483 & 0.00000 & 0.51355 & 1.4999 & 0.0000 & $\{9,11\}$ \\
0.95 & 0.01860 & 0.50321 & 0.07528 & 0.55595 & 1.4999 & 0.0754 & $\{9,11\}$ \\
0.96 & 0.01507 & 0.55397 & 0.17500 & 0.61203 & 1.4999 & 0.1768 & $\{9,11\}$ \\
0.97 & 0.01172 & 0.49142 & 0.13992 & 0.54292 & 1.5000 & 0.1408 & $\{11,13\}$ \\
0.98 & 0.00824 & 0.46318 & 0.15847 & 0.51172 & 1.5000 & 0.1598 & $\{13,15\}$ \\
0.99 & 0.00453 & 0.40766 & 0.16864 & 0.45038 & 1.5000 & 0.1703 & $\{17,19\}$ \\
\midrule
\multicolumn{8}{c}{ \textit{Squeezing constraint $|\xi_1|,|\xi_2|\leq1$} } \\
\midrule
0.90 & 0.00170 & 0.50405 & 0.00000 & 0.66184 & 1.0000 & 0.0000 & $\{7,9\}$ \\
0.91 & 0.00154 & 0.51143 & 0.00000 & 0.67153 & 1.0000 & 0.0000 & $\{7,9\}$ \\
0.92 & 0.00130 & 0.52384 & 0.00812 & 0.68783 & 1.0000 & 0.0081 & $\{7,9\}$ \\
0.93 & 0.00083 & 0.54100 & 0.01107 & 0.71036 & 1.0000 & 0.0111 & $\{7\}$ \\
0.94 & 0.00073 & 0.46483 & 0.00000 & 0.61034 & 1.0000 & 0.0000 & $\{9,11\}$ \\
0.95 & 0.00043 & 0.48620 & 0.01620 & 0.63840 & 1.0000 & 0.0162 & $\{9\}$ \\
0.96 & 0.00028 & 0.44303 & 0.04396 & 0.58172 & 1.0000 & 0.0440 & $\{11,13\}$ \\
0.97 & 0.00014 & 0.46951 & 0.07282 & 0.61649 & 1.0000 & 0.0729 & $\{11\}$ \\
0.98 & 0.00006 & 0.44021 & 0.09118 & 0.57802 & 1.0000 & 0.0914 & $\{13\}$ \\
0.99 & 0.00002 & 0.49863 & 0.20109 & 0.65472 & 1.0000 & 0.2039 & $\{13\}$ \\
\bottomrule
\end{tabular}

\caption{Optimized multibranch fidelity--success-probability trade-off for generating an odd cat state with $|\alpha|=3$ using the nonlinear-interferometer-type scheme. For each minimum fidelity $F_{\min}$, $P_{\mathrm{succ}}$ denotes the total probability of the accepted photon-number outcomes $\mathcal{S}$. A dagger denotes an asymptotic optimum approached in the infinite-squeezing limit. For each such entry, the immediately following row with an empty $F_{\min}$ field gives an approximative representative finite squeezing parameter realization using $|\tilde{\xi}_1|=2$, which closely approaches the corresponding asymptotic success probability while satisfying the stated fidelity constraint.}
\label{tab:NLI-multibranch-odd}
\end{table*}

\begin{table*}[h]
\centering
\setlength{\tabcolsep}{4pt}

\begin{tabular}{clccccccccc}
\multicolumn{11}{c}{\textbf{Even cat, alternating scheme, two squeezer, two subtractions}}\\[1mm]
\toprule
$F_{\min}$ & \multicolumn{1}{c}{$P_{\mathrm{succ}}$} & $|\tilde{\breve{\tau}}_2|$ & $|\tilde{\tau}_2|$ & $T_1$ & $T_2$ & $|\xi_1|$ & $|\xi_2|$ & $\varphi_1$ & $\varphi_2$ & $\mathcal{S}$ \\
\midrule
\multicolumn{11}{c}{\textit{Unconstrained squeezing}} \\
\midrule
0.90 & $0.09474^{\dagger}$ & 0.55499 & $\to1$ & $\to1$ & 1.00000 & $\to\infty$ & $\to\infty$ & $\pi$ & $0$ & $\{(4,0),(6,0)\}$ \\
\quad & 0.09455 & 0.56079 & 0.87778 & 0.96730 & 1.00000 & 3.1901 & 1.3660 & $\pi$ & $0$ & $\{(4,0),(6,0)\}$ \\[1mm]
0.91 & 0.07215 & 0.28426 & 0.94896 & 0.97260 & 1.00000 & 3.6288 & 1.82123 & $\pi$ & $0$ & $\{(8,0),(10,0),(12,0)\}$ \\[1mm]
0.92 & 0.07178 & 0.33863 & 0.58610 & 0.78024 & 1.00000 & 2.5949 & 0.67170 & $\pi$ & $0$ & $\{(8,0),(10,0),(12,0)\}$ \\[1mm]
0.93 & $0.06771^{\dagger}$ & 0.42867 & $\to1$ & $\to1$ & 1.00000 & $\to\infty$ & $\to\infty$ & $\pi$ & $0$ & $\{(6,0),(8,0)\}$ \\
\quad & 0.06762 & 0.43582 & 0.91094 & 0.96639 & 1.00000 & 3.3537 & 1.5330 & $\pi$ & $0$ & $\{(6,0),(8,0)\}$ \\[1mm]
0.94 & $0.06771^{\dagger}$ & 0.42867 & $\to1$ & $\to1$ & 1.00000 & $\to\infty$ & $\to\infty$ & $\pi$ & $0$ & $\{(6,0),(8,0)\}$ \\
\quad & 0.06762 & 0.43582 & 0.91094 & 0.96639 & 1.00000 & 3.3537 & 1.5330 & $\pi$ & $0$ & $\{(6,0),(8,0)\}$ \\[1mm]
0.95 & $0.06771^{\dagger}$ & 0.42867 & $\to1$ & $\to1$ & 1.00000 & $\to\infty$ & $\to\infty$ & $\pi$ & $0$ & $\{(6,0),(8,0)\}$ \\
\quad & 0.06762 & 0.43582 & 0.91094 & 0.96639 & 1.00000 & 3.3537 & 1.5330 & $\pi$ & $0$ & $\{(6,0),(8,0)\}$ \\[1mm]
0.96 & $0.05278^{\dagger}$ & 0.32194 & $\to1$ & $\to1$ & 1.00000 & $\to\infty$ & $\to\infty$ & $\pi$ & $0$ & $\{(8,0),(10,0)\}$ \\
\quad & 0.05272 & 0.32994 & 0.92985 & 0.96588 & 1.00000 & 3.4755 & 1.6573 & $\pi$ & $0$ & $\{(8,0),(10,0)\}$ \\[1mm]
0.97 & $0.05278^{\dagger}$ & 0.32194 & $\to1$ & $\to1$ & 1.00000 & $\to\infty$ & $\to\infty$ & $\pi$ & $0$ & $\{(8,0),(10,0)\}$ \\
\quad& 0.05272 & 0.32994 & 0.92985 & 0.96588 & 1.00000 & 3.4755 & 1.6573 & $\pi$ & $0$ & $\{(8,0),(10,0)\}$ \\[1mm]
0.98 & 0.04313 & 0.26598 & 0.77218 & 0.86652 & 1.00000 & 2.8982 & 1.02571 & $\pi$ & $0$ & $\{(10,0),(12,0)\}$ \\[1mm]
0.99 & 0.03163 & 0.21864 & 0.44264 & 0.61074 & 1.00000 & 2.5230 & 0.47551 & $\pi$ & $0$ & $\{(14,0),(16,0)\}$ \\
\midrule
\multicolumn{11}{c}{ \textit{Squeezing constraint $|\xi_1|,|\xi_2|\leq1.5$} } \\
\midrule
0.90 & 0.06663 & 0.55391 & 0.44578 & 1.00000 & 0.57096 & 1.0473 & 1.0473 & $\pi$ & $\pi$ & $\{(0,6),(0,8)\}$ \\
0.91 & 0.06383 & 0.37139 & 0.30909 & 1.00000 & 0.37917 & 1.1423 & 1.1423 & $\pi$ & $\pi$ & $\{(0,12),(0,14),(0,16),(0,18)\}$ \\
0.92 & 0.05934 & 0.42389 & 0.34915 & 1.00000 & 0.43395 & 1.1115 & 1.1115 & $\pi$ & $\pi$ & $\{(0,10),(0,12),(0,14)\}$ \\
0.93 & 0.05203 & 0.49160 & 0.39981 & 1.00000 & 0.50509 & 1.0756 & 1.0756 & $\pi$ & $\pi$ & $\{(0,8),(0,10)\}$ \\
0.94 & 0.04689 & 0.19928 & 0.17354 & 1.00000 & 0.20151 & 1.2979 & 1.2979 & $\pi$ & $\pi$ & $\{(0,30),(0,32),\ldots,(0,42)\}$ \\
0.95 & 0.04456 & 0.35583 & 0.29734 & 1.00000 & 0.36292 & 1.1547 & 1.1547 & $\pi$ & $\pi$ & $\{(0,14),(0,16),(0,18)\}$ \\
0.96 & 0.03827 & 0.26310 & 0.22498 & 1.00000 & 0.26696 & 1.2306 & 1.2306 & $\pi$ & $\pi$ & $\{(0,22),(0,24),(0,26),(0,28)\}$ \\
0.97 & 0.03334 & 0.19855 & 0.17296 & 1.00000 & 0.20076 & 1.2992 & 1.2992 & $\pi$ & $\pi$ & $\{(0,32),(0,34),(0,36),(0,38),(0,40)\}$ \\
0.98 & 0.02625 & 0.13997 & 0.12452 & 1.00000 & 0.14106 & 1.3881 & 1.3881 & $\pi$ & $\pi$ & $\{(0,50),(0,52),\ldots,(0,60)\}$ \\
0.99 & 0.01829 & 0.10187 & 0.09209 & 1.00000 & 0.10245 & 1.4661 & 1.4661 & $\pi$ & $\pi$ & $\{(0,74),(0,76),\ldots,(0,84)\}$ \\
\midrule
\multicolumn{11}{c}{ \textit{Squeezing constraint $|\xi_1|,|\xi_2|\leq1$} } \\
\midrule
0.90 & 0.06604 & 0.54939 & 0.43402 & 1.00000 & 0.56989 & 1.0000 & 1.0000 & $\pi$ & $\pi$ & $\{(0,6),(0,8)\}$ \\
0.91 & 0.05782 & 0.37080 & 0.29294 & 1.00000 & 0.38464 & 1.0000 & 1.0000 & $\pi$ & $\pi$ & $\{(0,12),(0,14),(0,16),(0,18)\}$ \\
0.92 & 0.05604 & 0.42075 & 0.33240 & 1.00000 & 0.43645 & 1.0000 & 1.0000 & $\pi$ & $\pi$ & $\{(0,10),(0,12),(0,14)\}$ \\
0.93 & 0.05072 & 0.49019 & 0.38726 & 1.00000 & 0.50848 & 1.0000 & 1.0000 & $\pi$ & $\pi$ & $\{(0,8),(0,10)\}$ \\
0.94 & 0.04086 & 0.43428 & 0.34308 & 1.00000 & 0.45048 & 1.0000 & 1.0000 & $\pi$ & $\pi$ & $\{(0,10),(0,12)\}$ \\
0.95 & 0.03955 & 0.35428 & 0.27989 & 1.00000 & 0.36750 & 1.0000 & 1.0000 & $\pi$ & $\pi$ & $\{(0,14),(0,16),(0,18)\}$ \\
0.96 & 0.02951 & 0.30242 & 0.23891 & 1.00000 & 0.31370 & 1.0000 & 1.0000 & $\pi$ & $\pi$ & $\{(0,18),(0,20),(0,22)\}$ \\
0.97 & 0.02572 & 0.28582 & 0.22581 & 1.00000 & 0.29649 & 1.0000 & 1.0000 & $\pi$ & $\pi$ & $\{(0,20),(0,22),(0,24)\}$ \\
0.98 & 0.01823 & 0.29452 & 0.23267 & 1.00000 & 0.30551 & 1.0000 & 1.0000 & $\pi$ & $\pi$ & $\{(0,20),(0,22)\}$ \\
0.99 & 0.00990 & 0.22220 & 0.17554 & 1.00000 & 0.23049 & 1.0000 & 1.0000 & $\pi$ & $\pi$ & $\{(0,30),(0,32)\}$ \\
\bottomrule
\end{tabular}

\caption{Optimized multibranch fidelity--success-probability trade-off for generating an even cat state with $|\alpha|=3$ using the alternating two-squeezer, two-subtraction scheme. The accepted heralding patterns are denoted by $\mathcal{S}=\{(N_1,N_2)\}$. The squeezing phases are specified relative to twice the phase of the cat-state parameter, $2\varphi_\alpha$, and the beam-splitter transmission phases set to zero. Thus, for a nonzero $\varphi_\alpha$, the listed squeezing phases are to be understood as $\varphi_i-2\varphi_\alpha$. A dagger denotes an asymptotic optimum approached in the infinite-squeezing limit. For each such entry, the immediately following row with an empty $F_{\min}$ field gives a representative finite-resource operating point approaching the corresponding asymptotic success probability. For the constrained optima, the global solutions occur for $N_1=0$ and $T_1=1$, such that the fidelity depends only on the combined effective squeezing. For $|\xi_i|\leq1.5$, this effective squeezing can be realized by a continuous family of decompositions between the two physical squeezers. The displayed equal distribution and aligned phases therefore represent one convenient choice. For the tighter constraint $|\xi_i|\leq1$, by contrast, the optimum requires the largest accessible effective squeezing. Consequently, both squeezers saturate the bound and are aligned.}
\label{tab:alternating_multibranch_even}
\end{table*}

\begin{table*}[h]
\centering
\setlength{\tabcolsep}{4pt}

\begin{tabular}{clccccccccc}
\multicolumn{11}{c}{\textbf{Odd cat, alternating scheme, two squeezer, two subtractions}}\\[1mm]
\toprule
$F_{\min}$ & \multicolumn{1}{c}{$P_{\mathrm{succ}}$} & $|\tilde{\breve{\tau}}_2|$ & $|\tilde{\tau}_2|$ & $T_1$ & $T_2$ & $|\xi_1|$ & $|\xi_2|$ & $\varphi_1$ & $\varphi_2$ & $\mathcal{S}$ \\
\midrule
\multicolumn{11}{c}{\textit{Unconstrained squeezing}} \\
\midrule
0.90 & 0.08017 & 0.34601 & 0.82679 & 0.91629 & 1.00000 & 3.0182 & 1.1779 & $\pi$ & $0$ & $\{(7,0),(9,0),(11,0)\}$ \\
0.91 & $0.07893^{\dagger}$ & 0.48903 & $\to1$ & $\to1$ & 1.00000 & $\to\infty$ & $\to\infty$ & $\pi$ & $0$ & $\{(5,0),(7,0)\}$ \\
\quad & 0.07881 & 0.49558 & 0.89699 & 0.96677 & 1.00000 & 3.2789 & 1.4566 & $\pi$ & $0$ & $\{(5,0),(7,0)\}$ \\
0.92 & $0.07893^{\dagger}$ & 0.48903 & $\to1$ & $\to1$ & 1.00000 & $\to\infty$ & $\to\infty$ & $\pi$ & $0$ & $\{(5,0),(7,0)\}$ \\
\quad & 0.07881 & 0.49558 & 0.89699 & 0.96677 & 1.00000 & 3.2789 & 1.4566 & $\pi$ & $0$ & $\{(5,0),(7,0)\}$ \\
0.93 & $0.07893^{\dagger}$ & 0.48903 & $\to1$ & $\to1$ & 1.00000 & $\to\infty$ & $\to\infty$ & $\pi$ & $0$ & $\{(5,0),(7,0)\}$ \\
\quad & 0.07881 & 0.49558 & 0.89699 & 0.96677 & 1.00000 & 3.2789 & 1.4566 & $\pi$ & $0$ & $\{(5,0),(7,0)\}$ \\
0.94 & $0.05931^{\dagger}$ & 0.37317 & $\to1$ & $\to1$ & 1.00000 & $\to\infty$ & $\to\infty$ & $\pi$ & $0$ & $\{(7,0),(9,0)\}$ \\
\quad & 0.05924 & 0.38082 & 0.92152 & 0.96610 & 1.00000 & 3.4189 & 1.5990 & $\pi$ & $0$ & $\{(7,0),(9,0)\}$ \\
0.95 & $0.05931^{\dagger}$ & 0.37317 & $\to1$ & $\to1$ & 1.00000 & $\to\infty$ & $\to\infty$ & $\pi$ & $0$ & $\{(7,0),(9,0)\}$ \\
\quad & 0.05924 & 0.38082 & 0.92152 & 0.96610 & 1.00000 & 3.4189 & 1.5990 & $\pi$ & $0$ & $\{(7,0),(9,0)\}$ \\
0.96 & $0.05931^{\dagger}$ & 0.37317 & $\to1$ & $\to1$ & 1.00000 & $\to\infty$ & $\to\infty$ & $\pi$ & $0$ & $\{(7,0),(9,0)\}$ \\
\quad & 0.05924 & 0.38082 & 0.92152 & 0.96610 & 1.00000 & 3.4189 & 1.5990 & $\pi$ & $0$ & $\{(7,0),(9,0)\}$ \\
0.97 & $0.04755^{\dagger}$ & 0.27448 & $\to1$ & $\to1$ & 1.00000 & $\to\infty$ & $\to\infty$ & $\pi$ & $0$ & $\{(9,0),(11,0)\}$ \\
\quad & 0.04751 & 0.28282 & 0.93656 & 0.96570 & 1.00000 & 3.5262 & 1.7093 & $\pi$ & $0$ & $\{(9,0),(11,0)\}$ \\
0.98 & 0.03958 & 0.22810 & 0.77599 & 0.85815 & 1.00000 & 2.9118 & 1.0352 & $\pi$ & $0$ & $\{(11,0),(13,0)\}$ \\
0.99 & $0.03337^{\dagger}$ & 0.42205 & $\to1$ & $\to1$ & 1.00000 & $\to\infty$ & $\to\infty$ & $\pi$ & $0$ & $\{(7,0)\}$ \\
\quad & 0.03333 & 0.42831 & 0.91249 & 0.96634 & 1.00000 & 3.3641 & 1.5422 & $\pi$ & $0$ & $\{(7,0)\}$ \\
\midrule
\multicolumn{11}{c}{ \textit{Squeezing constraint $|\xi_1|,|\xi_2|\leq1.5$} } \\
\midrule
0.90 & 0.06844 & 0.38741 & 0.32127 & 1.00000 & 0.39589 & 1.1314 & 1.1314 & $\pi$ & $\pi$ & $\{(0,11),(0,13),(0,15),(0,17)\}$ \\
0.91 & 0.06472 & 0.44458 & 0.36459 & 1.00000 & 0.45568 & 1.0989 & 1.0989 & $\pi$ & $\pi$ & $\{(0,9),(0,11),(0,13)\}$ \\
0.92 & 0.05722 & 0.29826 & 0.25260 & 1.00000 & 0.30326 & 1.1977 & 1.1977 & $\pi$ & $\pi$ & $\{(0,17),(0,19),(0,21),(0,23),(0,25)\}$ \\
0.93 & 0.05478 & 0.40434 & 0.33435 & 1.00000 & 0.41350 & 1.1229 & 1.1229 & $\pi$ & $\pi$ & $\{(0,11),(0,13),(0,15)\}$ \\
0.94 & 0.04801 & 0.26334 & 0.22512 & 1.00000 & 0.26722 & 1.2295 & 1.2295 & $\pi$ & $\pi$ & $\{(0,21),(0,23),(0,25),(0,27),(0,29)\}$ \\
0.95 & 0.04348 & 0.28856 & 0.24510 & 1.00000 & 0.29321 & 1.2075 & 1.2075 & $\pi$ & $\pi$ & $\{(0,19),(0,21),(0,23),(0,25)\}$ \\
0.96 & 0.03871 & 0.22351 & 0.19323 & 1.00000 & 0.22630 & 1.2703 & 1.2703 & $\pi$ & $\pi$ & $\{(0,27),(0,29),(0,31),(0,33),(0,35)\}$ \\
0.97 & 0.03280 & 0.16886 & 0.14859 & 1.00000 & 0.17045 & 1.3402 & 1.3402 & $\pi$ & $\pi$ & $\{(0,39),(0,41),\dots,(0,49)\}$ \\
0.98 & 0.02669 & 0.16561 & 0.14590 & 1.00000 & 0.16714 & 1.3453 & 1.3453 & $\pi$ & $\pi$ & $\{(0,41),(0,43),(0,45),(0,47),(0,49)\}$ \\
0.99 & 0.01851 & 0.12099 & 0.10846 & 1.00000 & 0.12181 & 1.4239 & 1.4239 & $\pi$ & $\pi$ & $\{(0,61),(0,63),(0,65),(0,67),(0,69)\}$ \\
\midrule
\multicolumn{11}{c}{ \textit{Squeezing constraint $|\xi_1|,|\xi_2|\leq1$} } \\
\midrule
0.90 & 0.06299 & 0.38722 & 0.30591 & 1.00000 & 0.40167 & 1.0000 & 1.0000 & $\pi$ & $\pi$ & $\{(0,11),(0,13),(0,15),(0,17)\}$ \\
0.91 & 0.06191 & 0.44237 & 0.34948 & 1.00000 & 0.45888 & 1.0000 & 1.0000 & $\pi$ & $\pi$ & $\{(0,9),(0,11),(0,13)\}$ \\
0.92 & 0.05115 & 0.39656 & 0.31329 & 1.00000 & 0.41136 & 1.0000 & 1.0000 & $\pi$ & $\pi$ & $\{(0,11),(0,13),(0,15)\}$ \\
0.93 & 0.05101 & 0.40203 & 0.31761 & 1.00000 & 0.41703 & 1.0000 & 1.0000 & $\pi$ & $\pi$ & $\{(0,11),(0,13),(0,15)\}$ \\
0.94 & 0.04525 & 0.46483 & 0.36722 & 1.00000 & 0.48218 & 1.0000 & 1.0000 & $\pi$ & $\pi$ & $\{(0,9),(0,11)\}$ \\
0.95 & 0.03699 & 0.41662 & 0.32914 & 1.00000 & 0.43217 & 1.0000 & 1.0000 & $\pi$ & $\pi$ & $\{(0,11),(0,13)\}$ \\
0.96 & 0.03161 & 0.31511 & 0.24894 & 1.00000 & 0.32687 & 1.0000 & 1.0000 & $\pi$ & $\pi$ & $\{(0,17),(0,19),(0,21)\}$ \\
0.97 & 0.02617 & 0.35369 & 0.27942 & 1.00000 & 0.36689 & 1.0000 & 1.0000 & $\pi$ & $\pi$ & $\{(0,15),(0,17)\}$ \\
0.98 & 0.01709 & 0.28343 & 0.22391 & 1.00000 & 0.29401 & 1.0000 & 1.0000 & $\pi$ & $\pi$ & $\{(0,21),(0,23)\}$ \\
0.99 & 0.00937 & 0.21608 & 0.17071 & 1.00000 & 0.22414 & 1.0000 & 1.0000 & $\pi$ & $\pi$ & $\{(0,31),(0,33)\}$ \\
\bottomrule
\end{tabular}

\caption{Optimized multibranch fidelity--success-probability trade-off for generating an odd cat state with $|\alpha|=3$ using the alternating two-squeezer, two-subtraction scheme. The accepted heralding patterns are denoted by $\mathcal{S}=\{(N_1,N_2)\}$. The squeezing phases are specified relative to twice the phase of the cat-state parameter, $2\varphi_\alpha$, and the beam-splitter transmission phases set to zero. Thus, for a nonzero $\varphi_\alpha$, the listed squeezing phases are to be understood as $\varphi_i-2\varphi_\alpha$. A dagger denotes an asymptotic optimum approached in the infinite-squeezing limit. For each such entry, the immediately following row with an empty $F_{\min}$ field gives a representative finite-resource operating point with $|\tilde{\xi}_1|=2$. For the constrained optima, the global solutions occur for $N_1=0$ and $T_1=1$, such that the fidelity depends only on the combined effective squeezing. For $|\xi_i|\leq1.5$, this effective squeezing can be realized by a continuous family of decompositions between the two physical squeezers. The displayed equal distribution and aligned phases therefore represent one convenient choice. For the tighter constraint $|\xi_i|\leq1$, by contrast, the optimum requires the largest accessible effective squeezing. Consequently, both squeezers saturate the bound and are aligned.}
\label{tab:alternating_multibranch_odd}
\end{table*}

\end{widetext}

\clearpage

\bibliography{bibliography}

\begin{thebibliography}{76}%
\makeatletter
\providecommand \@ifxundefined [1]{%
 \@ifx{#1\undefined}
}%
\providecommand \@ifnum [1]{%
 \ifnum #1\expandafter \@firstoftwo
 \else \expandafter \@secondoftwo
 \fi
}%
\providecommand \@ifx [1]{%
 \ifx #1\expandafter \@firstoftwo
 \else \expandafter \@secondoftwo
 \fi
}%
\providecommand \natexlab [1]{#1}%
\providecommand \enquote  [1]{``#1''}%
\providecommand \bibnamefont  [1]{#1}%
\providecommand \bibfnamefont [1]{#1}%
\providecommand \citenamefont [1]{#1}%
\providecommand \href@noop [0]{\@secondoftwo}%
\providecommand \href [0]{\begingroup \@sanitize@url \@href}%
\providecommand \@href[1]{\@@startlink{#1}\@@href}%
\providecommand \@@href[1]{\endgroup#1\@@endlink}%
\providecommand \@sanitize@url [0]{\catcode `\\12\catcode `\$12\catcode
  `\&12\catcode `\#12\catcode `\^12\catcode `\_12\catcode `\%12\relax}%
\providecommand \@@startlink[1]{}%
\providecommand \@@endlink[0]{}%
\providecommand \url  [0]{\begingroup\@sanitize@url \@url }%
\providecommand \@url [1]{\endgroup\@href {#1}{\urlprefix }}%
\providecommand \urlprefix  [0]{URL }%
\providecommand \Eprint [0]{\href }%
\providecommand \doibase [0]{http://dx.doi.org/}%
\providecommand \selectlanguage [0]{\@gobble}%
\providecommand \bibinfo  [0]{\@secondoftwo}%
\providecommand \bibfield  [0]{\@secondoftwo}%
\providecommand \translation [1]{[#1]}%
\providecommand \BibitemOpen [0]{}%
\providecommand \bibitemStop [0]{}%
\providecommand \bibitemNoStop [0]{.\EOS\space}%
\providecommand \EOS [0]{\spacefactor3000\relax}%
\providecommand \BibitemShut  [1]{\csname bibitem#1\endcsname}%
\let\auto@bib@innerbib\@empty
\bibitem [{\citenamefont {Dodonov}\ \emph {et~al.}(1974)\citenamefont
  {Dodonov}, \citenamefont {Malkin},\ and\ \citenamefont
  {Man'ko}}]{dodonov1974}%
  \BibitemOpen
  \bibfield  {author} {\bibinfo {author} {\bibfnamefont {V.}~\bibnamefont
  {Dodonov}}, \bibinfo {author} {\bibfnamefont {I.}~\bibnamefont {Malkin}}, \
  and\ \bibinfo {author} {\bibfnamefont {V.}~\bibnamefont {Man'ko}},\ }\href
  {\doibase 10.1016/0031-8914(74)90215-8} {\bibfield  {journal} {\bibinfo
  {journal} {Physica}\ }\textbf {\bibinfo {volume} {72}},\ \bibinfo {pages}
  {597} (\bibinfo {year} {1974})}\BibitemShut {NoStop}%
\bibitem [{\citenamefont {Glauber}(1963{\natexlab{a}})}]{Glauber1963}%
  \BibitemOpen
  \bibfield  {author} {\bibinfo {author} {\bibfnamefont {R.~J.}\ \bibnamefont
  {Glauber}},\ }\href {\doibase 10.1103/PhysRev.130.2529} {\bibfield  {journal}
  {\bibinfo  {journal} {Phys. Rev.}\ }\textbf {\bibinfo {volume} {130}},\
  \bibinfo {pages} {2529} (\bibinfo {year} {1963}{\natexlab{a}})}\BibitemShut
  {NoStop}%
\bibitem [{\citenamefont {Glauber}(1963{\natexlab{b}})}]{Glauber1963-2}%
  \BibitemOpen
  \bibfield  {author} {\bibinfo {author} {\bibfnamefont {R.~J.}\ \bibnamefont
  {Glauber}},\ }\href {\doibase 10.1103/PhysRev.131.2766} {\bibfield  {journal}
  {\bibinfo  {journal} {Phys. Rev.}\ }\textbf {\bibinfo {volume} {131}},\
  \bibinfo {pages} {2766} (\bibinfo {year} {1963}{\natexlab{b}})}\BibitemShut
  {NoStop}%
\bibitem [{\citenamefont {Sudarshan}(1963)}]{Sudarshan1963}%
  \BibitemOpen
  \bibfield  {author} {\bibinfo {author} {\bibfnamefont {E.~C.~G.}\
  \bibnamefont {Sudarshan}},\ }\href {\doibase 10.1103/PhysRevLett.10.277}
  {\bibfield  {journal} {\bibinfo  {journal} {Phys. Rev. Lett.}\ }\textbf
  {\bibinfo {volume} {10}},\ \bibinfo {pages} {277} (\bibinfo {year}
  {1963})}\BibitemShut {NoStop}%
\bibitem [{\citenamefont {Eisert}\ \emph {et~al.}(2002)\citenamefont {Eisert},
  \citenamefont {Scheel},\ and\ \citenamefont {Plenio}}]{Eisert2002}%
  \BibitemOpen
  \bibfield  {author} {\bibinfo {author} {\bibfnamefont {J.}~\bibnamefont
  {Eisert}}, \bibinfo {author} {\bibfnamefont {S.}~\bibnamefont {Scheel}}, \
  and\ \bibinfo {author} {\bibfnamefont {M.~B.}\ \bibnamefont {Plenio}},\
  }\href {\doibase 10.1103/PhysRevLett.89.137903} {\bibfield  {journal}
  {\bibinfo  {journal} {Physical Review Letters}\ }\textbf {\bibinfo {volume}
  {89}},\ \bibinfo {pages} {137903} (\bibinfo {year} {2002})}\BibitemShut
  {NoStop}%
\bibitem [{\citenamefont {Niset}\ \emph {et~al.}(2009)\citenamefont {Niset},
  \citenamefont {Fiur{\'a}{\v{s}}ek},\ and\ \citenamefont {Cerf}}]{Niset2009}%
  \BibitemOpen
  \bibfield  {author} {\bibinfo {author} {\bibfnamefont {J.}~\bibnamefont
  {Niset}}, \bibinfo {author} {\bibfnamefont {J.}~\bibnamefont
  {Fiur{\'a}{\v{s}}ek}}, \ and\ \bibinfo {author} {\bibfnamefont {N.~J.}\
  \bibnamefont {Cerf}},\ }\href {\doibase 10.1103/PhysRevLett.102.120501}
  {\bibfield  {journal} {\bibinfo  {journal} {Phys. Rev. Lett.}\ }\textbf
  {\bibinfo {volume} {102}},\ \bibinfo {pages} {120501} (\bibinfo {year}
  {2009})}\BibitemShut {NoStop}%
\bibitem [{\citenamefont {Bergmann}\ and\ \citenamefont
  {Van~Loock}(2016)}]{Bergmann2016}%
  \BibitemOpen
  \bibfield  {author} {\bibinfo {author} {\bibfnamefont {M.}~\bibnamefont
  {Bergmann}}\ and\ \bibinfo {author} {\bibfnamefont {P.}~\bibnamefont
  {Van~Loock}},\ }\href {\doibase 10.1103/PhysRevA.94.042332} {\bibfield
  {journal} {\bibinfo  {journal} {Physical Review A}\ }\textbf {\bibinfo
  {volume} {94}},\ \bibinfo {pages} {042332} (\bibinfo {year}
  {2016})}\BibitemShut {NoStop}%
\bibitem [{\citenamefont {Gomes}\ \emph {et~al.}(2009)\citenamefont {Gomes},
  \citenamefont {Salles}, \citenamefont {Toscano}, \citenamefont
  {Souto~Ribeiro},\ and\ \citenamefont {Walborn}}]{Gomes2009}%
  \BibitemOpen
  \bibfield  {author} {\bibinfo {author} {\bibfnamefont {R.~M.}\ \bibnamefont
  {Gomes}}, \bibinfo {author} {\bibfnamefont {A.}~\bibnamefont {Salles}},
  \bibinfo {author} {\bibfnamefont {F.}~\bibnamefont {Toscano}}, \bibinfo
  {author} {\bibfnamefont {P.~H.}\ \bibnamefont {Souto~Ribeiro}}, \ and\
  \bibinfo {author} {\bibfnamefont {S.~P.}\ \bibnamefont {Walborn}},\ }\href
  {\doibase 10.1073/pnas.0908329106} {\bibfield  {journal} {\bibinfo  {journal}
  {Proceedings of the National Academy of Sciences}\ }\textbf {\bibinfo
  {volume} {106}},\ \bibinfo {pages} {21517} (\bibinfo {year}
  {2009})}\BibitemShut {NoStop}%
\bibitem [{\citenamefont {Dell'Anno}\ \emph {et~al.}(2019)\citenamefont
  {Dell'Anno}, \citenamefont {Buono}, \citenamefont {Nocerino}, \citenamefont
  {De~Siena},\ and\ \citenamefont {Illuminati}}]{DellAnno2019}%
  \BibitemOpen
  \bibfield  {author} {\bibinfo {author} {\bibfnamefont {F.}~\bibnamefont
  {Dell'Anno}}, \bibinfo {author} {\bibfnamefont {D.}~\bibnamefont {Buono}},
  \bibinfo {author} {\bibfnamefont {G.}~\bibnamefont {Nocerino}}, \bibinfo
  {author} {\bibfnamefont {S.}~\bibnamefont {De~Siena}}, \ and\ \bibinfo
  {author} {\bibfnamefont {F.}~\bibnamefont {Illuminati}},\ }\href {\doibase
  10.1007/s11128-018-2133-1} {\bibfield  {journal} {\bibinfo  {journal}
  {Quantum Information Processing}\ }\textbf {\bibinfo {volume} {18}},\
  \bibinfo {pages} {20} (\bibinfo {year} {2019})}\BibitemShut {NoStop}%
\bibitem [{\citenamefont {Takahashi}\ \emph {et~al.}(2010)\citenamefont
  {Takahashi}, \citenamefont {{Neergaard-Nielsen}}, \citenamefont {Takeuchi},
  \citenamefont {Takeoka}, \citenamefont {Hayasaka}, \citenamefont {Furusawa},\
  and\ \citenamefont {Sasaki}}]{Takahashi2010}%
  \BibitemOpen
  \bibfield  {author} {\bibinfo {author} {\bibfnamefont {H.}~\bibnamefont
  {Takahashi}}, \bibinfo {author} {\bibfnamefont {J.~S.}\ \bibnamefont
  {{Neergaard-Nielsen}}}, \bibinfo {author} {\bibfnamefont {M.}~\bibnamefont
  {Takeuchi}}, \bibinfo {author} {\bibfnamefont {M.}~\bibnamefont {Takeoka}},
  \bibinfo {author} {\bibfnamefont {K.}~\bibnamefont {Hayasaka}}, \bibinfo
  {author} {\bibfnamefont {A.}~\bibnamefont {Furusawa}}, \ and\ \bibinfo
  {author} {\bibfnamefont {M.}~\bibnamefont {Sasaki}},\ }\href {\doibase
  10.1038/nphoton.2010.1} {\bibfield  {journal} {\bibinfo  {journal} {Nature
  Photonics}\ }\textbf {\bibinfo {volume} {4}},\ \bibinfo {pages} {178}
  (\bibinfo {year} {2010})}\BibitemShut {NoStop}%
\bibitem [{\citenamefont {Dell'Anno}\ \emph {et~al.}(2007)\citenamefont
  {Dell'Anno}, \citenamefont {De~Siena}, \citenamefont {Albano},\ and\
  \citenamefont {Illuminati}}]{DellAnno2007}%
  \BibitemOpen
  \bibfield  {author} {\bibinfo {author} {\bibfnamefont {F.}~\bibnamefont
  {Dell'Anno}}, \bibinfo {author} {\bibfnamefont {S.}~\bibnamefont {De~Siena}},
  \bibinfo {author} {\bibfnamefont {L.}~\bibnamefont {Albano}}, \ and\ \bibinfo
  {author} {\bibfnamefont {F.}~\bibnamefont {Illuminati}},\ }\href {\doibase
  10.1103/PhysRevA.76.022301} {\bibfield  {journal} {\bibinfo  {journal}
  {Physical Review A}\ }\textbf {\bibinfo {volume} {76}},\ \bibinfo {pages}
  {022301} (\bibinfo {year} {2007})}\BibitemShut {NoStop}%
\bibitem [{\citenamefont {Nha}\ \emph {et~al.}(2010)\citenamefont {Nha},
  \citenamefont {Milburn},\ and\ \citenamefont {Carmichael}}]{Nha2010}%
  \BibitemOpen
  \bibfield  {author} {\bibinfo {author} {\bibfnamefont {H.}~\bibnamefont
  {Nha}}, \bibinfo {author} {\bibfnamefont {G.~J.}\ \bibnamefont {Milburn}}, \
  and\ \bibinfo {author} {\bibfnamefont {H.~J.}\ \bibnamefont {Carmichael}},\
  }\href {\doibase 10.1088/1367-2630/12/10/103010} {\bibfield  {journal}
  {\bibinfo  {journal} {New Journal of Physics}\ }\textbf {\bibinfo {volume}
  {12}},\ \bibinfo {pages} {103010} (\bibinfo {year} {2010})}\BibitemShut
  {NoStop}%
\bibitem [{\citenamefont {He}\ and\ \citenamefont {Liu}(2024)}]{He2024}%
  \BibitemOpen
  \bibfield  {author} {\bibinfo {author} {\bibfnamefont {M.}~\bibnamefont
  {He}}\ and\ \bibinfo {author} {\bibfnamefont {S.}~\bibnamefont {Liu}},\
  }\href {\doibase 10.1103/PhysRevA.110.012425} {\bibfield  {journal} {\bibinfo
   {journal} {Physical Review A}\ }\textbf {\bibinfo {volume} {110}},\ \bibinfo
  {pages} {012425} (\bibinfo {year} {2024})}\BibitemShut {NoStop}%
\bibitem [{\citenamefont {Yurke}\ \emph {et~al.}(1986)\citenamefont {Yurke},
  \citenamefont {McCall},\ and\ \citenamefont {Klauder}}]{Yurke1986}%
  \BibitemOpen
  \bibfield  {author} {\bibinfo {author} {\bibfnamefont {B.}~\bibnamefont
  {Yurke}}, \bibinfo {author} {\bibfnamefont {S.~L.}\ \bibnamefont {McCall}}, \
  and\ \bibinfo {author} {\bibfnamefont {J.~R.}\ \bibnamefont {Klauder}},\
  }\href {\doibase 10.1103/PhysRevA.33.4033} {\bibfield  {journal} {\bibinfo
  {journal} {Physical Review A}\ }\textbf {\bibinfo {volume} {33}},\ \bibinfo
  {pages} {4033} (\bibinfo {year} {1986})}\BibitemShut {NoStop}%
\bibitem [{\citenamefont {Hudelist}\ \emph {et~al.}(2014)\citenamefont
  {Hudelist}, \citenamefont {Kong}, \citenamefont {Liu}, \citenamefont {Jing},
  \citenamefont {Ou},\ and\ \citenamefont {Zhang}}]{Hudelist2014}%
  \BibitemOpen
  \bibfield  {author} {\bibinfo {author} {\bibfnamefont {F.}~\bibnamefont
  {Hudelist}}, \bibinfo {author} {\bibfnamefont {J.}~\bibnamefont {Kong}},
  \bibinfo {author} {\bibfnamefont {C.}~\bibnamefont {Liu}}, \bibinfo {author}
  {\bibfnamefont {J.}~\bibnamefont {Jing}}, \bibinfo {author} {\bibfnamefont
  {Z.}~\bibnamefont {Ou}}, \ and\ \bibinfo {author} {\bibfnamefont
  {W.}~\bibnamefont {Zhang}},\ }\href {\doibase 10.1038/ncomms4049} {\bibfield
  {journal} {\bibinfo  {journal} {Nature Communications}\ }\textbf {\bibinfo
  {volume} {5}},\ \bibinfo {pages} {3049} (\bibinfo {year} {2014})}\BibitemShut
  {NoStop}%
\bibitem [{\citenamefont {Knott}\ \emph {et~al.}(2016)\citenamefont {Knott},
  \citenamefont {Proctor}, \citenamefont {Hayes}, \citenamefont {Cooling},\
  and\ \citenamefont {Dunningham}}]{Knott2016}%
  \BibitemOpen
  \bibfield  {author} {\bibinfo {author} {\bibfnamefont {P.~A.}\ \bibnamefont
  {Knott}}, \bibinfo {author} {\bibfnamefont {T.~J.}\ \bibnamefont {Proctor}},
  \bibinfo {author} {\bibfnamefont {A.~J.}\ \bibnamefont {Hayes}}, \bibinfo
  {author} {\bibfnamefont {J.~P.}\ \bibnamefont {Cooling}}, \ and\ \bibinfo
  {author} {\bibfnamefont {J.~A.}\ \bibnamefont {Dunningham}},\ }\href
  {\doibase 10.1103/PhysRevA.93.033859} {\bibfield  {journal} {\bibinfo
  {journal} {Physical Review A}\ }\textbf {\bibinfo {volume} {93}},\ \bibinfo
  {pages} {033859} (\bibinfo {year} {2016})}\BibitemShut {NoStop}%
\bibitem [{\citenamefont {Dakna}\ \emph {et~al.}(1997)\citenamefont {Dakna},
  \citenamefont {Anhut}, \citenamefont {Opatrn{\'y}}, \citenamefont
  {Kn{\"o}ll},\ and\ \citenamefont
  {Welsch}}]{daknaGeneratingSchrodingercatlikeStates1997}%
  \BibitemOpen
  \bibfield  {author} {\bibinfo {author} {\bibfnamefont {M.}~\bibnamefont
  {Dakna}}, \bibinfo {author} {\bibfnamefont {T.}~\bibnamefont {Anhut}},
  \bibinfo {author} {\bibfnamefont {T.}~\bibnamefont {Opatrn{\'y}}}, \bibinfo
  {author} {\bibfnamefont {L.}~\bibnamefont {Kn{\"o}ll}}, \ and\ \bibinfo
  {author} {\bibfnamefont {D.-G.}\ \bibnamefont {Welsch}},\ }\href {\doibase
  10.1103/PhysRevA.55.3184} {\bibfield  {journal} {\bibinfo  {journal}
  {Physical Review A}\ }\textbf {\bibinfo {volume} {55}},\ \bibinfo {pages}
  {3184} (\bibinfo {year} {1997})}\BibitemShut {NoStop}%
\bibitem [{\citenamefont {Ourjoumtsev}\ \emph {et~al.}(2006)\citenamefont
  {Ourjoumtsev}, \citenamefont {{Tualle-Brouri}}, \citenamefont {Laurat},\ and\
  \citenamefont {Grangier}}]{ourjoumtsevGeneratingOpticalSchrodinger2006}%
  \BibitemOpen
  \bibfield  {author} {\bibinfo {author} {\bibfnamefont {A.}~\bibnamefont
  {Ourjoumtsev}}, \bibinfo {author} {\bibfnamefont {R.}~\bibnamefont
  {{Tualle-Brouri}}}, \bibinfo {author} {\bibfnamefont {J.}~\bibnamefont
  {Laurat}}, \ and\ \bibinfo {author} {\bibfnamefont {P.}~\bibnamefont
  {Grangier}},\ }\href {\doibase 10.1126/science.1122858} {\bibfield  {journal}
  {\bibinfo  {journal} {Science}\ }\textbf {\bibinfo {volume} {312}},\ \bibinfo
  {pages} {83} (\bibinfo {year} {2006})}\BibitemShut {NoStop}%
\bibitem [{\citenamefont {Vahlbruch}\ \emph {et~al.}(2016)\citenamefont
  {Vahlbruch}, \citenamefont {Mehmet}, \citenamefont {Danzmann},\ and\
  \citenamefont {Schnabel}}]{vahlbruchDetection15DB2016}%
  \BibitemOpen
  \bibfield  {author} {\bibinfo {author} {\bibfnamefont {H.}~\bibnamefont
  {Vahlbruch}}, \bibinfo {author} {\bibfnamefont {M.}~\bibnamefont {Mehmet}},
  \bibinfo {author} {\bibfnamefont {K.}~\bibnamefont {Danzmann}}, \ and\
  \bibinfo {author} {\bibfnamefont {R.}~\bibnamefont {Schnabel}},\ }\href
  {\doibase 10.1103/PhysRevLett.117.110801} {\bibfield  {journal} {\bibinfo
  {journal} {Physical Review Letters}\ }\textbf {\bibinfo {volume} {117}},\
  \bibinfo {pages} {110801} (\bibinfo {year} {2016})}\BibitemShut {NoStop}%
\bibitem [{\citenamefont {Kalash}\ \emph {et~al.}(2025)\citenamefont {Kalash},
  \citenamefont {Sudharsanam}, \citenamefont {Passos}, \citenamefont {Parigi},\
  and\ \citenamefont {Chekhova}}]{Kalash2025}%
  \BibitemOpen
  \bibfield  {author} {\bibinfo {author} {\bibfnamefont {M.}~\bibnamefont
  {Kalash}}, \bibinfo {author} {\bibfnamefont {A.}~\bibnamefont {Sudharsanam}},
  \bibinfo {author} {\bibfnamefont {M.~H.~M.}\ \bibnamefont {Passos}}, \bibinfo
  {author} {\bibfnamefont {V.}~\bibnamefont {Parigi}}, \ and\ \bibinfo {author}
  {\bibfnamefont {M.}~\bibnamefont {Chekhova}},\ }\href {\doibase
  10.48550/arXiv.2503.07486} {\enquote {\bibinfo {title} {Real-{{Time
  Monitoring}} of {{Multimode Squeezing}}},}\ } (\bibinfo {year} {2025}),\
  \Eprint {http://arxiv.org/abs/2503.07486} {arXiv:2503.07486 [quant-ph]}
  \BibitemShut {NoStop}%
\bibitem [{\citenamefont {Endo}\ \emph {et~al.}(2025)\citenamefont {Endo},
  \citenamefont {Nomura}, \citenamefont {Sonoyama}, \citenamefont {Takahashi},
  \citenamefont {Takasu}, \citenamefont {Fukuda}, \citenamefont {Kashiwazaki},
  \citenamefont {Inoue}, \citenamefont {Umeki}, \citenamefont {Nehra},
  \citenamefont {Marek}, \citenamefont {Filip}, \citenamefont {Takase},
  \citenamefont {Asavanant},\ and\ \citenamefont {Furusawa}}]{Endo2025}%
  \BibitemOpen
  \bibfield  {author} {\bibinfo {author} {\bibfnamefont {M.}~\bibnamefont
  {Endo}}, \bibinfo {author} {\bibfnamefont {T.}~\bibnamefont {Nomura}},
  \bibinfo {author} {\bibfnamefont {T.}~\bibnamefont {Sonoyama}}, \bibinfo
  {author} {\bibfnamefont {K.}~\bibnamefont {Takahashi}}, \bibinfo {author}
  {\bibfnamefont {S.}~\bibnamefont {Takasu}}, \bibinfo {author} {\bibfnamefont
  {D.}~\bibnamefont {Fukuda}}, \bibinfo {author} {\bibfnamefont
  {T.}~\bibnamefont {Kashiwazaki}}, \bibinfo {author} {\bibfnamefont
  {A.}~\bibnamefont {Inoue}}, \bibinfo {author} {\bibfnamefont
  {T.}~\bibnamefont {Umeki}}, \bibinfo {author} {\bibfnamefont
  {R.}~\bibnamefont {Nehra}}, \bibinfo {author} {\bibfnamefont
  {P.}~\bibnamefont {Marek}}, \bibinfo {author} {\bibfnamefont
  {R.}~\bibnamefont {Filip}}, \bibinfo {author} {\bibfnamefont
  {K.}~\bibnamefont {Takase}}, \bibinfo {author} {\bibfnamefont
  {W.}~\bibnamefont {Asavanant}}, \ and\ \bibinfo {author} {\bibfnamefont
  {A.}~\bibnamefont {Furusawa}},\ }\href {\doibase 10.48550/ARXIV.2502.08952}
  {\enquote {\bibinfo {title} {High-{{Rate Four Photon Subtraction}} from
  {{Squeezed Vacuum}}: {{Preparing Cat State}} for {{Optical Quantum
  Computation}}},}\ } (\bibinfo {year} {2025})\BibitemShut {NoStop}%
\bibitem [{\citenamefont {Sauer}\ \emph {et~al.}(2023)\citenamefont {Sauer},
  \citenamefont {Kolarczik}, \citenamefont {Gomez}, \citenamefont {Conrad},\
  and\ \citenamefont {Steinlechner}}]{Sauer2023}%
  \BibitemOpen
  \bibfield  {author} {\bibinfo {author} {\bibfnamefont {G.}~\bibnamefont
  {Sauer}}, \bibinfo {author} {\bibfnamefont {M.}~\bibnamefont {Kolarczik}},
  \bibinfo {author} {\bibfnamefont {R.}~\bibnamefont {Gomez}}, \bibinfo
  {author} {\bibfnamefont {J.}~\bibnamefont {Conrad}}, \ and\ \bibinfo {author}
  {\bibfnamefont {F.}~\bibnamefont {Steinlechner}},\ }\href {\doibase
  10.48550/ARXIV.2310.12472} {\enquote {\bibinfo {title} {Resolving {{Photon
  Numbers Using Ultra-High-Resolution Timing}} of a {{Single Low-Jitter
  Superconducting Nanowire Detector}}},}\ } (\bibinfo {year}
  {2023})\BibitemShut {NoStop}%
\bibitem [{\citenamefont {Schapeler}\ \emph {et~al.}(2024)\citenamefont
  {Schapeler}, \citenamefont {Lamberty}, \citenamefont {Hummel}, \citenamefont
  {Schlue}, \citenamefont {Stefszky}, \citenamefont {Brecht}, \citenamefont
  {Silberhorn},\ and\ \citenamefont {Bartley}}]{Schapeler2024}%
  \BibitemOpen
  \bibfield  {author} {\bibinfo {author} {\bibfnamefont {T.}~\bibnamefont
  {Schapeler}}, \bibinfo {author} {\bibfnamefont {N.}~\bibnamefont {Lamberty}},
  \bibinfo {author} {\bibfnamefont {T.}~\bibnamefont {Hummel}}, \bibinfo
  {author} {\bibfnamefont {F.}~\bibnamefont {Schlue}}, \bibinfo {author}
  {\bibfnamefont {M.}~\bibnamefont {Stefszky}}, \bibinfo {author}
  {\bibfnamefont {B.}~\bibnamefont {Brecht}}, \bibinfo {author} {\bibfnamefont
  {C.}~\bibnamefont {Silberhorn}}, \ and\ \bibinfo {author} {\bibfnamefont
  {T.~J.}\ \bibnamefont {Bartley}},\ }\href {\doibase
  10.1103/PhysRevApplied.22.014024} {\bibfield  {journal} {\bibinfo  {journal}
  {Physical Review Applied}\ }\textbf {\bibinfo {volume} {22}},\ \bibinfo
  {pages} {014024} (\bibinfo {year} {2024})}\BibitemShut {NoStop}%
\bibitem [{\citenamefont {Stasi}\ \emph {et~al.}(2023)\citenamefont {Stasi},
  \citenamefont {Caspar}, \citenamefont {Brydges}, \citenamefont {Zbinden},
  \citenamefont {Bussi{\`e}res},\ and\ \citenamefont {Thew}}]{Stasi2023}%
  \BibitemOpen
  \bibfield  {author} {\bibinfo {author} {\bibfnamefont {L.}~\bibnamefont
  {Stasi}}, \bibinfo {author} {\bibfnamefont {P.}~\bibnamefont {Caspar}},
  \bibinfo {author} {\bibfnamefont {T.}~\bibnamefont {Brydges}}, \bibinfo
  {author} {\bibfnamefont {H.}~\bibnamefont {Zbinden}}, \bibinfo {author}
  {\bibfnamefont {F.}~\bibnamefont {Bussi{\`e}res}}, \ and\ \bibinfo {author}
  {\bibfnamefont {R.}~\bibnamefont {Thew}},\ }\href {\doibase
  10.1088/2058-9565/ace54b} {\bibfield  {journal} {\bibinfo  {journal} {Quantum
  Science and Technology}\ }\textbf {\bibinfo {volume} {8}},\ \bibinfo {pages}
  {045006} (\bibinfo {year} {2023})}\BibitemShut {NoStop}%
\bibitem [{\citenamefont {Hao}\ \emph {et~al.}(2024)\citenamefont {Hao},
  \citenamefont {Zhao}, \citenamefont {Huang}, \citenamefont {Deng},
  \citenamefont {Yang}, \citenamefont {Ru}, \citenamefont {Liu}, \citenamefont
  {Wan}, \citenamefont {Liu}, \citenamefont {Li}, \citenamefont {Wang},
  \citenamefont {Tu}, \citenamefont {Zhang}, \citenamefont {Jia}, \citenamefont
  {Wu}, \citenamefont {Chen}, \citenamefont {Kang},\ and\ \citenamefont
  {Wu}}]{Hao2024}%
  \BibitemOpen
  \bibfield  {author} {\bibinfo {author} {\bibfnamefont {H.}~\bibnamefont
  {Hao}}, \bibinfo {author} {\bibfnamefont {Q.-Y.}\ \bibnamefont {Zhao}},
  \bibinfo {author} {\bibfnamefont {Y.-H.}\ \bibnamefont {Huang}}, \bibinfo
  {author} {\bibfnamefont {J.}~\bibnamefont {Deng}}, \bibinfo {author}
  {\bibfnamefont {F.}~\bibnamefont {Yang}}, \bibinfo {author} {\bibfnamefont
  {S.-Y.}\ \bibnamefont {Ru}}, \bibinfo {author} {\bibfnamefont
  {Z.}~\bibnamefont {Liu}}, \bibinfo {author} {\bibfnamefont {C.}~\bibnamefont
  {Wan}}, \bibinfo {author} {\bibfnamefont {H.}~\bibnamefont {Liu}}, \bibinfo
  {author} {\bibfnamefont {Z.-J.}\ \bibnamefont {Li}}, \bibinfo {author}
  {\bibfnamefont {H.-B.}\ \bibnamefont {Wang}}, \bibinfo {author}
  {\bibfnamefont {X.-C.}\ \bibnamefont {Tu}}, \bibinfo {author} {\bibfnamefont
  {L.-B.}\ \bibnamefont {Zhang}}, \bibinfo {author} {\bibfnamefont {X.-Q.}\
  \bibnamefont {Jia}}, \bibinfo {author} {\bibfnamefont {X.-L.}\ \bibnamefont
  {Wu}}, \bibinfo {author} {\bibfnamefont {J.}~\bibnamefont {Chen}}, \bibinfo
  {author} {\bibfnamefont {L.}~\bibnamefont {Kang}}, \ and\ \bibinfo {author}
  {\bibfnamefont {P.-H.}\ \bibnamefont {Wu}},\ }\href {\doibase
  10.1038/s41377-023-01374-1} {\bibfield  {journal} {\bibinfo  {journal}
  {Light: Science \& Applications}\ }\textbf {\bibinfo {volume} {13}},\
  \bibinfo {pages} {25} (\bibinfo {year} {2024})}\BibitemShut {NoStop}%
\bibitem [{\citenamefont {Schmidt}\ \emph {et~al.}(2018)\citenamefont
  {Schmidt}, \citenamefont {Von~Helversen}, \citenamefont {L{\'o}pez},
  \citenamefont {Gericke}, \citenamefont {Schlottmann}, \citenamefont
  {Heindel}, \citenamefont {K{\"u}ck}, \citenamefont {Reitzenstein},\ and\
  \citenamefont {Beyer}}]{Schmidt2018}%
  \BibitemOpen
  \bibfield  {author} {\bibinfo {author} {\bibfnamefont {M.}~\bibnamefont
  {Schmidt}}, \bibinfo {author} {\bibfnamefont {M.}~\bibnamefont
  {Von~Helversen}}, \bibinfo {author} {\bibfnamefont {M.}~\bibnamefont
  {L{\'o}pez}}, \bibinfo {author} {\bibfnamefont {F.}~\bibnamefont {Gericke}},
  \bibinfo {author} {\bibfnamefont {E.}~\bibnamefont {Schlottmann}}, \bibinfo
  {author} {\bibfnamefont {T.}~\bibnamefont {Heindel}}, \bibinfo {author}
  {\bibfnamefont {S.}~\bibnamefont {K{\"u}ck}}, \bibinfo {author}
  {\bibfnamefont {S.}~\bibnamefont {Reitzenstein}}, \ and\ \bibinfo {author}
  {\bibfnamefont {J.}~\bibnamefont {Beyer}},\ }\href {\doibase
  10.1007/s10909-018-1932-1} {\bibfield  {journal} {\bibinfo  {journal}
  {Journal of Low Temperature Physics}\ }\textbf {\bibinfo {volume} {193}},\
  \bibinfo {pages} {1243} (\bibinfo {year} {2018})}\BibitemShut {NoStop}%
\bibitem [{\citenamefont {Li}\ \emph {et~al.}(2024)\citenamefont {Li},
  \citenamefont {Kendall}, \citenamefont {Machado}, \citenamefont {Zhu},
  \citenamefont {Mer}, \citenamefont {Zhan}, \citenamefont {Zhang},
  \citenamefont {Yu}, \citenamefont {Walmsley},\ and\ \citenamefont
  {Patel}}]{Li2024}%
  \BibitemOpen
  \bibfield  {author} {\bibinfo {author} {\bibfnamefont {Z.}~\bibnamefont
  {Li}}, \bibinfo {author} {\bibfnamefont {M.~J.~H.}\ \bibnamefont {Kendall}},
  \bibinfo {author} {\bibfnamefont {G.~J.}\ \bibnamefont {Machado}}, \bibinfo
  {author} {\bibfnamefont {R.}~\bibnamefont {Zhu}}, \bibinfo {author}
  {\bibfnamefont {E.}~\bibnamefont {Mer}}, \bibinfo {author} {\bibfnamefont
  {H.}~\bibnamefont {Zhan}}, \bibinfo {author} {\bibfnamefont {A.}~\bibnamefont
  {Zhang}}, \bibinfo {author} {\bibfnamefont {S.}~\bibnamefont {Yu}}, \bibinfo
  {author} {\bibfnamefont {I.~A.}\ \bibnamefont {Walmsley}}, \ and\ \bibinfo
  {author} {\bibfnamefont {R.~B.}\ \bibnamefont {Patel}},\ }\href {\doibase
  10.48550/ARXIV.2411.15360} {\enquote {\bibinfo {title} {Boosting
  {{Photon-Number-Resolved Detection Rates}} of {{Transition-Edge Sensors}} by
  {{Machine Learning}}},}\ } (\bibinfo {year} {2024})\BibitemShut {NoStop}%
\bibitem [{\citenamefont {De~Lucia}\ \emph {et~al.}(2024)\citenamefont
  {De~Lucia}, \citenamefont {Dal~Bo}, \citenamefont {Di~Giorgi}, \citenamefont
  {Lari}, \citenamefont {Puglia},\ and\ \citenamefont
  {Paolucci}}]{DeLucia2024}%
  \BibitemOpen
  \bibfield  {author} {\bibinfo {author} {\bibfnamefont {M.}~\bibnamefont
  {De~Lucia}}, \bibinfo {author} {\bibfnamefont {P.}~\bibnamefont {Dal~Bo}},
  \bibinfo {author} {\bibfnamefont {E.}~\bibnamefont {Di~Giorgi}}, \bibinfo
  {author} {\bibfnamefont {T.}~\bibnamefont {Lari}}, \bibinfo {author}
  {\bibfnamefont {C.}~\bibnamefont {Puglia}}, \ and\ \bibinfo {author}
  {\bibfnamefont {F.}~\bibnamefont {Paolucci}},\ }\href {\doibase
  10.3390/instruments8040047} {\bibfield  {journal} {\bibinfo  {journal}
  {Instruments}\ }\textbf {\bibinfo {volume} {8}},\ \bibinfo {pages} {47}
  (\bibinfo {year} {2024})}\BibitemShut {NoStop}%
\bibitem [{\citenamefont {{Neergaard-Nielsen}}\ \emph
  {et~al.}(2006)\citenamefont {{Neergaard-Nielsen}}, \citenamefont {Nielsen},
  \citenamefont {Hettich}, \citenamefont {M{\o}lmer},\ and\ \citenamefont
  {Polzik}}]{Neergaard-Nielsen2006}%
  \BibitemOpen
  \bibfield  {author} {\bibinfo {author} {\bibfnamefont {J.~S.}\ \bibnamefont
  {{Neergaard-Nielsen}}}, \bibinfo {author} {\bibfnamefont {B.~M.}\
  \bibnamefont {Nielsen}}, \bibinfo {author} {\bibfnamefont {C.}~\bibnamefont
  {Hettich}}, \bibinfo {author} {\bibfnamefont {K.}~\bibnamefont {M{\o}lmer}},
  \ and\ \bibinfo {author} {\bibfnamefont {E.~S.}\ \bibnamefont {Polzik}},\
  }\href {\doibase 10.1103/PhysRevLett.97.083604} {\bibfield  {journal}
  {\bibinfo  {journal} {Physical Review Letters}\ }\textbf {\bibinfo {volume}
  {97}},\ \bibinfo {pages} {083604} (\bibinfo {year} {2006})}\BibitemShut
  {NoStop}%
\bibitem [{\citenamefont {Gerrits}\ \emph {et~al.}(2010)\citenamefont
  {Gerrits}, \citenamefont {Glancy}, \citenamefont {Clement}, \citenamefont
  {Calkins}, \citenamefont {Lita}, \citenamefont {Miller}, \citenamefont
  {Migdall}, \citenamefont {Nam}, \citenamefont {Mirin},\ and\ \citenamefont
  {Knill}}]{Gerrits2010}%
  \BibitemOpen
  \bibfield  {author} {\bibinfo {author} {\bibfnamefont {T.}~\bibnamefont
  {Gerrits}}, \bibinfo {author} {\bibfnamefont {S.}~\bibnamefont {Glancy}},
  \bibinfo {author} {\bibfnamefont {T.~S.}\ \bibnamefont {Clement}}, \bibinfo
  {author} {\bibfnamefont {B.}~\bibnamefont {Calkins}}, \bibinfo {author}
  {\bibfnamefont {A.~E.}\ \bibnamefont {Lita}}, \bibinfo {author}
  {\bibfnamefont {A.~J.}\ \bibnamefont {Miller}}, \bibinfo {author}
  {\bibfnamefont {A.~L.}\ \bibnamefont {Migdall}}, \bibinfo {author}
  {\bibfnamefont {S.~W.}\ \bibnamefont {Nam}}, \bibinfo {author} {\bibfnamefont
  {R.~P.}\ \bibnamefont {Mirin}}, \ and\ \bibinfo {author} {\bibfnamefont
  {E.}~\bibnamefont {Knill}},\ }\href {\doibase 10.1103/PhysRevA.82.031802}
  {\bibfield  {journal} {\bibinfo  {journal} {Physical Review A}\ }\textbf
  {\bibinfo {volume} {82}},\ \bibinfo {pages} {031802} (\bibinfo {year}
  {2010})}\BibitemShut {NoStop}%
\bibitem [{\citenamefont {Takahashi}\ \emph {et~al.}(2008)\citenamefont
  {Takahashi}, \citenamefont {Wakui}, \citenamefont {Suzuki}, \citenamefont
  {Takeoka}, \citenamefont {Hayasaka}, \citenamefont {Furusawa},\ and\
  \citenamefont {Sasaki}}]{Takahashi2008}%
  \BibitemOpen
  \bibfield  {author} {\bibinfo {author} {\bibfnamefont {H.}~\bibnamefont
  {Takahashi}}, \bibinfo {author} {\bibfnamefont {K.}~\bibnamefont {Wakui}},
  \bibinfo {author} {\bibfnamefont {S.}~\bibnamefont {Suzuki}}, \bibinfo
  {author} {\bibfnamefont {M.}~\bibnamefont {Takeoka}}, \bibinfo {author}
  {\bibfnamefont {K.}~\bibnamefont {Hayasaka}}, \bibinfo {author}
  {\bibfnamefont {A.}~\bibnamefont {Furusawa}}, \ and\ \bibinfo {author}
  {\bibfnamefont {M.}~\bibnamefont {Sasaki}},\ }\href {\doibase
  10.1103/PhysRevLett.101.233605} {\bibfield  {journal} {\bibinfo  {journal}
  {Physical Review Letters}\ }\textbf {\bibinfo {volume} {101}},\ \bibinfo
  {pages} {233605} (\bibinfo {year} {2008})}\BibitemShut {NoStop}%
\bibitem [{\citenamefont {Kuts}\ \emph {et~al.}(2022)\citenamefont {Kuts},
  \citenamefont {Podoshvedov}, \citenamefont {Nguyen},\ and\ \citenamefont
  {Podoshvedov}}]{Kuts2022}%
  \BibitemOpen
  \bibfield  {author} {\bibinfo {author} {\bibfnamefont {D.~A.}\ \bibnamefont
  {Kuts}}, \bibinfo {author} {\bibfnamefont {M.~S.}\ \bibnamefont
  {Podoshvedov}}, \bibinfo {author} {\bibfnamefont {B.~A.}\ \bibnamefont
  {Nguyen}}, \ and\ \bibinfo {author} {\bibfnamefont {S.~A.}\ \bibnamefont
  {Podoshvedov}},\ }\href {\doibase 10.1088/1402-4896/ac955f} {\bibfield
  {journal} {\bibinfo  {journal} {Physica Scripta}\ }\textbf {\bibinfo {volume}
  {97}},\ \bibinfo {pages} {115002} (\bibinfo {year} {2022})}\BibitemShut
  {NoStop}%
\bibitem [{\citenamefont {Podoshvedov}\ \emph {et~al.}(2023)\citenamefont
  {Podoshvedov}, \citenamefont {Podoshvedov},\ and\ \citenamefont
  {Kulik}}]{Podoshvedov2023}%
  \BibitemOpen
  \bibfield  {author} {\bibinfo {author} {\bibfnamefont {M.~S.}\ \bibnamefont
  {Podoshvedov}}, \bibinfo {author} {\bibfnamefont {S.~A.}\ \bibnamefont
  {Podoshvedov}}, \ and\ \bibinfo {author} {\bibfnamefont {S.~P.}\ \bibnamefont
  {Kulik}},\ }\href {\doibase 10.1038/s41598-023-30218-6} {\bibfield  {journal}
  {\bibinfo  {journal} {Scientific Reports}\ }\textbf {\bibinfo {volume}
  {13}},\ \bibinfo {pages} {3965} (\bibinfo {year} {2023})}\BibitemShut
  {NoStop}%
\bibitem [{\citenamefont {Lee}\ \emph {et~al.}(2012)\citenamefont {Lee},
  \citenamefont {Lee}, \citenamefont {Nha},\ and\ \citenamefont
  {Jeong}}]{Lee2012}%
  \BibitemOpen
  \bibfield  {author} {\bibinfo {author} {\bibfnamefont {C.-W.}\ \bibnamefont
  {Lee}}, \bibinfo {author} {\bibfnamefont {J.}~\bibnamefont {Lee}}, \bibinfo
  {author} {\bibfnamefont {H.}~\bibnamefont {Nha}}, \ and\ \bibinfo {author}
  {\bibfnamefont {H.}~\bibnamefont {Jeong}},\ }\href {\doibase
  10.1103/PhysRevA.85.063815} {\bibfield  {journal} {\bibinfo  {journal}
  {Physical Review A}\ }\textbf {\bibinfo {volume} {85}},\ \bibinfo {pages}
  {063815} (\bibinfo {year} {2012})}\BibitemShut {NoStop}%
\bibitem [{\citenamefont {Ourjoumtsev}\ \emph {et~al.}(2007)\citenamefont
  {Ourjoumtsev}, \citenamefont {Jeong}, \citenamefont {{Tualle-Brouri}},\ and\
  \citenamefont {Grangier}}]{Ourjoumtsev2007}%
  \BibitemOpen
  \bibfield  {author} {\bibinfo {author} {\bibfnamefont {A.}~\bibnamefont
  {Ourjoumtsev}}, \bibinfo {author} {\bibfnamefont {H.}~\bibnamefont {Jeong}},
  \bibinfo {author} {\bibfnamefont {R.}~\bibnamefont {{Tualle-Brouri}}}, \ and\
  \bibinfo {author} {\bibfnamefont {P.}~\bibnamefont {Grangier}},\ }\href
  {\doibase 10.1038/nature06054} {\bibfield  {journal} {\bibinfo  {journal}
  {Nature}\ }\textbf {\bibinfo {volume} {448}},\ \bibinfo {pages} {784}
  (\bibinfo {year} {2007})}\BibitemShut {NoStop}%
\bibitem [{\citenamefont {Ulanov}\ \emph {et~al.}(2016)\citenamefont {Ulanov},
  \citenamefont {Fedorov}, \citenamefont {Sychev}, \citenamefont {Grangier},\
  and\ \citenamefont {Lvovsky}}]{ulanovLosstolerantStateEngineering2016}%
  \BibitemOpen
  \bibfield  {author} {\bibinfo {author} {\bibfnamefont {A.~E.}\ \bibnamefont
  {Ulanov}}, \bibinfo {author} {\bibfnamefont {I.~A.}\ \bibnamefont {Fedorov}},
  \bibinfo {author} {\bibfnamefont {D.}~\bibnamefont {Sychev}}, \bibinfo
  {author} {\bibfnamefont {P.}~\bibnamefont {Grangier}}, \ and\ \bibinfo
  {author} {\bibfnamefont {A.~I.}\ \bibnamefont {Lvovsky}},\ }\href {\doibase
  10.1038/ncomms11925} {\bibfield  {journal} {\bibinfo  {journal} {Nature
  Communications}\ }\textbf {\bibinfo {volume} {7}},\ \bibinfo {pages} {11925}
  (\bibinfo {year} {2016})}\BibitemShut {NoStop}%
\bibitem [{\citenamefont {Huang}\ \emph {et~al.}(2015)\citenamefont {Huang},
  \citenamefont {Le~Jeannic}, \citenamefont {Ruaudel}, \citenamefont {Verma},
  \citenamefont {Shaw}, \citenamefont {Marsili}, \citenamefont {Nam},
  \citenamefont {Wu}, \citenamefont {Zeng}, \citenamefont {Jeong},
  \citenamefont {Filip}, \citenamefont {Morin},\ and\ \citenamefont
  {Laurat}}]{Huang2015}%
  \BibitemOpen
  \bibfield  {author} {\bibinfo {author} {\bibfnamefont {K.}~\bibnamefont
  {Huang}}, \bibinfo {author} {\bibfnamefont {H.}~\bibnamefont {Le~Jeannic}},
  \bibinfo {author} {\bibfnamefont {J.}~\bibnamefont {Ruaudel}}, \bibinfo
  {author} {\bibfnamefont {V.~B.}\ \bibnamefont {Verma}}, \bibinfo {author}
  {\bibfnamefont {M.~D.}\ \bibnamefont {Shaw}}, \bibinfo {author}
  {\bibfnamefont {F.}~\bibnamefont {Marsili}}, \bibinfo {author} {\bibfnamefont
  {S.~W.}\ \bibnamefont {Nam}}, \bibinfo {author} {\bibfnamefont
  {E.}~\bibnamefont {Wu}}, \bibinfo {author} {\bibfnamefont {H.}~\bibnamefont
  {Zeng}}, \bibinfo {author} {\bibfnamefont {Y.-C.}\ \bibnamefont {Jeong}},
  \bibinfo {author} {\bibfnamefont {R.}~\bibnamefont {Filip}}, \bibinfo
  {author} {\bibfnamefont {O.}~\bibnamefont {Morin}}, \ and\ \bibinfo {author}
  {\bibfnamefont {J.}~\bibnamefont {Laurat}},\ }\href {\doibase
  10.1103/PhysRevLett.115.023602} {\bibfield  {journal} {\bibinfo  {journal}
  {Physical Review Letters}\ }\textbf {\bibinfo {volume} {115}},\ \bibinfo
  {pages} {023602} (\bibinfo {year} {2015})}\BibitemShut {NoStop}%
\bibitem [{\citenamefont {Takase}\ \emph {et~al.}(2021)\citenamefont {Takase},
  \citenamefont {Yoshikawa}, \citenamefont {Asavanant}, \citenamefont {Endo},\
  and\ \citenamefont {Furusawa}}]{Takase2021}%
  \BibitemOpen
  \bibfield  {author} {\bibinfo {author} {\bibfnamefont {K.}~\bibnamefont
  {Takase}}, \bibinfo {author} {\bibfnamefont {J.-i.}\ \bibnamefont
  {Yoshikawa}}, \bibinfo {author} {\bibfnamefont {W.}~\bibnamefont
  {Asavanant}}, \bibinfo {author} {\bibfnamefont {M.}~\bibnamefont {Endo}}, \
  and\ \bibinfo {author} {\bibfnamefont {A.}~\bibnamefont {Furusawa}},\ }\href
  {\doibase 10.1103/PhysRevA.103.013710} {\bibfield  {journal} {\bibinfo
  {journal} {Physical Review A}\ }\textbf {\bibinfo {volume} {103}},\ \bibinfo
  {pages} {013710} (\bibinfo {year} {2021})}\BibitemShut {NoStop}%
\bibitem [{\citenamefont {Mikheev}\ \emph {et~al.}(2019)\citenamefont
  {Mikheev}, \citenamefont {Pugin}, \citenamefont {Kuts}, \citenamefont
  {Podoshvedov},\ and\ \citenamefont {An}}]{Mikheev2019}%
  \BibitemOpen
  \bibfield  {author} {\bibinfo {author} {\bibfnamefont {E.~V.}\ \bibnamefont
  {Mikheev}}, \bibinfo {author} {\bibfnamefont {A.~S.}\ \bibnamefont {Pugin}},
  \bibinfo {author} {\bibfnamefont {D.~A.}\ \bibnamefont {Kuts}}, \bibinfo
  {author} {\bibfnamefont {S.~A.}\ \bibnamefont {Podoshvedov}}, \ and\ \bibinfo
  {author} {\bibfnamefont {N.~B.}\ \bibnamefont {An}},\ }\href {\doibase
  10.1038/s41598-019-50703-1} {\bibfield  {journal} {\bibinfo  {journal}
  {Scientific Reports}\ }\textbf {\bibinfo {volume} {9}},\ \bibinfo {pages}
  {14301} (\bibinfo {year} {2019})}\BibitemShut {NoStop}%
\bibitem [{\citenamefont {Chen}\ \emph {et~al.}(2024)\citenamefont {Chen},
  \citenamefont {Hsieh}, \citenamefont {Ning}, \citenamefont {Wu},
  \citenamefont {Chen}, \citenamefont {Shi}, \citenamefont {Yang},
  \citenamefont {Steuernagel}, \citenamefont {Wu},\ and\ \citenamefont
  {Lee}}]{Chen2024}%
  \BibitemOpen
  \bibfield  {author} {\bibinfo {author} {\bibfnamefont {Y.-R.}\ \bibnamefont
  {Chen}}, \bibinfo {author} {\bibfnamefont {H.-Y.}\ \bibnamefont {Hsieh}},
  \bibinfo {author} {\bibfnamefont {J.}~\bibnamefont {Ning}}, \bibinfo {author}
  {\bibfnamefont {H.-C.}\ \bibnamefont {Wu}}, \bibinfo {author} {\bibfnamefont
  {H.~L.}\ \bibnamefont {Chen}}, \bibinfo {author} {\bibfnamefont {Z.-H.}\
  \bibnamefont {Shi}}, \bibinfo {author} {\bibfnamefont {P.}~\bibnamefont
  {Yang}}, \bibinfo {author} {\bibfnamefont {O.}~\bibnamefont {Steuernagel}},
  \bibinfo {author} {\bibfnamefont {C.-M.}\ \bibnamefont {Wu}}, \ and\ \bibinfo
  {author} {\bibfnamefont {R.-K.}\ \bibnamefont {Lee}},\ }\href {\doibase
  10.1103/PhysRevA.110.023703} {\bibfield  {journal} {\bibinfo  {journal}
  {Physical Review A}\ }\textbf {\bibinfo {volume} {110}},\ \bibinfo {pages}
  {023703} (\bibinfo {year} {2024})}\BibitemShut {NoStop}%
\bibitem [{\citenamefont {Lund}\ \emph {et~al.}(2004)\citenamefont {Lund},
  \citenamefont {Jeong}, \citenamefont {Ralph},\ and\ \citenamefont
  {Kim}}]{Lund2004}%
  \BibitemOpen
  \bibfield  {author} {\bibinfo {author} {\bibfnamefont {A.~P.}\ \bibnamefont
  {Lund}}, \bibinfo {author} {\bibfnamefont {H.}~\bibnamefont {Jeong}},
  \bibinfo {author} {\bibfnamefont {T.~C.}\ \bibnamefont {Ralph}}, \ and\
  \bibinfo {author} {\bibfnamefont {M.~S.}\ \bibnamefont {Kim}},\ }\href
  {\doibase 10.1103/PhysRevA.70.020101} {\bibfield  {journal} {\bibinfo
  {journal} {Physical Review A}\ }\textbf {\bibinfo {volume} {70}},\ \bibinfo
  {pages} {020101} (\bibinfo {year} {2004})}\BibitemShut {NoStop}%
\bibitem [{\citenamefont {Adam}\ \emph {et~al.}(2010)\citenamefont {Adam},
  \citenamefont {Kiss}, \citenamefont {Dar{\'a}zs},\ and\ \citenamefont
  {Jex}}]{Adam2010}%
  \BibitemOpen
  \bibfield  {author} {\bibinfo {author} {\bibfnamefont {P.}~\bibnamefont
  {Adam}}, \bibinfo {author} {\bibfnamefont {T.}~\bibnamefont {Kiss}}, \bibinfo
  {author} {\bibfnamefont {Z.}~\bibnamefont {Dar{\'a}zs}}, \ and\ \bibinfo
  {author} {\bibfnamefont {I.}~\bibnamefont {Jex}},\ }\href {\doibase
  10.1088/0031-8949/2010/T140/014011} {\bibfield  {journal} {\bibinfo
  {journal} {Physica Scripta}\ }\textbf {\bibinfo {volume} {T140}},\ \bibinfo
  {pages} {014011} (\bibinfo {year} {2010})}\BibitemShut {NoStop}%
\bibitem [{\citenamefont {Etesse}\ \emph {et~al.}(2015)\citenamefont {Etesse},
  \citenamefont {Bouillard}, \citenamefont {Kanseri},\ and\ \citenamefont
  {{Tualle-Brouri}}}]{etesseExperimentalGenerationSqueezed2015}%
  \BibitemOpen
  \bibfield  {author} {\bibinfo {author} {\bibfnamefont {J.}~\bibnamefont
  {Etesse}}, \bibinfo {author} {\bibfnamefont {M.}~\bibnamefont {Bouillard}},
  \bibinfo {author} {\bibfnamefont {B.}~\bibnamefont {Kanseri}}, \ and\
  \bibinfo {author} {\bibfnamefont {R.}~\bibnamefont {{Tualle-Brouri}}},\
  }\href {\doibase 10.1103/PhysRevLett.114.193602} {\bibfield  {journal}
  {\bibinfo  {journal} {Physical Review Letters}\ }\textbf {\bibinfo {volume}
  {114}},\ \bibinfo {pages} {193602} (\bibinfo {year} {2015})}\BibitemShut
  {NoStop}%
\bibitem [{\citenamefont {Sychev}\ \emph {et~al.}(2017)\citenamefont {Sychev},
  \citenamefont {Ulanov}, \citenamefont {Pushkina}, \citenamefont {Richards},
  \citenamefont {Fedorov},\ and\ \citenamefont {Lvovsky}}]{Sychev2017}%
  \BibitemOpen
  \bibfield  {author} {\bibinfo {author} {\bibfnamefont {D.~V.}\ \bibnamefont
  {Sychev}}, \bibinfo {author} {\bibfnamefont {A.~E.}\ \bibnamefont {Ulanov}},
  \bibinfo {author} {\bibfnamefont {A.~A.}\ \bibnamefont {Pushkina}}, \bibinfo
  {author} {\bibfnamefont {M.~W.}\ \bibnamefont {Richards}}, \bibinfo {author}
  {\bibfnamefont {I.~A.}\ \bibnamefont {Fedorov}}, \ and\ \bibinfo {author}
  {\bibfnamefont {A.~I.}\ \bibnamefont {Lvovsky}},\ }\href {\doibase
  10.1038/nphoton.2017.57} {\bibfield  {journal} {\bibinfo  {journal} {Nature
  Photonics}\ }\textbf {\bibinfo {volume} {11}},\ \bibinfo {pages} {379}
  (\bibinfo {year} {2017})}\BibitemShut {NoStop}%
\bibitem [{\citenamefont {Sychev}\ \emph {et~al.}(2018)\citenamefont {Sychev},
  \citenamefont {Ulanov}, \citenamefont {Pushkina}, \citenamefont {Fedorov},
  \citenamefont {Richards}, \citenamefont {Grangier},\ and\ \citenamefont
  {Lvovsky}}]{Sychev2018}%
  \BibitemOpen
  \bibfield  {author} {\bibinfo {author} {\bibfnamefont {D.~V.}\ \bibnamefont
  {Sychev}}, \bibinfo {author} {\bibfnamefont {A.~E.}\ \bibnamefont {Ulanov}},
  \bibinfo {author} {\bibfnamefont {A.~A.}\ \bibnamefont {Pushkina}}, \bibinfo
  {author} {\bibfnamefont {I.~A.}\ \bibnamefont {Fedorov}}, \bibinfo {author}
  {\bibfnamefont {M.~W.}\ \bibnamefont {Richards}}, \bibinfo {author}
  {\bibfnamefont {P.}~\bibnamefont {Grangier}}, \ and\ \bibinfo {author}
  {\bibfnamefont {A.~I.}\ \bibnamefont {Lvovsky}},\ }in\ \href {\doibase
  10.1063/1.5025456} {\emph {\bibinfo {booktitle} {{{FOURTH INTERNATIONAL
  CONFERENCE ON QUANTUM TECHNOLOGIES}} ({{ICQT-2017}})}}}\ (\bibinfo {address}
  {Moscow, Russia},\ \bibinfo {year} {2018})\ p.\ \bibinfo {pages}
  {020018}\BibitemShut {NoStop}%
\bibitem [{\citenamefont {Eaton}\ \emph
  {et~al.}(2019{\natexlab{a}})\citenamefont {Eaton}, \citenamefont {Nehra},\
  and\ \citenamefont {Pfister}}]{Eaton2019}%
  \BibitemOpen
  \bibfield  {author} {\bibinfo {author} {\bibfnamefont {M.}~\bibnamefont
  {Eaton}}, \bibinfo {author} {\bibfnamefont {R.}~\bibnamefont {Nehra}}, \ and\
  \bibinfo {author} {\bibfnamefont {O.}~\bibnamefont {Pfister}},\ }\href
  {\doibase 10.1088/1367-2630/ab5330} {\bibfield  {journal} {\bibinfo
  {journal} {New Journal of Physics}\ }\textbf {\bibinfo {volume} {21}},\
  \bibinfo {pages} {113034} (\bibinfo {year} {2019}{\natexlab{a}})}\BibitemShut
  {NoStop}%
\bibitem [{\citenamefont {Wang}\ and\ \citenamefont {Duan}(2005)}]{Wang2005}%
  \BibitemOpen
  \bibfield  {author} {\bibinfo {author} {\bibfnamefont {B.}~\bibnamefont
  {Wang}}\ and\ \bibinfo {author} {\bibfnamefont {L.-M.}\ \bibnamefont
  {Duan}},\ }\href {\doibase 10.1103/PhysRevA.72.022320} {\bibfield  {journal}
  {\bibinfo  {journal} {Physical Review A}\ }\textbf {\bibinfo {volume} {72}},\
  \bibinfo {pages} {022320} (\bibinfo {year} {2005})}\BibitemShut {NoStop}%
\bibitem [{\citenamefont {Hacker}\ \emph {et~al.}(2019)\citenamefont {Hacker},
  \citenamefont {Welte}, \citenamefont {Daiss}, \citenamefont {Shaukat},
  \citenamefont {Ritter}, \citenamefont {Li},\ and\ \citenamefont
  {Rempe}}]{Hacker2019}%
  \BibitemOpen
  \bibfield  {author} {\bibinfo {author} {\bibfnamefont {B.}~\bibnamefont
  {Hacker}}, \bibinfo {author} {\bibfnamefont {S.}~\bibnamefont {Welte}},
  \bibinfo {author} {\bibfnamefont {S.}~\bibnamefont {Daiss}}, \bibinfo
  {author} {\bibfnamefont {A.}~\bibnamefont {Shaukat}}, \bibinfo {author}
  {\bibfnamefont {S.}~\bibnamefont {Ritter}}, \bibinfo {author} {\bibfnamefont
  {L.}~\bibnamefont {Li}}, \ and\ \bibinfo {author} {\bibfnamefont
  {G.}~\bibnamefont {Rempe}},\ }\href {\doibase 10.1038/s41566-018-0339-5}
  {\bibfield  {journal} {\bibinfo  {journal} {Nature Photonics}\ }\textbf
  {\bibinfo {volume} {13}},\ \bibinfo {pages} {110} (\bibinfo {year}
  {2019})}\BibitemShut {NoStop}%
\bibitem [{\citenamefont {Su}\ \emph {et~al.}(2019)\citenamefont {Su},
  \citenamefont {Myers},\ and\ \citenamefont {Sabapathy}}]{Su:2019vdv}%
  \BibitemOpen
  \bibfield  {author} {\bibinfo {author} {\bibfnamefont {D.}~\bibnamefont
  {Su}}, \bibinfo {author} {\bibfnamefont {C.~R.}\ \bibnamefont {Myers}}, \
  and\ \bibinfo {author} {\bibfnamefont {K.~K.}\ \bibnamefont {Sabapathy}},\
  }\href {\doibase 10.1103/PhysRevA.100.052301} {\bibfield  {journal} {\bibinfo
   {journal} {Phys. Rev. A}\ }\textbf {\bibinfo {volume} {100}},\ \bibinfo
  {pages} {052301} (\bibinfo {year} {2019})},\ \Eprint
  {http://arxiv.org/abs/1902.02323} {arXiv:1902.02323 [quant-ph]} \BibitemShut
  {NoStop}%
\bibitem [{\citenamefont {Nauth}\ \emph {et~al.}(2026)\citenamefont {Nauth},
  \citenamefont {Walk}, \citenamefont {Datta}, \citenamefont {Busch},
  \citenamefont {Eisert}, \citenamefont {Benson},\ and\ \citenamefont
  {K{\"o}gler}}]{Nauth:2026jyy}%
  \BibitemOpen
  \bibfield  {author} {\bibinfo {author} {\bibfnamefont {J.~K.}\ \bibnamefont
  {Nauth}}, \bibinfo {author} {\bibfnamefont {N.}~\bibnamefont {Walk}},
  \bibinfo {author} {\bibfnamefont {A.~M.}\ \bibnamefont {Datta}}, \bibinfo
  {author} {\bibfnamefont {K.}~\bibnamefont {Busch}}, \bibinfo {author}
  {\bibfnamefont {J.}~\bibnamefont {Eisert}}, \bibinfo {author} {\bibfnamefont
  {O.}~\bibnamefont {Benson}}, \ and\ \bibinfo {author} {\bibfnamefont {R.~A.}\
  \bibnamefont {K{\"o}gler}},\ }\href@noop {} {\  (\bibinfo {year} {2026})},\
  \Eprint {http://arxiv.org/abs/2607.02427} {arXiv:2607.02427 [quant-ph]}
  \BibitemShut {NoStop}%
\bibitem [{\citenamefont {Luo}\ and\ \citenamefont
  {Mahmoodian}(2025)}]{luo2025}%
  \BibitemOpen
  \bibfield  {author} {\bibinfo {author} {\bibfnamefont {H.}~\bibnamefont
  {Luo}}\ and\ \bibinfo {author} {\bibfnamefont {S.}~\bibnamefont
  {Mahmoodian}},\ }\href {\doibase 10.1103/t96h-488y} {\bibfield  {journal}
  {\bibinfo  {journal} {Phys. Rev. Applied}\ } (\bibinfo {year} {2025}),\
  10.1103/t96h-488y},\ \Eprint {http://arxiv.org/abs/2412.14798}
  {arXiv:2412.14798} \BibitemShut {NoStop}%
\bibitem [{\citenamefont {Podoshvedov}(2011)}]{Podoshvedov2011}%
  \BibitemOpen
  \bibfield  {author} {\bibinfo {author} {\bibfnamefont {S.~A.}\ \bibnamefont
  {Podoshvedov}},\ }\href {\doibase 10.1134/S1063776111020208} {\bibfield
  {journal} {\bibinfo  {journal} {Journal of Experimental and Theoretical
  Physics}\ }\textbf {\bibinfo {volume} {112}},\ \bibinfo {pages} {551}
  (\bibinfo {year} {2011})}\BibitemShut {NoStop}%
\bibitem [{\citenamefont {Takase}\ \emph {et~al.}(2022)\citenamefont {Takase},
  \citenamefont {Kawasaki}, \citenamefont {Jeong}, \citenamefont {Endo},
  \citenamefont {Kashiwazaki}, \citenamefont {Kazama}, \citenamefont {Enbutsu},
  \citenamefont {Watanabe}, \citenamefont {Umeki}, \citenamefont {Miki},
  \citenamefont {Terai}, \citenamefont {Yabuno}, \citenamefont {China},
  \citenamefont {Asavanant}, \citenamefont {Yoshikawa},\ and\ \citenamefont
  {Furusawa}}]{takaseGenerationSchrodingerCat2022}%
  \BibitemOpen
  \bibfield  {author} {\bibinfo {author} {\bibfnamefont {K.}~\bibnamefont
  {Takase}}, \bibinfo {author} {\bibfnamefont {A.}~\bibnamefont {Kawasaki}},
  \bibinfo {author} {\bibfnamefont {B.~K.}\ \bibnamefont {Jeong}}, \bibinfo
  {author} {\bibfnamefont {M.}~\bibnamefont {Endo}}, \bibinfo {author}
  {\bibfnamefont {T.}~\bibnamefont {Kashiwazaki}}, \bibinfo {author}
  {\bibfnamefont {T.}~\bibnamefont {Kazama}}, \bibinfo {author} {\bibfnamefont
  {K.}~\bibnamefont {Enbutsu}}, \bibinfo {author} {\bibfnamefont
  {K.}~\bibnamefont {Watanabe}}, \bibinfo {author} {\bibfnamefont
  {T.}~\bibnamefont {Umeki}}, \bibinfo {author} {\bibfnamefont
  {S.}~\bibnamefont {Miki}}, \bibinfo {author} {\bibfnamefont {H.}~\bibnamefont
  {Terai}}, \bibinfo {author} {\bibfnamefont {M.}~\bibnamefont {Yabuno}},
  \bibinfo {author} {\bibfnamefont {F.}~\bibnamefont {China}}, \bibinfo
  {author} {\bibfnamefont {W.}~\bibnamefont {Asavanant}}, \bibinfo {author}
  {\bibfnamefont {J.-i.}\ \bibnamefont {Yoshikawa}}, \ and\ \bibinfo {author}
  {\bibfnamefont {A.}~\bibnamefont {Furusawa}},\ }\href {\doibase
  10.1364/OE.454123} {\bibfield  {journal} {\bibinfo  {journal} {Optics
  Express}\ }\textbf {\bibinfo {volume} {30}},\ \bibinfo {pages} {14161}
  (\bibinfo {year} {2022})}\BibitemShut {NoStop}%
\bibitem [{\citenamefont {Winnel}\ \emph {et~al.}(2023)\citenamefont {Winnel},
  \citenamefont {Guanzon}, \citenamefont {Singh},\ and\ \citenamefont
  {Ralph}}]{Winnel:2023tzb}%
  \BibitemOpen
  \bibfield  {author} {\bibinfo {author} {\bibfnamefont {M.~S.}\ \bibnamefont
  {Winnel}}, \bibinfo {author} {\bibfnamefont {J.~J.}\ \bibnamefont {Guanzon}},
  \bibinfo {author} {\bibfnamefont {D.}~\bibnamefont {Singh}}, \ and\ \bibinfo
  {author} {\bibfnamefont {T.~C.}\ \bibnamefont {Ralph}},\ }\href@noop {} {\
  (\bibinfo {year} {2023})},\ \Eprint {http://arxiv.org/abs/2311.10510}
  {arXiv:2311.10510 [quant-ph]} \BibitemShut {NoStop}%
\bibitem [{\citenamefont {Crescimanna}\ \emph {et~al.}(2025)\citenamefont
  {Crescimanna}, \citenamefont {Yu}, \citenamefont {Heshami},\ and\
  \citenamefont {Patel}}]{Crescimanna:2025lwf}%
  \BibitemOpen
  \bibfield  {author} {\bibinfo {author} {\bibfnamefont {V.}~\bibnamefont
  {Crescimanna}}, \bibinfo {author} {\bibfnamefont {S.}~\bibnamefont {Yu}},
  \bibinfo {author} {\bibfnamefont {K.}~\bibnamefont {Heshami}}, \ and\
  \bibinfo {author} {\bibfnamefont {R.~B.}\ \bibnamefont {Patel}},\ }\href
  {\doibase 10.1103/jhkz-84dz} {\bibfield  {journal} {\bibinfo  {journal}
  {Phys. Rev. A}\ }\textbf {\bibinfo {volume} {112}},\ \bibinfo {pages}
  {053705} (\bibinfo {year} {2025})},\ \Eprint
  {http://arxiv.org/abs/2502.14967} {arXiv:2502.14967 [quant-ph]} \BibitemShut
  {NoStop}%
\bibitem [{\citenamefont {Ismailzadeh}\ and\ \citenamefont
  {Ravan}(2026)}]{Ismailzadeh:2026yye}%
  \BibitemOpen
  \bibfield  {author} {\bibinfo {author} {\bibfnamefont {S.}~\bibnamefont
  {Ismailzadeh}}\ and\ \bibinfo {author} {\bibfnamefont {B.~A.}\ \bibnamefont
  {Ravan}},\ }\href {\doibase 10.1103/zljk-9wfx} {\bibfield  {journal}
  {\bibinfo  {journal} {Phys. Rev. A}\ }\textbf {\bibinfo {volume} {114}},\
  \bibinfo {pages} {043511} (\bibinfo {year} {2026})},\ \Eprint
  {http://arxiv.org/abs/2603.18303} {arXiv:2603.18303 [quant-ph]} \BibitemShut
  {NoStop}%
\bibitem [{\citenamefont {Campos}\ \emph {et~al.}(1989)\citenamefont {Campos},
  \citenamefont {Saleh},\ and\ \citenamefont {Teich}}]{Campos1989}%
  \BibitemOpen
  \bibfield  {author} {\bibinfo {author} {\bibfnamefont {R.~A.}\ \bibnamefont
  {Campos}}, \bibinfo {author} {\bibfnamefont {B.~E.~A.}\ \bibnamefont
  {Saleh}}, \ and\ \bibinfo {author} {\bibfnamefont {M.~C.}\ \bibnamefont
  {Teich}},\ }\href {\doibase 10.1103/physreva.40.1371} {\bibfield  {journal}
  {\bibinfo  {journal} {Physical Review A}\ }\textbf {\bibinfo {volume} {40}},\
  \bibinfo {pages} {1371} (\bibinfo {year} {1989})}\BibitemShut {NoStop}%
\bibitem [{\citenamefont {Zimmermann}\ \emph {et~al.}(2026)\citenamefont
  {Zimmermann}, \citenamefont {Steinlechenr},\ and\ \citenamefont
  {Sondenheimer}}]{Zimmermann-InPreparation}%
  \BibitemOpen
  \bibfield  {author} {\bibinfo {author} {\bibfnamefont {G.~O.}\ \bibnamefont
  {Zimmermann}}, \bibinfo {author} {\bibfnamefont {F.}~\bibnamefont
  {Steinlechenr}}, \ and\ \bibinfo {author} {\bibfnamefont {R.}~\bibnamefont
  {Sondenheimer}},\ }\href@noop {} {\bibfield  {journal} {\bibinfo  {journal}
  {In preparation}\ } (\bibinfo {year} {2026})}\BibitemShut {NoStop}%
\bibitem [{\citenamefont {Jing}\ \emph {et~al.}(2011)\citenamefont {Jing},
  \citenamefont {Liu}, \citenamefont {Zhou}, \citenamefont {Ou},\ and\
  \citenamefont {Zhang}}]{Jing2011}%
  \BibitemOpen
  \bibfield  {author} {\bibinfo {author} {\bibfnamefont {J.}~\bibnamefont
  {Jing}}, \bibinfo {author} {\bibfnamefont {C.}~\bibnamefont {Liu}}, \bibinfo
  {author} {\bibfnamefont {Z.}~\bibnamefont {Zhou}}, \bibinfo {author}
  {\bibfnamefont {Z.~Y.}\ \bibnamefont {Ou}}, \ and\ \bibinfo {author}
  {\bibfnamefont {W.}~\bibnamefont {Zhang}},\ }\href {\doibase
  10.1063/1.3606549} {\bibfield  {journal} {\bibinfo  {journal} {Applied
  Physics Letters}\ }\textbf {\bibinfo {volume} {99}},\ \bibinfo {pages}
  {011110} (\bibinfo {year} {2011})}\BibitemShut {NoStop}%
\bibitem [{\citenamefont {Zou}\ \emph {et~al.}(1991)\citenamefont {Zou},
  \citenamefont {Wang},\ and\ \citenamefont {Mandel}}]{zou1991}%
  \BibitemOpen
  \bibfield  {author} {\bibinfo {author} {\bibfnamefont {X.~Y.}\ \bibnamefont
  {Zou}}, \bibinfo {author} {\bibfnamefont {L.~J.}\ \bibnamefont {Wang}}, \
  and\ \bibinfo {author} {\bibfnamefont {L.}~\bibnamefont {Mandel}},\ }\href
  {\doibase 10.1103/PhysRevLett.67.318} {\bibfield  {journal} {\bibinfo
  {journal} {Physical Review Letters}\ }\textbf {\bibinfo {volume} {67}},\
  \bibinfo {pages} {318} (\bibinfo {year} {1991})}\BibitemShut {NoStop}%
\bibitem [{\citenamefont {Burlakov}\ \emph {et~al.}(1998)\citenamefont
  {Burlakov}, \citenamefont {Kulik}, \citenamefont {Penin},\ and\ \citenamefont
  {Chekhova}}]{Burlakov1998}%
  \BibitemOpen
  \bibfield  {author} {\bibinfo {author} {\bibfnamefont {A.~V.}\ \bibnamefont
  {Burlakov}}, \bibinfo {author} {\bibfnamefont {S.~P.}\ \bibnamefont {Kulik}},
  \bibinfo {author} {\bibfnamefont {A.~N.}\ \bibnamefont {Penin}}, \ and\
  \bibinfo {author} {\bibfnamefont {M.~V.}\ \bibnamefont {Chekhova}},\ }\href
  {\doibase 10.1134/1.558576} {\bibfield  {journal} {\bibinfo  {journal}
  {Journal of Experimental and Theoretical Physics}\ }\textbf {\bibinfo
  {volume} {86}},\ \bibinfo {pages} {1090} (\bibinfo {year}
  {1998})}\BibitemShut {NoStop}%
\bibitem [{\citenamefont {Lemos}\ \emph {et~al.}(2014)\citenamefont {Lemos},
  \citenamefont {Borish}, \citenamefont {Cole}, \citenamefont {Ramelow},
  \citenamefont {Lapkiewicz},\ and\ \citenamefont {Zeilinger}}]{lemos2014}%
  \BibitemOpen
  \bibfield  {author} {\bibinfo {author} {\bibfnamefont {G.~B.}\ \bibnamefont
  {Lemos}}, \bibinfo {author} {\bibfnamefont {V.}~\bibnamefont {Borish}},
  \bibinfo {author} {\bibfnamefont {G.~D.}\ \bibnamefont {Cole}}, \bibinfo
  {author} {\bibfnamefont {S.}~\bibnamefont {Ramelow}}, \bibinfo {author}
  {\bibfnamefont {R.}~\bibnamefont {Lapkiewicz}}, \ and\ \bibinfo {author}
  {\bibfnamefont {A.}~\bibnamefont {Zeilinger}},\ }\href {\doibase
  10.1038/nature13586} {\bibfield  {journal} {\bibinfo  {journal} {Nature}\
  }\textbf {\bibinfo {volume} {512}},\ \bibinfo {pages} {409} (\bibinfo {year}
  {2014})}\BibitemShut {NoStop}%
\bibitem [{\citenamefont {Fuenzalida}\ \emph {et~al.}(2022)\citenamefont
  {Fuenzalida}, \citenamefont {Hochrainer}, \citenamefont {Lemos},
  \citenamefont {Ortega}, \citenamefont {Lapkiewicz}, \citenamefont {Lahiri},\
  and\ \citenamefont {Zeilinger}}]{Fuenzalida:2020bgu}%
  \BibitemOpen
  \bibfield  {author} {\bibinfo {author} {\bibfnamefont {J.}~\bibnamefont
  {Fuenzalida}}, \bibinfo {author} {\bibfnamefont {A.}~\bibnamefont
  {Hochrainer}}, \bibinfo {author} {\bibfnamefont {G.~B.}\ \bibnamefont
  {Lemos}}, \bibinfo {author} {\bibfnamefont {E.~A.}\ \bibnamefont {Ortega}},
  \bibinfo {author} {\bibfnamefont {R.}~\bibnamefont {Lapkiewicz}}, \bibinfo
  {author} {\bibfnamefont {M.}~\bibnamefont {Lahiri}}, \ and\ \bibinfo {author}
  {\bibfnamefont {A.}~\bibnamefont {Zeilinger}},\ }\href {\doibase
  10.22331/q-2022-02-09-646} {\bibfield  {journal} {\bibinfo  {journal}
  {Quantum}\ }\textbf {\bibinfo {volume} {6}},\ \bibinfo {pages} {646}
  (\bibinfo {year} {2022})},\ \Eprint {http://arxiv.org/abs/2010.07712}
  {arXiv:2010.07712 [quant-ph]} \BibitemShut {NoStop}%
\bibitem [{\citenamefont {Gilaberte~Basset}\ \emph {et~al.}(2021)\citenamefont
  {Gilaberte~Basset}, \citenamefont {Hochrainer}, \citenamefont {Töpfer},
  \citenamefont {Riexinger}, \citenamefont {Bickert}, \citenamefont
  {León-Torres}, \citenamefont {Steinlechner},\ and\ \citenamefont
  {Gräfe}}]{GilaberteBasset-Video2021}%
  \BibitemOpen
  \bibfield  {author} {\bibinfo {author} {\bibfnamefont {M.}~\bibnamefont
  {Gilaberte~Basset}}, \bibinfo {author} {\bibfnamefont {A.}~\bibnamefont
  {Hochrainer}}, \bibinfo {author} {\bibfnamefont {S.}~\bibnamefont {Töpfer}},
  \bibinfo {author} {\bibfnamefont {F.}~\bibnamefont {Riexinger}}, \bibinfo
  {author} {\bibfnamefont {P.}~\bibnamefont {Bickert}}, \bibinfo {author}
  {\bibfnamefont {J.~R.}\ \bibnamefont {León-Torres}}, \bibinfo {author}
  {\bibfnamefont {F.}~\bibnamefont {Steinlechner}}, \ and\ \bibinfo {author}
  {\bibfnamefont {M.}~\bibnamefont {Gräfe}},\ }\href {\doibase
  https://doi.org/10.1002/lpor.202000327} {\bibfield  {journal} {\bibinfo
  {journal} {Laser \& Photonics Reviews}\ }\textbf {\bibinfo {volume} {15}},\
  \bibinfo {pages} {2000327} (\bibinfo {year} {2021})},\ \Eprint
  {http://arxiv.org/abs/https://onlinelibrary.wiley.com/doi/pdf/10.1002/lpor.202000327}
  {https://onlinelibrary.wiley.com/doi/pdf/10.1002/lpor.202000327} \BibitemShut
  {NoStop}%
\bibitem [{\citenamefont {Fuenzalida}\ \emph {et~al.}(2023)\citenamefont
  {Fuenzalida}, \citenamefont {Basset}, \citenamefont {T{\"o}pfer},
  \citenamefont {Torres},\ and\ \citenamefont
  {Gr{\"a}fe}}]{Fuenzalida:2023anr}%
  \BibitemOpen
  \bibfield  {author} {\bibinfo {author} {\bibfnamefont {J.}~\bibnamefont
  {Fuenzalida}}, \bibinfo {author} {\bibfnamefont {M.~G.}\ \bibnamefont
  {Basset}}, \bibinfo {author} {\bibfnamefont {S.}~\bibnamefont {T{\"o}pfer}},
  \bibinfo {author} {\bibfnamefont {J.~P.}\ \bibnamefont {Torres}}, \ and\
  \bibinfo {author} {\bibfnamefont {M.}~\bibnamefont {Gr{\"a}fe}},\ }\href
  {\doibase 10.1126/sciadv.adg9573} {\bibfield  {journal} {\bibinfo  {journal}
  {Sci. Adv.}\ }\textbf {\bibinfo {volume} {9}},\ \bibinfo {pages} {adg9573}
  (\bibinfo {year} {2023})},\ \Eprint {http://arxiv.org/abs/2301.02529}
  {arXiv:2301.02529 [quant-ph]} \BibitemShut {NoStop}%
\bibitem [{\citenamefont {Basset}\ \emph {et~al.}(2023)\citenamefont {Basset},
  \citenamefont {Sondenheimer}, \citenamefont {Fuenzalida}, \citenamefont
  {Vega}, \citenamefont {T{\"o}pfer}, \citenamefont {Santos}, \citenamefont
  {Saravi}, \citenamefont {Setzpfandt}, \citenamefont {Steinlechner},\ and\
  \citenamefont {Gr{\"a}fe}}]{Basset:2023hlg}%
  \BibitemOpen
  \bibfield  {author} {\bibinfo {author} {\bibfnamefont {M.~G.}\ \bibnamefont
  {Basset}}, \bibinfo {author} {\bibfnamefont {R.}~\bibnamefont
  {Sondenheimer}}, \bibinfo {author} {\bibfnamefont {J.}~\bibnamefont
  {Fuenzalida}}, \bibinfo {author} {\bibfnamefont {A.}~\bibnamefont {Vega}},
  \bibinfo {author} {\bibfnamefont {S.}~\bibnamefont {T{\"o}pfer}}, \bibinfo
  {author} {\bibfnamefont {E.~A.}\ \bibnamefont {Santos}}, \bibinfo {author}
  {\bibfnamefont {S.}~\bibnamefont {Saravi}}, \bibinfo {author} {\bibfnamefont
  {F.}~\bibnamefont {Setzpfandt}}, \bibinfo {author} {\bibfnamefont
  {F.}~\bibnamefont {Steinlechner}}, \ and\ \bibinfo {author} {\bibfnamefont
  {M.}~\bibnamefont {Gr{\"a}fe}},\ }\href {\doibase
  10.1103/PhysRevA.108.052610} {\bibfield  {journal} {\bibinfo  {journal}
  {Phys. Rev. A}\ }\textbf {\bibinfo {volume} {108}},\ \bibinfo {pages}
  {052610} (\bibinfo {year} {2023})},\ \Eprint
  {http://arxiv.org/abs/2306.01884} {arXiv:2306.01884 [quant-ph]} \BibitemShut
  {NoStop}%
\bibitem [{\citenamefont {Machado}\ \emph {et~al.}(2026)\citenamefont {Machado}
  \emph {et~al.}}]{Machado:2026aip}%
  \BibitemOpen
  \bibfield  {author} {\bibinfo {author} {\bibfnamefont {G.~J.}\ \bibnamefont
  {Machado}} \emph {et~al.},\ }\href@noop {} {\  (\bibinfo {year} {2026})},\
  \Eprint {http://arxiv.org/abs/2606.30761} {arXiv:2606.30761 [quant-ph]}
  \BibitemShut {NoStop}%
\bibitem [{\citenamefont {Chekhova}\ and\ \citenamefont
  {Ou}(2016)}]{Chekhova2016}%
  \BibitemOpen
  \bibfield  {author} {\bibinfo {author} {\bibfnamefont {M.~V.}\ \bibnamefont
  {Chekhova}}\ and\ \bibinfo {author} {\bibfnamefont {Z.~Y.}\ \bibnamefont
  {Ou}},\ }\href {\doibase 10.1364/AOP.8.000104} {\bibfield  {journal}
  {\bibinfo  {journal} {Advances in Optics and Photonics}\ }\textbf {\bibinfo
  {volume} {8}},\ \bibinfo {pages} {104} (\bibinfo {year} {2016})}\BibitemShut
  {NoStop}%
\bibitem [{\citenamefont {Steuernagel}\ \emph {et~al.}(2025)\citenamefont
  {Steuernagel}, \citenamefont {Hu},\ and\ \citenamefont
  {Lee}}]{Steuernagel:2024seq}%
  \BibitemOpen
  \bibfield  {author} {\bibinfo {author} {\bibfnamefont {O.}~\bibnamefont
  {Steuernagel}}, \bibinfo {author} {\bibfnamefont {S.-H.}\ \bibnamefont {Hu}},
  \ and\ \bibinfo {author} {\bibfnamefont {R.-K.}\ \bibnamefont {Lee}},\ }\href
  {\doibase 10.1063/5.0246230} {\bibfield  {journal} {\bibinfo  {journal} {APL
  Quantum}\ }\textbf {\bibinfo {volume} {2}},\ \bibinfo {pages} {016117}
  (\bibinfo {year} {2025})},\ \Eprint {http://arxiv.org/abs/2410.21907}
  {arXiv:2410.21907 [quant-ph]} \BibitemShut {NoStop}%
\bibitem [{\citenamefont {Endo}\ \emph {et~al.}(2026)\citenamefont {Endo},
  \citenamefont {Takase}, \citenamefont {Nomura}, \citenamefont {Sonoyama},
  \citenamefont {Takahashi}, \citenamefont {Takasu}, \citenamefont {Fukuda},
  \citenamefont {Kashiwazaki}, \citenamefont {Inoue}, \citenamefont {Umeki},
  \citenamefont {van Loock}, \citenamefont {Marek}, \citenamefont {Filip},
  \citenamefont {Asavanant},\ and\ \citenamefont {Furusawa}}]{endo2026}%
  \BibitemOpen
  \bibfield  {author} {\bibinfo {author} {\bibfnamefont {M.}~\bibnamefont
  {Endo}}, \bibinfo {author} {\bibfnamefont {K.}~\bibnamefont {Takase}},
  \bibinfo {author} {\bibfnamefont {T.}~\bibnamefont {Nomura}}, \bibinfo
  {author} {\bibfnamefont {T.}~\bibnamefont {Sonoyama}}, \bibinfo {author}
  {\bibfnamefont {K.}~\bibnamefont {Takahashi}}, \bibinfo {author}
  {\bibfnamefont {S.}~\bibnamefont {Takasu}}, \bibinfo {author} {\bibfnamefont
  {D.}~\bibnamefont {Fukuda}}, \bibinfo {author} {\bibfnamefont
  {T.}~\bibnamefont {Kashiwazaki}}, \bibinfo {author} {\bibfnamefont
  {A.}~\bibnamefont {Inoue}}, \bibinfo {author} {\bibfnamefont
  {T.}~\bibnamefont {Umeki}}, \bibinfo {author} {\bibfnamefont
  {P.}~\bibnamefont {van Loock}}, \bibinfo {author} {\bibfnamefont
  {P.}~\bibnamefont {Marek}}, \bibinfo {author} {\bibfnamefont
  {R.}~\bibnamefont {Filip}}, \bibinfo {author} {\bibfnamefont
  {W.}~\bibnamefont {Asavanant}}, \ and\ \bibinfo {author} {\bibfnamefont
  {A.}~\bibnamefont {Furusawa}},\ }\href@noop {} {\  (\bibinfo {year}
  {2026})},\ \bibinfo {note} {preprint},\ \Eprint
  {http://arxiv.org/abs/2606.24002} {arXiv:2606.24002} \BibitemShut {NoStop}%
\bibitem [{\citenamefont {Caron}\ \emph {et~al.}(2026)\citenamefont {Caron},
  \citenamefont {Simon}, \citenamefont {Basset}, \citenamefont {Journet},\ and\
  \citenamefont {Tualle-Brouri}}]{caron2026}%
  \BibitemOpen
  \bibfield  {author} {\bibinfo {author} {\bibfnamefont {L.}~\bibnamefont
  {Caron}}, \bibinfo {author} {\bibfnamefont {H.}~\bibnamefont {Simon}},
  \bibinfo {author} {\bibfnamefont {H.}~\bibnamefont {Basset}}, \bibinfo
  {author} {\bibfnamefont {R.}~\bibnamefont {Journet}}, \ and\ \bibinfo
  {author} {\bibfnamefont {R.}~\bibnamefont {Tualle-Brouri}},\ }\href {\doibase
  10.1103/q2h3-58pz} {\bibfield  {journal} {\bibinfo  {journal} {Phys. Rev. X}\
  } (\bibinfo {year} {2026}),\ 10.1103/q2h3-58pz},\ \Eprint
  {http://arxiv.org/abs/2601.09672} {arXiv:2601.09672} \BibitemShut {NoStop}%
\bibitem [{\citenamefont {Gottesman}\ \emph {et~al.}(2001)\citenamefont
  {Gottesman}, \citenamefont {Kitaev},\ and\ \citenamefont
  {Preskill}}]{Gottesman:2000di}%
  \BibitemOpen
  \bibfield  {author} {\bibinfo {author} {\bibfnamefont {D.}~\bibnamefont
  {Gottesman}}, \bibinfo {author} {\bibfnamefont {A.}~\bibnamefont {Kitaev}}, \
  and\ \bibinfo {author} {\bibfnamefont {J.}~\bibnamefont {Preskill}},\ }\href
  {\doibase 10.1103/PhysRevA.64.012310} {\bibfield  {journal} {\bibinfo
  {journal} {Phys. Rev. A}\ }\textbf {\bibinfo {volume} {64}},\ \bibinfo
  {pages} {012310} (\bibinfo {year} {2001})},\ \Eprint
  {http://arxiv.org/abs/quant-ph/0008040} {arXiv:quant-ph/0008040} \BibitemShut
  {NoStop}%
\bibitem [{\citenamefont {Eaton}\ \emph
  {et~al.}(2019{\natexlab{b}})\citenamefont {Eaton}, \citenamefont {Nehra},\
  and\ \citenamefont {Pfister}}]{Eaton:2019eyl}%
  \BibitemOpen
  \bibfield  {author} {\bibinfo {author} {\bibfnamefont {M.}~\bibnamefont
  {Eaton}}, \bibinfo {author} {\bibfnamefont {R.}~\bibnamefont {Nehra}}, \ and\
  \bibinfo {author} {\bibfnamefont {O.}~\bibnamefont {Pfister}},\ }\href
  {\doibase 10.1088/1367-2630/ab5330} {\bibfield  {journal} {\bibinfo
  {journal} {New J. Phys.}\ }\textbf {\bibinfo {volume} {21}},\ \bibinfo
  {pages} {113034} (\bibinfo {year} {2019}{\natexlab{b}})},\ \Eprint
  {http://arxiv.org/abs/1903.01925} {arXiv:1903.01925 [quant-ph]} \BibitemShut
  {NoStop}%
\bibitem [{\citenamefont {Takase}\ \emph {et~al.}(2024)\citenamefont {Takase},
  \citenamefont {Hanamura}, \citenamefont {Nagayoshi}, \citenamefont
  {Bourassa}, \citenamefont {Alexander}, \citenamefont {Kawasaki},
  \citenamefont {Asavanant}, \citenamefont {Endo},\ and\ \citenamefont
  {Furusawa}}]{Takase:2024fia}%
  \BibitemOpen
  \bibfield  {author} {\bibinfo {author} {\bibfnamefont {K.}~\bibnamefont
  {Takase}}, \bibinfo {author} {\bibfnamefont {F.}~\bibnamefont {Hanamura}},
  \bibinfo {author} {\bibfnamefont {H.}~\bibnamefont {Nagayoshi}}, \bibinfo
  {author} {\bibfnamefont {J.~E.}\ \bibnamefont {Bourassa}}, \bibinfo {author}
  {\bibfnamefont {R.~N.}\ \bibnamefont {Alexander}}, \bibinfo {author}
  {\bibfnamefont {A.}~\bibnamefont {Kawasaki}}, \bibinfo {author}
  {\bibfnamefont {W.}~\bibnamefont {Asavanant}}, \bibinfo {author}
  {\bibfnamefont {M.}~\bibnamefont {Endo}}, \ and\ \bibinfo {author}
  {\bibfnamefont {A.}~\bibnamefont {Furusawa}},\ }\href {\doibase
  10.1103/PhysRevA.110.012436} {\bibfield  {journal} {\bibinfo  {journal}
  {Phys. Rev. A}\ }\textbf {\bibinfo {volume} {110}},\ \bibinfo {pages}
  {012436} (\bibinfo {year} {2024})},\ \Eprint
  {http://arxiv.org/abs/2401.07287} {arXiv:2401.07287 [quant-ph]} \BibitemShut
  {NoStop}%
\bibitem [{\citenamefont {Michael}\ \emph {et~al.}(2016)\citenamefont
  {Michael}, \citenamefont {Silveri}, \citenamefont {Brierley}, \citenamefont
  {Albert}, \citenamefont {Salmilehto}, \citenamefont {Jiang},\ and\
  \citenamefont {Girvin}}]{Michael:2016eaq}%
  \BibitemOpen
  \bibfield  {author} {\bibinfo {author} {\bibfnamefont {M.~H.}\ \bibnamefont
  {Michael}}, \bibinfo {author} {\bibfnamefont {M.}~\bibnamefont {Silveri}},
  \bibinfo {author} {\bibfnamefont {R.~{\,}.}\ \bibnamefont {Brierley}},
  \bibinfo {author} {\bibfnamefont {V.~V.}\ \bibnamefont {Albert}}, \bibinfo
  {author} {\bibfnamefont {J.}~\bibnamefont {Salmilehto}}, \bibinfo {author}
  {\bibfnamefont {L.}~\bibnamefont {Jiang}}, \ and\ \bibinfo {author}
  {\bibfnamefont {S.~{\,}.}\ \bibnamefont {Girvin}},\ }\href {\doibase
  10.1103/PhysRevX.6.031006} {\bibfield  {journal} {\bibinfo  {journal} {Phys.
  Rev. X}\ }\textbf {\bibinfo {volume} {6}},\ \bibinfo {pages} {031006}
  (\bibinfo {year} {2016})},\ \Eprint {http://arxiv.org/abs/1602.00008}
  {arXiv:1602.00008 [quant-ph]} \BibitemShut {NoStop}%
\bibitem [{\citenamefont {Nehra}\ \emph {et~al.}(2021)\citenamefont {Nehra},
  \citenamefont {Eaton}, \citenamefont {Pfister},\ and\ \citenamefont
  {Marandi}}]{Nehra:2021hxw}%
  \BibitemOpen
  \bibfield  {author} {\bibinfo {author} {\bibfnamefont {R.}~\bibnamefont
  {Nehra}}, \bibinfo {author} {\bibfnamefont {M.}~\bibnamefont {Eaton}},
  \bibinfo {author} {\bibfnamefont {O.}~\bibnamefont {Pfister}}, \ and\
  \bibinfo {author} {\bibfnamefont {A.}~\bibnamefont {Marandi}},\ }\href@noop
  {} {\  (\bibinfo {year} {2021})},\ \Eprint {http://arxiv.org/abs/2105.11035}
  {arXiv:2105.11035 [quant-ph]} \BibitemShut {NoStop}%
\end{thebibliography}%

\end{document}